\documentclass[a4paper,11pt]{article}
\pdfoutput=1 

\usepackage{jheppub} 

\usepackage[T1]{fontenc} 
\newcommand{\be}{\begin{eqnarray}}
\newcommand{\ee}{\end{eqnarray}}

\def\nue{{\nu_e}}

\def\numu{{\nu_{\mu}}}

\def\nutau{{\nu_{\tau}}}

\def\gsim{\:\raisebox{-0.5ex}{$\stackrel{\textstyle>}{\sim}$}\:}
\newcommand{\ms}{\Delta m^2_{21}}
\newcommand{\ma}{\Delta m^2_{31}}

\def\gs{\mathrel{
   \rlap{\raise 0.511ex \hbox{$>$}}{\lower 0.511ex \hbox{$\sim$}}}}
\def\ls{\mathrel{
   \rlap{\raise 0.511ex \hbox{$<$}}{\lower 0.511ex \hbox{$\sim$}}}}
\newcommand{\bea}{\begin{equation} \begin{array}{c}}
\newcommand{\bead}{\begin{equation} \begin{array}{cccc}}
\newcommand{\eea}{ \end{array} \end{equation}}

\title{\boldmath Invisible neutrino  decay : First vs  second oscillation maximum
}

\author[a]{Kaustav Chakraborty}
\author[b]{Debajyoti Dutta}
\author[a]{Srubabati Goswami}
\author[a,c]{Dipyaman Pramanik}

\affiliation[a]{Physical Research Laboratory, Navrangpura,Ahmedabad-380009, India}
\affiliation[b]{Assam Don Bosco University, Tapesia Campus, Sonapur, Assam, 782402, India}
\affiliation[c]{Instituto de Física Gleb Wataghin - UNICAMP, 13083-859, Campinas, São Paulo, Brazil}

\emailAdd{kaustav@prl.res.in}
\emailAdd{sruba@prl.res.in}
\emailAdd{debajyoti.dutta@dbuniversity.ac.in}
\emailAdd{dipyaman@unicamp.br}

\abstract{We study the physics potential of the long-baseline experiments T2HK, T2HKK and ESS$\nu$SB in the context of invisible neutrino decay. 
We consider normal mass ordering and assume that the state $\nu_{3}$ as unstable, decaying into  sterile states during the flight and obtain constraints on the neutrino decay lifetime ($\tau_3$). 
 We find that T2HK, T2HKK and ESS$\nu$SB are sensitive to the decay-rate of
  $\nu_{3}$ for $\tau_{3}/m_{3} \leq 2.72\times10^{-11}$s/eV, $\tau_{3}/m_{3} \leq 4.36\times10^{-11}$s/eV and $\tau_{3}/m_{3} \leq 2.43\times10^{-11}$s/eV respectively at 3$\sigma$ C.L.  We compare and contrast the sensitivities of the three experiments and   specially investigate  the  role played by the mixing angle $\theta_{23}$. 
 It is seen that for  experiments with flux peak near the  second oscillation maxima,  the  poorer  sensitivity to $\theta_{23}$ results in weaker constraints on the decay lifetime.   
 Although, T2HKK  has one detector close to the second oscillation maxima,  having another detector at the first oscillation maxima  results in superior sensitivity to decay.
  In addition, we find a  synergy between the two baselines of the T2HKK experiment which  helps in giving a better sensitivity for $\theta_{23}$ in the  higher octant.
 We discuss the octant sensitivity in presence of decay and show that there is an enhancement in sensitivity which occurs due to the contribution from the survival probability $P_{\mu\mu}$ which is more pronounced for the experiments at the second oscillation maxima.
  We also obtain 
 the combined sensitivity of T2HK+ESS$\nu$SB and T2HKK+ESS$\nu$SB as $\tau_{3}/m_{3} \leq 4.36\times10^{-11}$s/eV and $\tau_{3}/m_{3} \leq 5.53\times10^{-11}$s/eV respectively at 3$\sigma$ C.L.  
 }

\begin{document}
\maketitle
\flushbottom
\section{Introduction}
Neutrinos are one of the most fascinating particles in nature. Observation of oscillation of neutrinos signifies evidence of neutrino mass. The very existence of neutrino mass is the reason of neutrinos being so special since in the Standard Model (SM)  of particle physics there is no explanation of neutrino mass. 
 The standard formalism of neutrino oscillation requires three mixing angles $\theta_{12}$, $\theta_{13}$ and $\theta_{23}$, two mass-squared differences $\ms$ and $\ma$ and one CP phase $\delta_{CP}$. The parameters $\theta_{12}$, $\theta_{13}$, $\ms$ and the absolute value of $\ma$ have been measured very precisely by oscillation experiments. However, the octant of $\theta_{23}$, the sign of $\ma$ (mass-ordering) and $\delta_{CP}$ are not yet determined with certainty. Global analysis results \cite{Esteban:2020cvm, deSalas:2020pgw, Capozzi:2017ipn} indicate towards normal hierarchy of $\ma$ and $\theta_{23}$ in the higher octant. The global analysis also disfavours $\delta_{CP} = \frac{\pi}{2}$ with more than 3$\sigma$ CL. 
 
 Many experiments are planned for the near future with the goal of determining these quantities. Some examples of future experiments are DUNE \cite{Acciarri:2015uup,Acciarri:2016crz,Acciarri:2016ooe,Abi:2018dnh,Strait:2016mof}, T2HK/T2HKK \cite{Abe:2015zbg, Abe:2016ero}, ESS$\nu$SB \cite{Baussan:2013zcy}, JUNO \cite{An:2015jdp}, INO \cite{Kumar:2017sdq}, PINGU\cite{Aartsen:2014oha}, KM3Net-ORCA\cite{Margiotta:2014gza} etc. It is a well known fact that if there is a new physics beyond the SM, then that can result in  modification of the oscillation probabilities in these experiments. So, these experiments can be sensitive to the new physics. Also, the new physics can affect the sensitivity of these experiments to the standard parameters. Invisible neutrino decay during the flight is one such new physics idea.

The idea of neutrino decay was first proposed to explain the solar neutrino problem in the very early days \cite{Bahcall:1972my}. Later, neutrino oscillation with decay solutions were studied as an explanation of the depletion of solar neutrinos \cite{Acker:1993sz,Berezhiani:1991vk,Berezhiani:1992xg,Choubey:2000an,Bandyopadhyay:2001ct,Joshipura:2002fb,Bandyopadhyay:2002qg,Picoreti:2015ika}. These assumed $\nu_{2}$  to be the unstable state and were able to put bound on the lifetime of $\nu_{2}$. 
The bound from the solar neutrino data is $\tau_{2}/m_{2} > 8.5\times10^{-7}$ s/eV \cite{Bandyopadhyay:2002qg}. A recent study analysed low energy solar neutrino data and put bounds on both $\tau_{2}$ and $\tau_{1}$ \cite{Berryman:2014qha}. Supernova neutrinos also give bounds on $\tau_{1}$ and $\tau_{2}$. SN1987A data gives the bound of $\tau/m > 10^{5}$ s/eV \cite{Frieman:1987as}. $\tau_{1}$ and $\tau_{2}$ can also be constrained from high resolution multi-ton Xenon detector. Recently, in ref. \cite{Huang:2018nxj}, the authors show that such a detector can give very strong bounds $\tau_1/m_{1} \gsim 3\times10^{-2}$ s/eV and $\tau_{2}/m_{2} \gsim 8\times10^{-3}$ s/eV at 2$\sigma$ level using solar neutrinos.

Atmospheric and long-baseline experiments give bounds on the $\nu_{3}$ lifetime. A neutrino decay solution (without any oscillation) was proposed in ref. \cite{LoSecco:1998cd} to explain the atmospheric neutrino problem but this solution  fitted the data poorly. Ref.~\cite{Barger:1998xk, Lipari:1999vh} considered neutrino decay and neutrino mixing together. 
 This was successful to reproduce the $L/E$ distribution of the Super-Kamiokande (SK) data. However, when zenith angle dependence was used instead of L/E distribution, this model was found to give poorer fit to the SK data \cite{Fogli:1999qt}. Ref.~\cite{Barger:1998xk,Fogli:1999qt} assumed $\Delta m^{2} > 0.1$ eV$^{2}$ to comply with the K-decay bounds \cite{Barger:1998xk}. Therefore $\Delta m^{2}$ dependent terms were averaged out. These constraints can be relaxed if the unstable state decays to some invisible state with which it has no mixing. There are two scenarios in the literature. The first kept $\Delta m^{2}$ unconstrained and it explicitly appeared in the probabilities \cite{Choubey:1999ir}. This fits the SK data with a best-fit value of non-zero decay parameter and $\Delta m^{2} \sim 0.003$ eV$^{2}$. Ref.~\cite{Barger:1999bg} considered $\Delta m^{2} << 10^{-4}$ eV$^{2}$. Here, the probability does not contain $\Delta m^{2}$ explicitly. This was able to fit the SK data. However, independent analysis by SK collaboration showed that this scenario gives a poorer fit to the data than only oscillation \cite{Ashie:2004mr}. Global analysis of atmospheric and long-baseline experiments were performed in ref.  \cite{GonzalezGarcia:2008ru}. Only oscillation gave best-fit to the SK data and the fit for the oscillation plus decay was not bad. But addition of LBL data from MINOS reduced the fit quality. This analysis put bound on $\tau_{3}/m_{3}\geq 2.9\times10^{-10}$ s/eV at the 90 \% C.L. Ref. \cite{Gomes:2014yua} studied the oscillation plus decay scenario with unconstrained $\Delta m^{2}$ for MINOS and T2K data and it found $\tau_{3}/m_{3} > 2.8\times10^{-12}$ s/eV at 90 \% C.L. Most of these analyses are done using two generation approximation without matter effect. Authors of ref. \cite{Choubey:2018cfz} performed a complete three generation study of the oscillation plus decay scenario assuming matter effect for the NOvA and T2K preliminary data. This study put a bound on the lifetime on the $\tau_{3}/m_{3}$ as $\tau_{3}/m_{3}\geq1.50\times10^{-12}$ s/eV at 3$\sigma$. Recently, ref. \cite{Denton:2018aml} addressed the IceCUBE track and cascade tension using invisible neutrino decay. They showed that an unstable neutrino with $\tau/m = 100$ s/eV solves the track vs cascade tension.

There have been many studies in literature which discuss the potential of future experiments to the invisible decay. Medium baseline reactor neutrino experiment like JUNO can give a bound on  $\tau_{3}/m_{3} > 7.5\times10^{-11}$ s/eV (95 \% C.L.) \cite{Abrahao:2015rba}. Decay of ultrahigh energy astrophysical neutrinos can give constraints on decay \cite{Beacom:2002vi,Maltoni:2008jr,Pakvasa:2012db,Pagliaroli:2015rca}. Recently, ref \cite{Bustamante:2016ciw} showed that IceCUBE can probe $\tau/m$ upto 10 s/eV for both mass-orderings for 100 TeV neutrinos coming from a source at a distance of 1 Gpc. The ref. \cite{Choubey:2017dyu} studied the invisible neutrino decay in the context of DUNE. It showed that using charged-current electron type and muon type events, DUNE after 5+5 years of running can put a bound $\tau_{3}/m_{3}>4.50\times10^{-11}$ s/eV at 90 \% confidence level. More recently ref. \cite{Ghoshal:2020hyo} performed a multi-channel analysis using charged-current, neutral current and tau-channel events and showed that DUNE's sensitivity would be $\tau_{3}/m_{3} > 5.2\times10^{-11}$ s/eV at 90 \% confidence level. For the potential of MOMENT experiment to probe invisible neutrino decay see ref \cite{Tang:2018rer}. There are also studies on the expected sensitivities from the future atmospheric experiments. See ref. \cite{Choubey:2017eyg,Mohan:2020tbi} for INO and ref. \cite{deSalas:2018kri} for KM3Net-ORCA. 

Invisible neutrino decay can happen for both Dirac and Majorana neutrinos. If neutrinos are Dirac, there can be a coupling between the neutrinos and a light scalar boson \cite{Acker:1991ej,Acker:1993sz}. This gives the decay channel $\nu_{j}\rightarrow\bar{\nu}_{iR}+\chi$, where $\bar{\nu}_{iR}$ is a right-handed singlet and $\chi$ is an iso-singlet scalar. If neutrinos are Majorana particles, neutrino can couple with a Majoron and a sterile neutrino via a pseudo-scalar coupling \cite{Gelmini:1980re, Chikashige:1980ui}. This gives $\nu_{j}\rightarrow\nu_{s}+J$. LEP data on the Z-decay to invisible particles constraints the Majoron to be dominantly singlet \cite{Pakvasa:1999ta}. Neutrinos can also decay to another active state \cite{Kim:1990km,Acker:1992eh,Lindner:2001fx}.This is called the visible neutrino decay scenario. This type of decay can happen in the following way. $\nu_{j}\rightarrow\bar{\nu}_{i}+J$ or $\nu_{j}\rightarrow\nu_{i}+J$. If neutrinos are Majorana, the decay product can be observed in the detector. Ref.~\cite{Coloma:2017zpg,Gago:2017zzy,Ascencio-Sosa:2018lbk} discusses the visible decay in the context of long-baseline experiments. In ref. \cite{Porto-Silva:2020gma}, the authors discuss the visible neutrino decay for reactor experiments KAMLAND and  JUNO. For visible decay of the astrophysical neutrinos at IceCUBE, see ref \cite{Abdullahi:2020rge}.

Invisible neutrino decay can be constrained using Cosmological observations. In ref. \cite{Escudero:2019gfk}, the authors put a bound on neutrino lifetime $\tau_{\nu} > (1.3-0.3)\times10^{9}$ s ($m_{\nu}/0.05$ eV)$^{3}$ at 95\% C.L. using Planck2018 data. 


In this paper we study the  constraints on invisible neutrino decay which can come from future planned/proposed long baseline experiments -- T2HK/T2HKK \cite{Abe:2015zbg,Abe:2016ero} and ESS$\nu$SB \cite{Baussan:2013zcy}.  We perform a full three flavour study using matter effect and obtain the sensitivity to $\tau_3 / m_3$ for these experiments.  The salient feature of the  T2HKK and ESS$\nu$SB experiment is that they are both designed to have energy peak near the second-oscillation maximum.  Since the second oscillation maximum occurs at a lower energy for a particular baseline the effect of decay is expected to be more at the second  oscillation maximum. We examine this aspect and delve into  the detail of  whether the experiments at the second oscillation maximum stand to gain in sensitivity in presence of decay. 
 We also check if the determination of $\theta_{23}$  can get affected if we assume decay in the data, while the fit does not not assume any decay to be present.  In particular, we investigate how the experiments at first and second  oscillation maximum fare in this respect and what are the important factors on which the measurement of $\theta_{23}$ can depend in presence of decay. We also explore how  the octant sensitivity of these experiments change in presence of neutrino decay. 


The paper is organized in the following way. The next section discusses the neutrino oscillation in presence of invisible neutrino decay. In Section \ref{sec:num} we give our experimental and numerical details, in section \ref{sec:res} we present our results and in section \ref{sec:concl}, we finally draw our conclusions.

\section{Neutrino oscillation in presence of invisible decay of neutrino}\label{sec:theory}
In this section we discuss the propagation of neutrinos in presence of invisible neutrino decay. We assume that the $\nu_{3}$ is unstable and it decays into a sterile neutrino and a singlet scalar ($\nu_{3}\rightarrow\bar{\nu}_{4}+J$) with lifetime $\tau_{3}$. In this case we can extend the mass and flavour bases by $\big( \nu_{i} \quad \nu_{4}\big)^{T}$ and $\big(\nu_{\alpha} \quad \nu_{s}\big)^{T}$, where $i=1,2,3$ and $\alpha=\nue,\numu,\nutau$. They are related by the following unitary relation.
\begin{equation}
\begin{pmatrix}
\nu_{\alpha}\\
\nu_{s}
\end{pmatrix}
= 
\begin{pmatrix}
U & 0\\
0 & 1
\end{pmatrix}
\begin{pmatrix}
\nu_{i}\\
\nu_{4}
\end{pmatrix}.
\end{equation}
$U$ is the standard PMNS matrix describing the standard 3 neutrino oscillation. We assume normal hierarchy and $m_{4}$ to be the least massive state. We assume that the decay eigenstates and the mass-eigenstates are same. Under these assumptions, we can write the neutrino evolution in presence of matter in the following way.
\begin{equation}\label{eq:prop}
i\frac{d}{dx}\nu_{f}
= \frac{1}{2E}\big[U\tilde{H}U^{\dagger}+ A\big]\nu_{f}.
\end{equation} 
Where,
\begin{equation}
\tilde{H}=
\begin{pmatrix}
0 & 0 & 0\\
0 & \ms & 0\\
0 & 0 & \ma - i\frac{m_{3}}{\tau_{3}}
\end{pmatrix},
\end{equation}
and
\begin{equation}
A = 2\sqrt{2}G_{F}n_{e}E.
\end{equation}
Here, A is the matter potential, $G_{F}$ is the Fermi constant, $E$ is the energy and $n_{e}$ density of electrons in the earth. We define $\alpha_{3}=m_{3}/\tau_{3}$ as the decay rate of the $\nu_{3}$ state. 
Since we are considering the decay of $\nu_{3}$ only, from here onward, we use $\alpha$ instead of $\alpha_{3}$ for the $\nu_{3}$ state. The probability of getting a neutrino in the flavour state $\nu_{b}$ for the initial flavour state of $\nu_{a}$ is given by
\begin{equation}
P_{a b} = \mid\langle\nu_{b}\mid\nu_{a}\rangle\mid^{2}.
\end{equation} 
Here, a,b denote the flavour states e,$\mu,\tau$.


The effect of decay comes as the $\exp(-\alpha L/E)$ factor in the probability. 
So an experiment is sensitive to the values of $\alpha$ where, $\alpha\sim E/L$. 
It is clear that, smaller values of $\alpha$ and hence a higher sensitivity 
can be achieved for longer baseline 
and for a particular baseline the sensitivity is more for a
lower energy. Thus sensitivity is expected to be more at the 
second oscillation maxima as compared to the first oscillation maxima.  

\section{Experimental and simulation details}\label{sec:num} 

In this section we describe various experiments and the specifications used in our analysis with the summary of the experiments presented in Tab.~\ref{tab:expt-details}. First we give the brief descriptions of the experiments followed by details of our numerical simulations.
\subsection{Experimental details}

\subsubsection{T2HK}
T2HK \cite{Abe:2015zbg} is a proposed upgradation plan of the currently running T2K experiment in Japan. The neutrinos will be generated at Tokai by an upgraded version of the J-PARC beam. Currently J-PARC gives a beam power of 470 KW, but before T2HK becomes operational, the beam power will be increased beyond 1.3 MW. Although near detector of T2HK is yet to be finalized, there are several ideas. Some of these are like upgradation of the current near detector ND280, building a water \v{C}erenkov detector similar to the far detector but in a smaller scale. The far detector will be the upgradation of the currently running Super-Kamiokande (SK) \cite{Fukuda:2002uc} detector. SK is situated at 295 km away from Tokai at the Kamioka village and it is slightly (2.5$^{\circ}$) off-axis to the beam-axis. Thus this will give a narrow beam centered around 0.56 GeV which is at the first oscillation maximum of the neutrino oscillation. The upgraded SK, called Hyper-Kamiokande (HK), will consist of two identical cylindrical tanks placed upright. Each tank will contain pure water which will have a fiducial mass of 187 kt each thus giving a total fiducial mass of 374 kt. The new detector will also have better resolution and efficiency compared to SK. 

\subsubsection{T2HKK}
One of the main aim of the future long-baseline experiments is to measure the CP-violation, and T2HK is no exception. T2HK has a very good sensitivity to the CP-violation if the mass-hierarchy is known. However, the CP-violation sensitivity of T2HK drops if mass-hierarchy is not known beforehand. This is due to a degeneracy between mass-hierarchy and $\delta_{CP}$ \cite{Kajita:2006bt}. In order to solve this problem, there is a proposal to shift one of the water tanks of HK to Korea at about a distance of 1100 km away from the source at Tokai. This proposal is called the T2HKK \cite{Abe:2016ero} (the second K is for Korea.). This will have more matter effect. Thus the mass-hierarchy sensitivity will increase significantly breaking the CP-hierarchy degeneracy. However, the second detector will be at the second oscillation maximum and as the baseline is larger and the flux will also decrease. 

The Korean detector site is not yet final and there are many possibilities \cite{Abe:2016ero}. However, all these sites are in the southern part of the Korean peninsula and lie within a range of 1-3$^{\circ}$ off-axis angle with the J-PARC beam line. 

\subsubsection{ESS$\nu$B}
ESS$\nu$B \cite{Baussan:2013zcy} is another future long-baseline super-beam experiment. The neutrino beam will be generated from the ESS facility at Lund, Sweden. This will have a 2 GeV proton beam with 5 MW beam power. The neutrino beam created by this will have a peak around 0.25 GeV. The far detector will be situated at a distance of about 540 km away from Lund at a mine in Garpenberg. The far detector is proposed to be 500 kt water \v{C}erenkov detector. At 540 km, the second oscillation maximum occurs at 0.35 GeV. Thus the peak of the neutrino beam will be close to the second oscillation maximum for ESS$\nu$B.

\begin{table}[h!]
\begin{center}
\begin{tabular}{|c|c|c|c|}
\hline
Experiment   & Baseline (L in km)  & L/E (in km/GeV) & Fiducial Volume (in kton) \\       
\hline
\hline
T2HK         & 295 km  & 527  & 187 $\times$ 2  \\
\hline
T2HKK         & 295 km; 1100 km  & 527(295 km); 1964(1100 km)  & 187(295 km) + 187(1100 km) \\
\hline
ESS$\nu$SB        & 540 km & 1543    & 500  \\
\hline
\end{tabular}
\end{center}
\caption{{\footnotesize The baselines, L/E and fiducial volumes of each detector for T2HK(L1), T2HK(L2) and ESS$\nu$SB. The energies for T2HK and ESS$\nu$SB are 0.56 GeV and 0.35 GeV respectively.}}
\label{tab:expt-details}
\end{table}

\subsection{Simulation details}

 We have used Global Long-Baseline Experiments Simulator (GLoBES)\cite{globes1, globes2} for simulating the long-baseline experiments. 
We present our results in terms of statistical $\chi^2$, where, 
\begin{eqnarray}
 \chi^2_{{\rm stat}} = 2 \sum_i \lbrace N_i^{{\rm test}}-N_i^{{\rm true}}+N_i^{{\rm true}} \ln \frac{N_i^{{\rm true}}}{N_i^{{\rm test}}} \rbrace, 
 \label{mh-eq}
\end{eqnarray} 
where $(N)_i^{{\rm true}}$ corresponds to the simulated data 
and  $(N)_i^{{\rm test}}$ is the number of events predicted by the theoretical model. 
The effect of systematic errors are included by  the ``pull'' method through the ``pull'' variables $\xi$. 
We have incorporated signal normalization error, background normalization error,
signal ``tilt'' error \& background ``tilt'' error in our analysis. 
Incorporating the errors, the signal and background events can be written as, 
 \begin{equation}
 N^{\rm test}_i = \sum_{s(b)}  N^{s(b)}_i \bigg(1+ {c_i^{s(b)}}^{norm} {\xi^{s(b)}}^{norm} +  {c_i^{s(b)}}^{tilt} { \xi^{s(b)}}^{tilt} \frac{E_i - \bar{E}}{E_{max} - E_{min}}
\bigg)
 \end{equation} 
where, $s(b)$ denotes signal(background). 
$c_i^{norm}$(${c_i}^{tilt}$) denotes the change in number of events by the
 variation of the ``pull'' variable $\xi^{norm}$(${ \xi}^{tilt}$). The signal and background normalization errors are presented in Tab. \ref{norm-uncertain-exp}.
 In the above equation $E_i$ represents the mean reconstructed energy of the $i^{th}$ bin. The maximum and minimum energy of the energy range are $E_{min}$ and $E_{max}$ respectively. 
The mean energy is given by, $\bar{E} = ({E_{max} +E_{min}})/{2}$. 

\begin{table}[h!]
\begin{center}
\begin{tabular}{|c|c|c|c|}
\hline
Channel   & T2HK(295 km) & T2HK(1100 km)  & ESS$\nu$SB  \\       
\hline
\hline
$\nu_e$ appearance         & 3.2\%(5\%)  & 3.8\%(5\%)   & 3.2\%(5\%)   \\
\hline
$\nu_{\bar{e}}$ appearance           & 3.9\%(5\%)  & 4.1\%(5\%)    & 3.9\%(5\%)  \\
\hline
$\nu_\mu$ disappearance         & 3.6\%(5\%)  & 3.8\%(5\%)    & 3.6\%(5\%)    \\
\hline
$\nu_{\bar{\mu}}$ disappearance         & 3.6\%(5\%)  & 3.8\%(5\%)   & 3.6\%(5\%)   \\
\hline
\end{tabular}
\end{center}
\caption{{\footnotesize The signal(background) normalization uncertainties of the experiments for the various channels for T2HK, T2HKK and ESS$\nu$SB.}}
\label{norm-uncertain-exp}
\end{table}

The detector type proposed for T2HK, T2HKK and ESS$\nu$SB are water \v{C}erenkov detectors. T2HK \& T2HKK have been simulated based on \cite{Abe:2016ero} and the ESS$\nu$SB experiment has been simulated according to \cite{Baussan:2013zcy}. The detector of ESS$\nu$SB is based on the MEMPHYS detector \cite{Agostino:2012fd}, which is a water \v{C}erenkov detector. The neutrino (antineutrino) appearance channels for signal events are $\nu_e$ ($\bar{\nu_e}$) and the background events in this channel are neutral current background, intrinsic beam background, mis-identified muons and wrong-sign signal errors. 
The neutrino (antineutrino) disappearance channels are  $\nu_\mu$ ($\bar{\nu_\mu}$), the background channels are neutral current background and wrong-sign muons. The signal and background normalization uncertainties have been considered as presented in Tab. \ref{norm-uncertain-exp}. Additionally, a 5\% (10\%) signal (background) ``tilt'' error have been considered for T2HK \& T2HKK and 0.1\% (0.1\%) in case of ESS$\nu$SB.

\begin{table}[h!]
\begin{center}
\begin{tabular}{|c|c|c|}
\hline
Oscillation parameters & True Values & Marginalization Range \\
\hline \hline
$\theta_{13}$ & $8.6^{\circ}$ & Fixed \\
\hline
$\theta_{12}$ & $33.82^{\circ}$ & Fixed   \\
\hline
$\sin^2 \theta_{23}$ & $0.447$(LO), $0.56$(HO) & $0.413$ ~---~ $0.671$  \\
\hline
$\Delta m^2_{21}$ (eV$^2$) & $7.39\times 10^{-5}$ & Fixed   \\
\hline
$|\Delta m^2_{31}|$ (eV$^2$) & $2.52\times 10^{-3}$ &  $2.3 \times 10^{-3}$ ~---~ $2.6\times 10^{-3}$  \\
\hline
$\delta$  & $-90^\circ$ &  $-180^\circ$ ~ --- ~ $180^\circ$ \\
\hline \hline

\end{tabular}
\caption{The values of the 3 neutrino oscillation parameters \cite{nufit,Esteban:2020cvm} used in the present analysis unless stated otherwise.. }
\label{tab:oscparam}  
\end{center}
\end{table}

For the calculation of probability in presence of invisible neutrino decay we modified the probability code of the GLoBES. We used $\theta_{12}=33.82^{\circ}$, $\theta_{13}=8.6^{\circ}$, $\sin^{2}\theta_{23} = 0.447 (0.56)$ for lower (higher) octant, $\delta_{CP}=-90^{\circ}$,$\ms=7.39\times10^{-5}$ eV$^{2}$ and $\ma=2.52\times10^{-3}$ eV$^{2}$ for generating the simulated data (true values) as shown in the Tab.~\ref{tab:oscparam}. These are consistent with the current ranges allowed by the latest global fits \cite{Esteban:2020cvm}. For the statistical studies we have marginalized over $\delta_{CP}$ and $\theta_{23}$ in their allowed ranges. We kept other standard parameters fixed as we found they have little effects on the marginalization. We also assumed the mass-hierarchy is known and it is normal hierarchy. 
 
\section{Results}\label{sec:res}

In this section, we present the results of our study. 
First, we give the plots at the probability level in presence of decay for T2HK, T2HKK and ESS$\nu$B. 
These figures can be helpful in understanding many of the results presented in the next sections. 
Then we give the sensitivity to the decay rate $\alpha$ for each experiment.
We also discuss a particular synergy which we observed between the two 
detectors in T2HKK experiment -- one at the first oscillation maximum and 
one at the second oscillation maximum.  
We also present the results  on the effect of  decay on the measurement of 
$\theta_{23}$ and octant sensitivity for these experiments 

 \subsection{Probability at T2HK/T2HKK and ESS$\nu$SB baselines}\label{sec:prob}
%
%

\begin{figure}
\begin{tabular}{lr}
\includegraphics[scale=0.45]{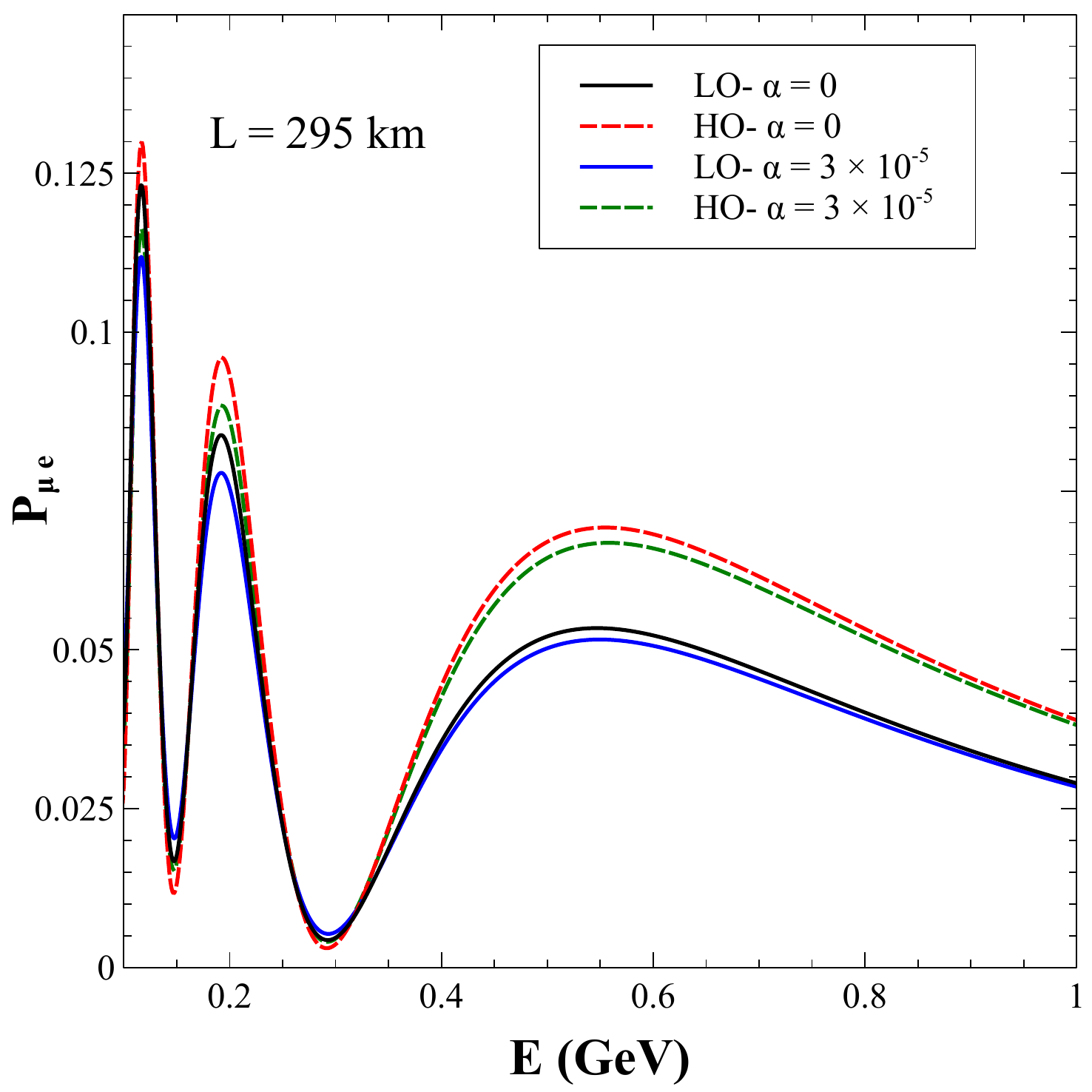}
\includegraphics[scale=0.45]{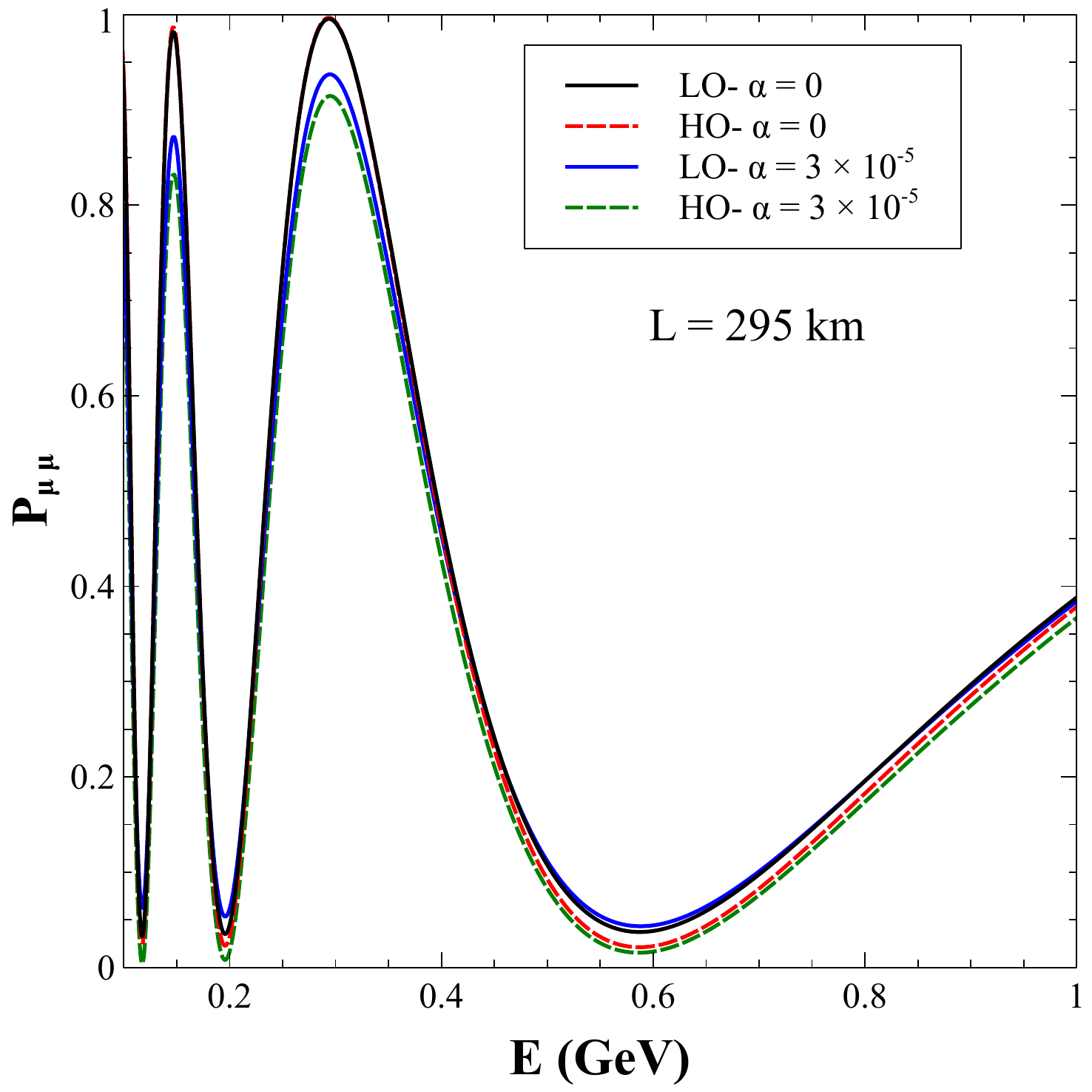} \\
\includegraphics[scale=0.45]{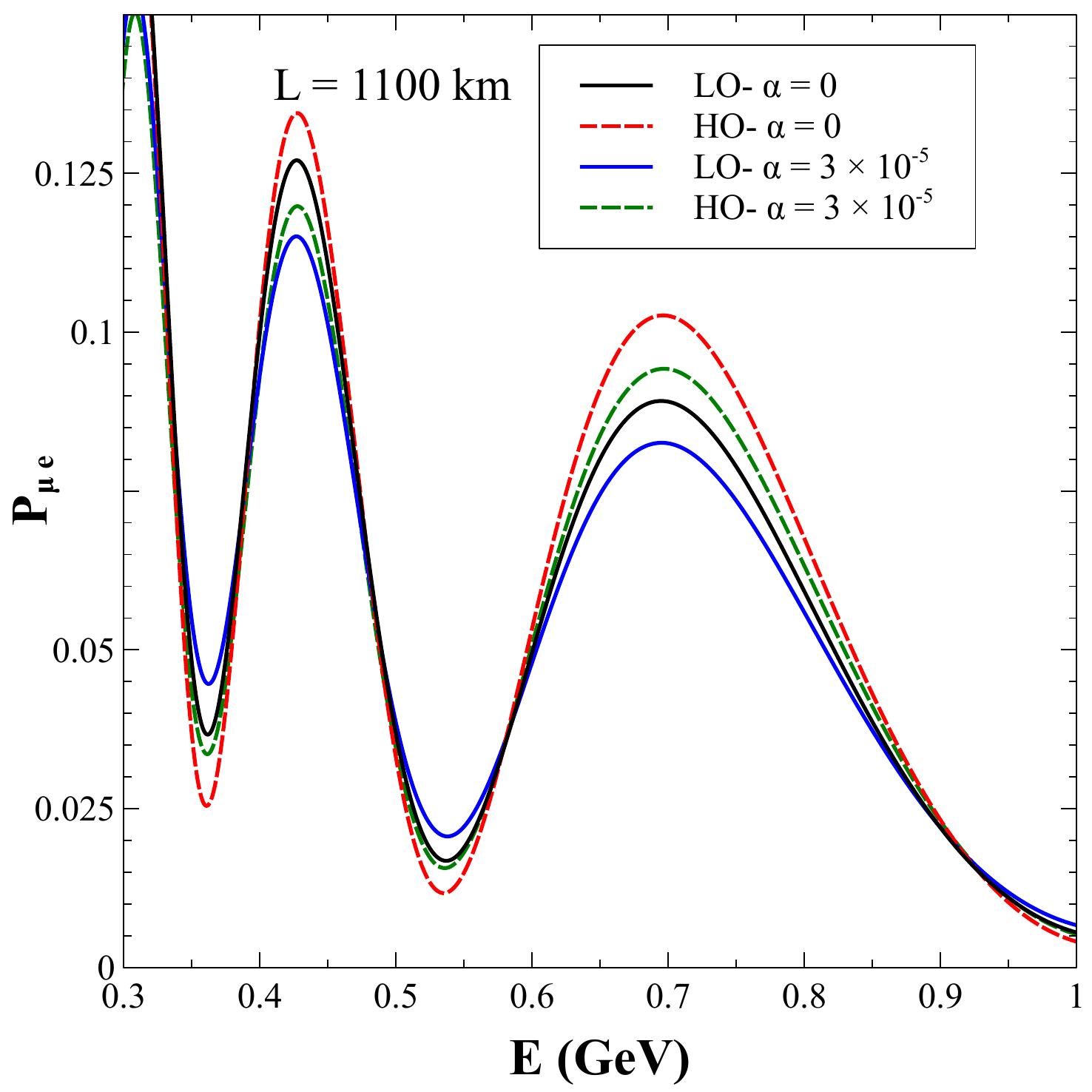}
\includegraphics[scale=0.45]{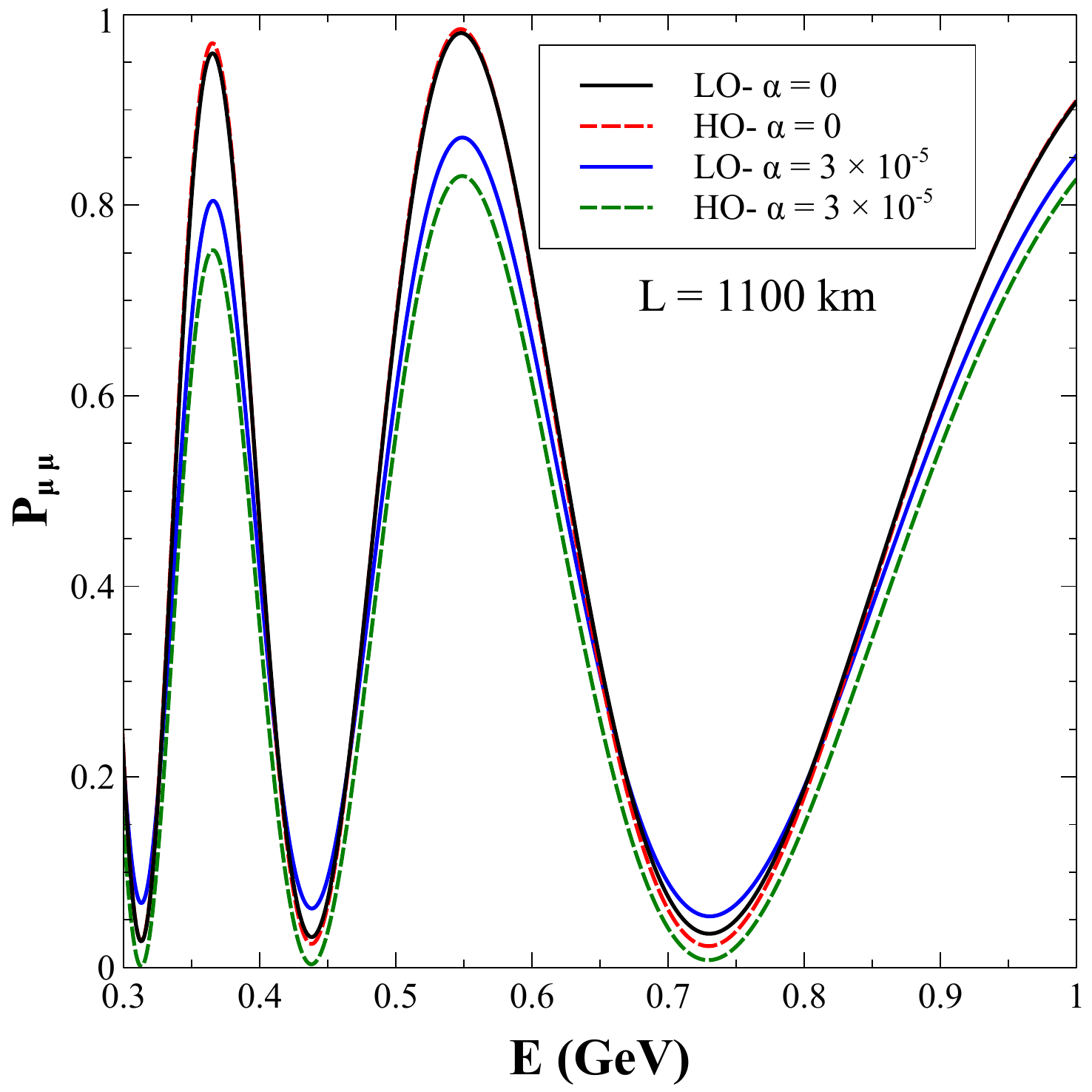} \\
\includegraphics[scale=0.45]{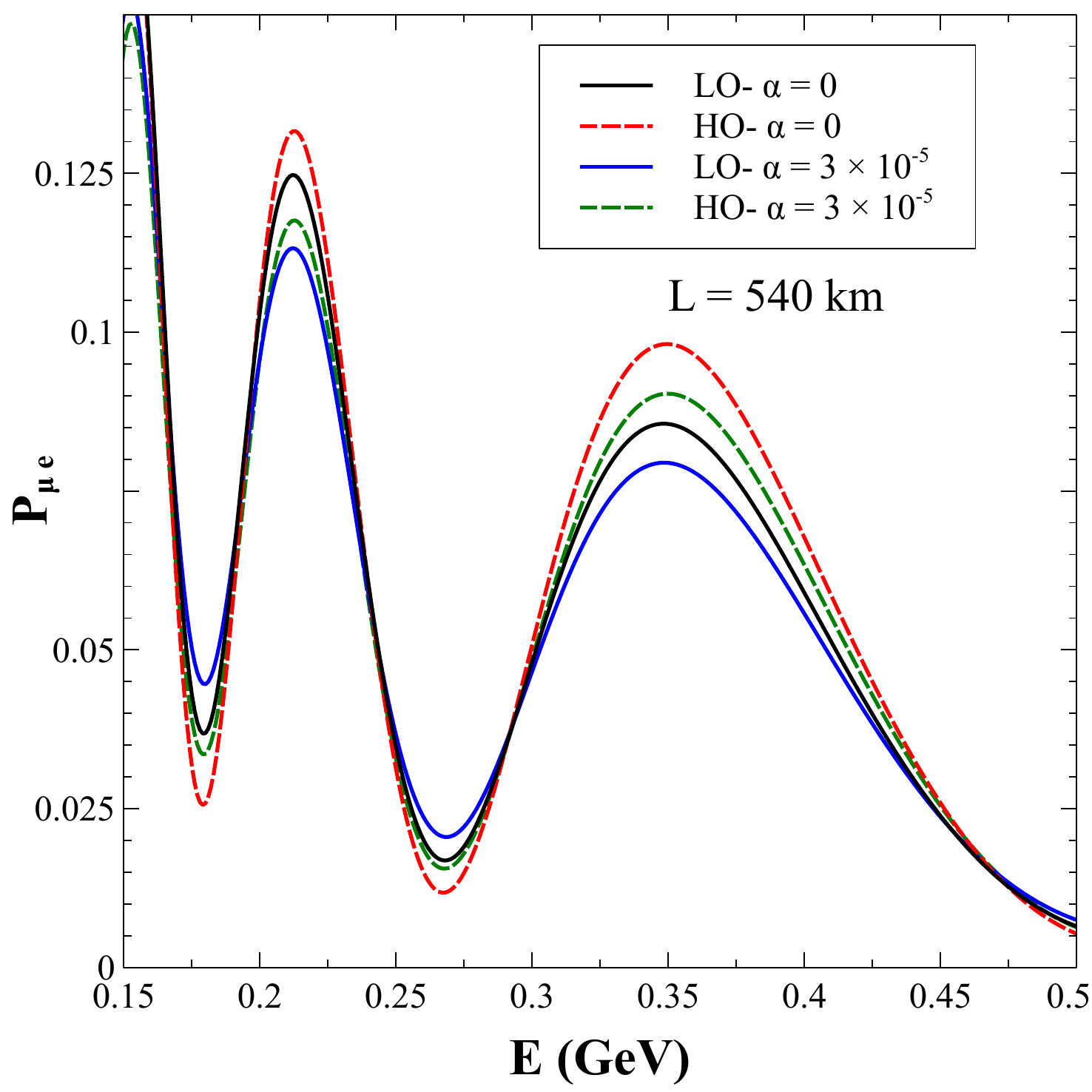}
\includegraphics[scale=0.45]{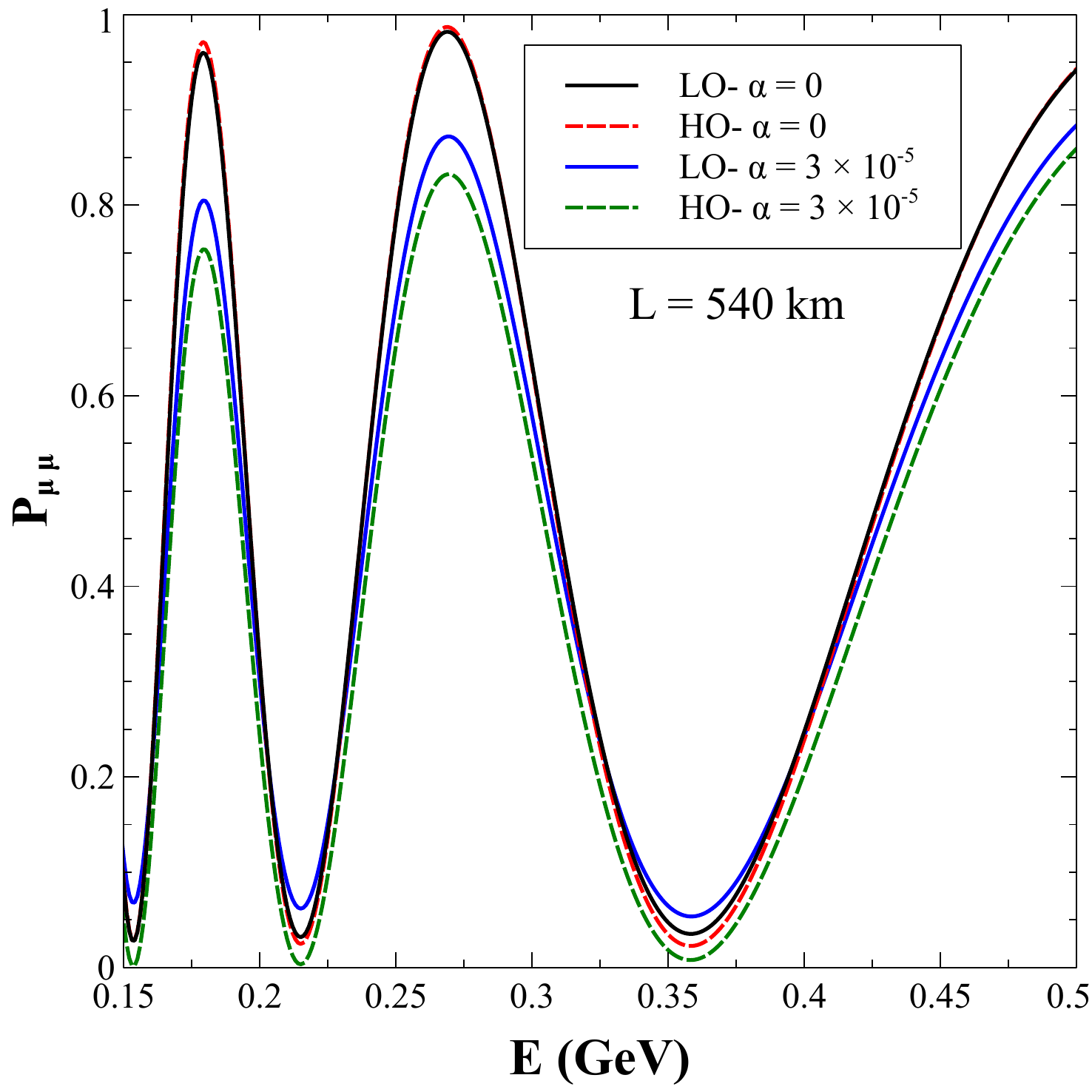} 
\end{tabular}
\caption{\label{fig:prob_oct} The probabilities  as a function of energy at L = 295 km, 1100 km and 540 km from top to bottom. The left (right) panel shows  ($\numu\rightarrow\numu$)$\numu\rightarrow\nue$ probabilities. The black (red dashed) curves are for $\alpha=0$ and $\theta_{23}$ in the lower (higher) octant. The blue solid (green dashed) curves are for $\alpha=3\times10^{-5}$ eV$^{2}$ and $\theta_{23}$ in the lower (higher) octant. Where, $\theta_{23}$ = $40^\circ$ for LO and $\theta_{23}$ = $51^\circ$ for HO.} 
\end{figure}

In fig.\ref{fig:prob_oct}, we   present the 
appearance and disappearance probabilities as a function of energy for the 
three different baselines under investigation. 
We give the plots for the no-decay case ($\alpha =0$) and also for 
a representative value of $\alpha = 3 \times 10^{-5}$ eV$^2$. 
The choice of this $\alpha$ is motivated from the current bounds as obtained from  
T2K and NO$\nu$A data analysis \cite{Choubey:2018cfz}.  
In both cases we give the plots for two values of $\theta_{23}$  -- one in lower octant $(40^{\circ})$ and one in the higher octant ($51^{\circ}$). 
The probabilities for other values of $\theta_{23}$ within this range 
will lie between these two curves.  
The figures 
show the interplay between $\theta_{23}$ and $\alpha$ for the various baselines. The left (right) panel shows $\numu\rightarrow\nue$ ($\numu\rightarrow\numu$) probabilities. 
The black (blue)  solid and red (green)  dashed curves are for 
$\alpha=0$ ($\alpha=3\times10^{-5}$ eV$^{2}$) and $\theta_{23}$ in the lower  octant and higher octant respectively. 
The top, middle and the bottom panel are for the baselines 295 km, 1100 km and 540 km respectively. 


In the left panels of fig.~\ref{fig:prob_oct}, 
we see that for all the  baselines, at the oscillation maxima,
 $P_{\mu e}$ for the 
 the no decay case is higher 
for a fixed  value of $\theta_{23}$. 
As decay is introduced, the probabilities reduce. On the other hand,  for a fixed value of 
$\alpha$, lower octant gives smaller probabilities compared to the higher octant.  
Thus  the  peak appearance probability can get reduced due to 
increase in the   value of $\alpha$  and/or decrease in the value of the 
mixing angle $\theta_{23}$. 

The top panel  is for T2HK and here the first oscillation maximum is 
at $\sim$0.6 GeV which is also where the flux peaks. 
For the baseline and energies involved the effect of decay for T2HK is 
small for the sample value of $\alpha$ considered in the plot. 
However, we see that the higher  and lower octant bands are well separated 
for T2HK. 

Next we focus on the middle and the bottom panels which are relevant baselines for 
T2HKK (second detector) and ESS$\nu$SB.     
For these experiments the flux peaks near the second oscillation maxima 
-- $\sim$0.7 GeV  for the baselines 1100 km  and  $\sim$0.35 GeV 540 km. 
Since these are higher baselines and/or lower energies
the effect of decay is more pronounced.  The HO and LO bands are much closer in these cases and a 
reduction in probability due to non-zero values of $\alpha$ can be compensated by 
increasing $\theta_{23}$. 

In the right panels of fig.~\ref{fig:prob_oct} we show the 
$\numu\rightarrow\numu$ disappearance probability  
$P_{\mu \mu}$ which plays a crucial role in determining the value of $\theta_{23}$ 
in long baseline experiments.   
In this case,  the 
curves with $\alpha = 0$ are overlapping irrespective of the octant of 
$\theta_{23}$  which shows  that the $P_{\mu \mu}$ channel do not have any octant sensitivity in absence of neutrino  decay, being 
dependent on $\sin^2 2\theta_{23}$. So, the red-dashed and black-solid curves (no-decay) are well separated from the blue-solid
and green-dashed curves which reveals  the effect of decay. 
 This feature is true for all the three baselines considered in our study.
 In case of the 295 km baseline the blue curve is closer to the black and red curves compared to the 1100 km and 540 km cases 
 since the effect of decay is less for the 295 km baseline for the $\alpha$ value chosen. 
Also, it has to be noted that, 
the otherwise octant degenerate $P_{\mu \mu}$ channel begins to have octant sensitivity once 
$\alpha \ne 0$, resulting in a gap between the blue and green curves.

This can be  simply understood if we consider the expression for  
two-generation survival probability in  vacuum

\begin{equation}
P_{\mu\mu} = \left [ 1 - \sin^2\theta_{23} ( 1 - e^{-\frac{\alpha L}{ E}} \right ]^2 - 
\sin^22\theta_{23} e^{-\frac{\alpha L}{2 E}}\,\sin^2\left(\frac{\Delta m_{31}^2 L}{4E}\right)\
\label{eq:pmumu}
\end{equation}

In absence of decay the  $e^{-\frac{\alpha L}{ E}}$ term is 1 and the probability depends on $\sin^2 2\theta_{23}$. 
But in presence of decay octant sensitivity ensues 
due to the $e^{-\frac{\alpha L}{ E}}$ factor.

\begin{figure}
\begin{tabular}{lr}
\includegraphics[scale=0.45]{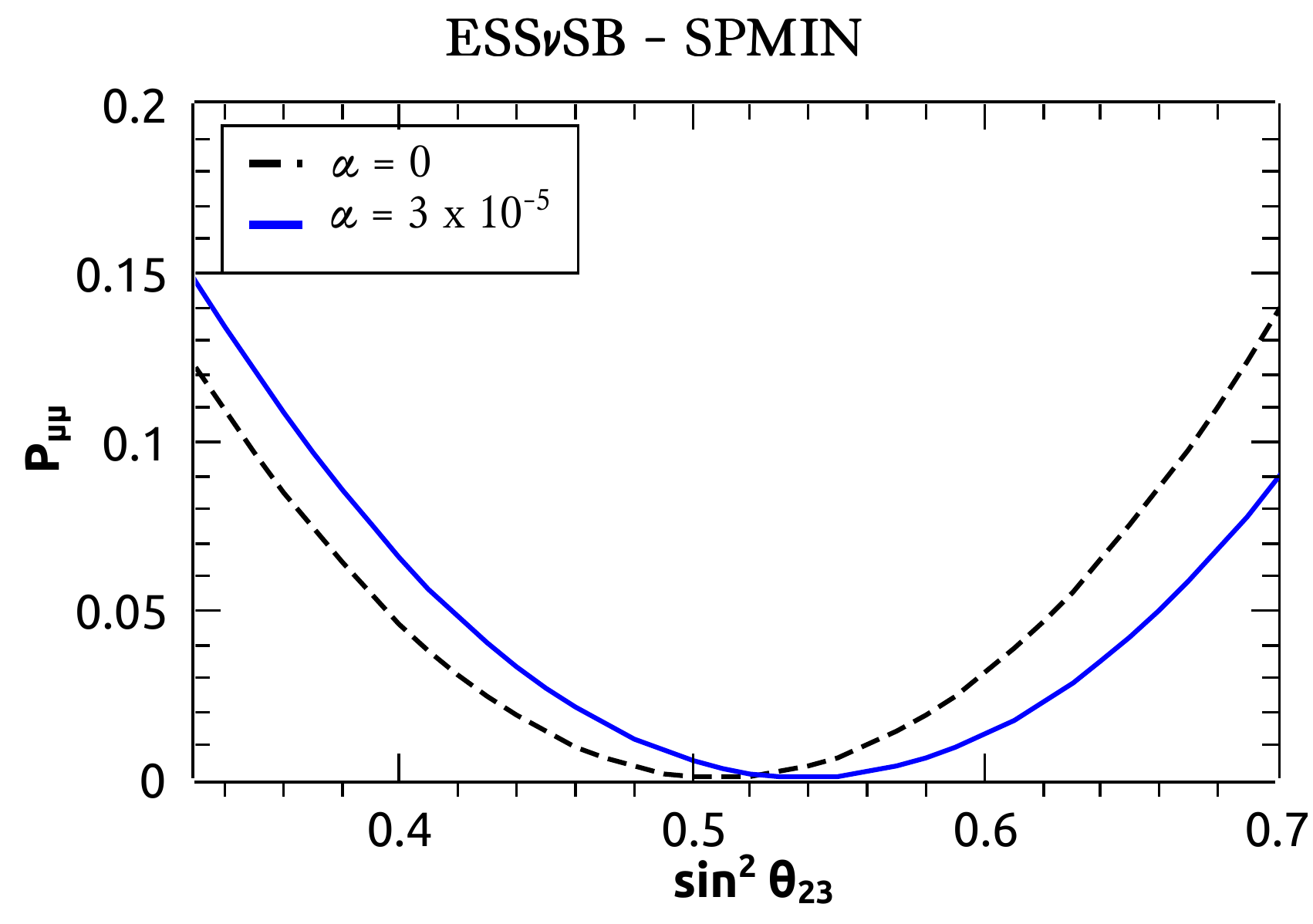}
\includegraphics[scale=0.45]{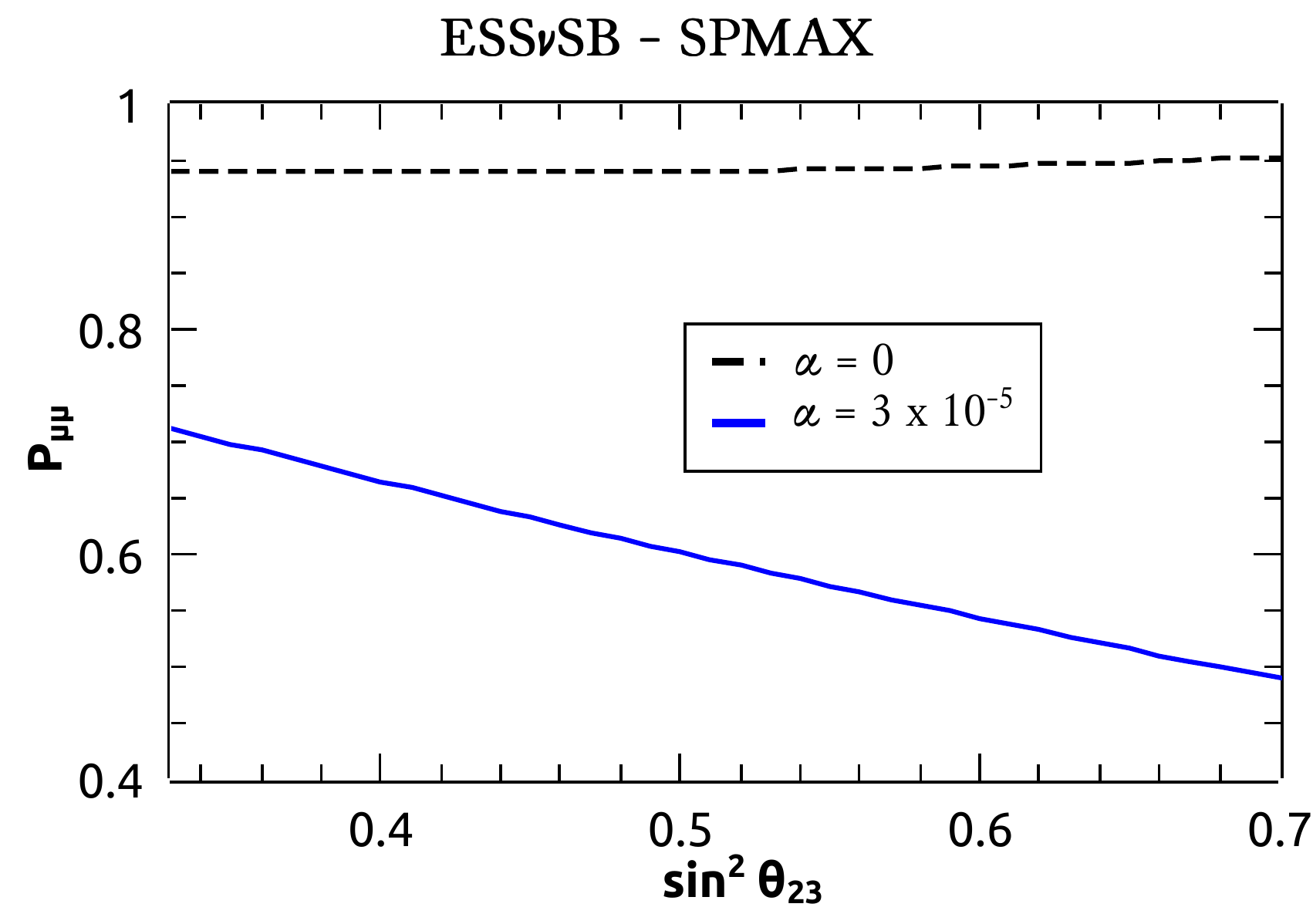}
\end{tabular}
\caption{\label{fig:prob_th23_dep} The probabilities  as a function of $\sin^2 \theta_{23}$ at L = 540 km for ESS$\nu$SB. The left (right) panel shows  $\numu\rightarrow\nue$($\numu\rightarrow\numu$) probabilities. The black dashed (blue) curves are for $\alpha=0$ and $\alpha=3\times10^{-5}$ eV$^{2}$  respectively.}
\end{figure}

For our later analysis, it is important to understand the dependence 
of the probability on  $\theta_{23}$ at the survival probability 
maxima (SPMAX) and minima (SPMIN).  At SPMIN, though the probability value is 
small, the flux  peaks near this energy since it corresponds to the 
oscillation maxima.   
In figure~\ref{fig:prob_th23_dep} we show the behaviour of the probability for the 540 km
baseline as a function of $\theta_{23}$ for energies corresponding to 
SPMAX  and SPMIN. Similar behaviour is also true for the two other baselines
-- 295 km and 1100 km.   
It is seen that at SPMAX, the no-decay probability is always higher than 
the decay probability. The no-decay probability has no dependence on 
$\theta_{23}$ while the decay probability reduces with increasing $\theta_{23}$ monotonically. This can be easily understood from the expression 
~\ref{eq:pmumu}. At SPMAX 
$\sin^2\left(\frac{\Delta m_{31}^2 L}{4E}\right)\ =0$  and the remaining term 
decreases monotonically with increasing $\theta_{23}$ for non-zero $\alpha$. 
At SPMIN, the $\theta_{23}$ dependence is more complicated. 
For lower octant the decay probability is larger as compared to the no-decay 
probability and the probability reduces with increasing $\theta_{23}$ till
a certain value (depending on $\alpha$), after which in higher octant 
it increases with $\theta_{23}$. For higher octant, no-decay probability is 
higher than the decay probability.   
As we will see, these features of the $P_{\mu \mu}$ probability 
will play a role in the $\chi^2$ fit.

\subsection{Sensitivity to the decay}
\begin{figure}[htb]
\begin{tabular}{cc}
\includegraphics[width=0.45\textwidth]{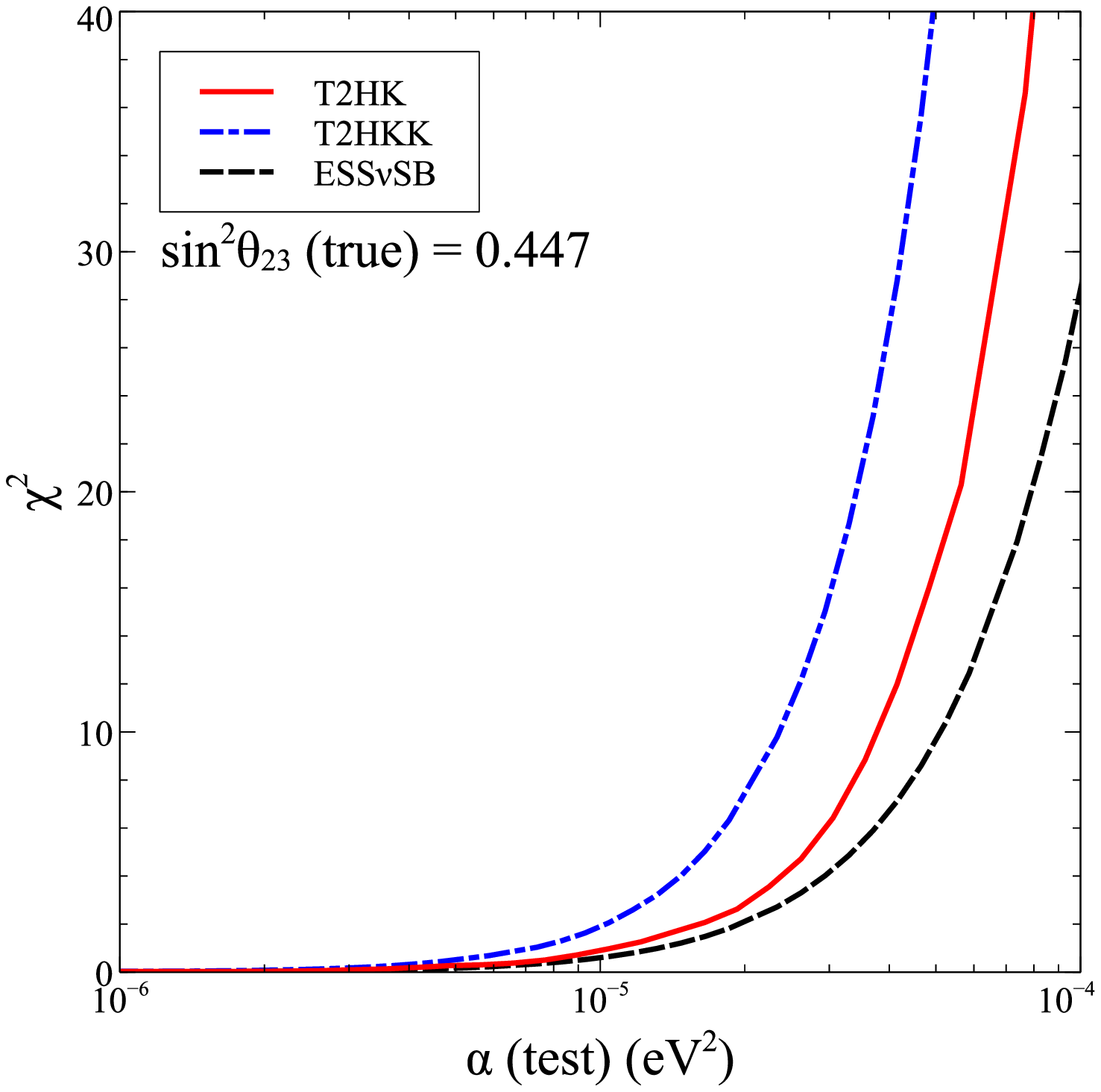}
\includegraphics[width=0.45\textwidth]{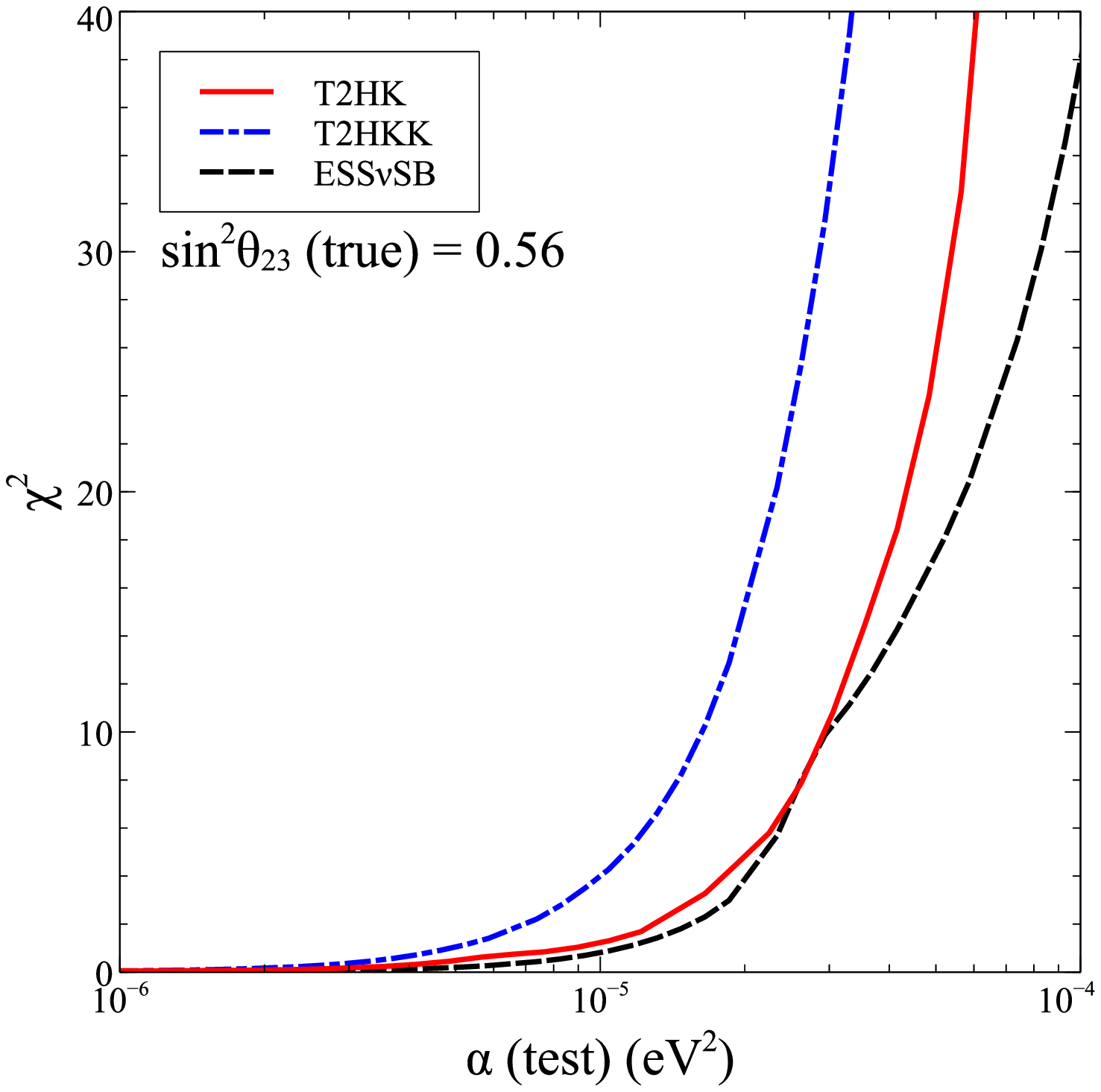}

\end{tabular}
\caption{\label{fig:sens} The $\chi^{2}$ as a function of $\alpha$, assuming no decay in the simulated data. The left (right) panel is for $\theta_{23}$ (true) in the lower (higher) octant. The red solid, blue dashed dotted and the black dashed curves are for T2HK, T2HKK and ESS$\nu$B respectively.}
\end{figure}

In this section  we study the capability of the  T2HK, T2HKK and ESS$\nu$B  experiments 
in constraining  invisible neutrino decay. 
We assume the neutrinos to be  stable while generating the simulated data. 
 In the fit we have marginalized over $|\Delta m^2_{31}$, $\theta_{23}$ and $\delta_{CP}$ in the range given in 
 Tab.~\ref{tab:oscparam}.
 The fig.~\ref{fig:sens} shows the value of $\chi^{2}$ for different values of $\alpha$  taken in the fit (test). A higher value of $\chi^{2}$ is more disfavoured compared to a lower values of $\chi^{2}$. 

%
%
 
In the left panel we assumed $\sin^{2}\theta_{23}\rm{(true)}=0.447$ (lower octant) while generating the data. In the right panel we assumed $\sin^{2}\theta_{23}\rm{(true)}=0.56$ (higher octant). The red solid, blue dashed dotted and black dashed curves are for T2HK, T2HKK and ESS$\nu$B respectively.
Naively, the sensitivity to decay depends on the $L/E$. From Tab.~\ref{tab:expt-details}, we see that the second detector of T2HKK has highest $L/E$ and the detector at 295 km has lowest $L/E$ (See Tab.~\ref{tab:expt-details}). So, we expect T2HKK to have the 
best sensitivity and T2HK to have the worst, but we observe that ESS$\nu$SB has a lower sensitivity than T2HK. 
 Here, marginalization over $\theta_{23}$ plays an important role. 
Determination  of $\theta_{23}$ is governed mainly by the 
$P_{\mu \mu}$ channel. 
In  absence of decay there is no sensitivity to $\theta_{23}$ at the maxima of survival probability which corresponds to 
${\sin^2 {\Delta m^2_{31} L} / {4 E} } =0$.   The  maximum precision of 
$\theta_{23}$  comes at the oscillation maximum or the survival probability minimum (SPMIN) where, 
${\sin^2 {\Delta m^2_{31} L} / {4 E} } =1$ and can be expressed as,  
\begin{equation} 
\Delta({\sin^2\theta_{23}}) = - \frac  {\Delta(P_{\mu \mu})}  
{4 \cos^2 2 \theta_{23} }
\end{equation} 
In presence of decay, the survival probability maxima also 
acquires some sensitivity to $\theta_{23}$ which increases 
with increasing value of the decay constant $\alpha$ for a fixed baseline.  

As seen in the earlier section, for T2HK the octant bands are well separated 
while the effect of decay  is not very 
significant. Hence marginalization  over $\theta_{23}$ does not play much role. 
For ESS$\nu$SB on the other hand, the octant bands are   not so well separated and the effect of decay 
is more.    As a result,  marginalization over $\theta_{23}$ tends to make 
the probabilities closer by shifting 
$\theta_{23}$ to a different value and the sensitivity reduces. 
The direction of shift  depends on the initial value of $\theta_{23}$.
For SPMIN, decay in the fit, can give same probability as no-decay in
data for an increased $\theta_{23}$ for both lower and higher 
octant. This can be seen by drawing a horizontal line in fig.~\ref{fig:prob_th23_dep}. 
For SPMAX, on the other hand, if we fit decay with no-decay in data, then
shift of $\theta_{23}$ towards lower values make the probabilities closer, 
giving a lower $\chi^2$. 
Thus, depending on which energy bins contribute most, the $\theta_{23}$ in the fit comes at a 
higher or lower value, reducing the $\chi^2$.  
We have checked that if we keep $\theta_{23}$ as fixed then ESS$\nu$SB gives 
a better sensitivity. 

In case of T2HKK, other than a detector close to the second oscillation maximum, 
there is also a detector at the first oscillation maximum which can measure  
$\theta_{23}$ with a good precision. So, the combination of the two detectors, 
one at the first oscillation maximum and other at the second oscillation maximum 
gives the best sensitivity.  This will be discussed further in the next section. 
There is also another interesting feature for ESS$\nu$SB, which is that for true octant as 
higher octant  we observe a sudden fall of sensitivity for
$\alpha$ above $\alpha\sim3\times10^{-5}$ eV$^{2}$. 
We will see in the next section that this is a unique feature of second oscillation maxima and 
we will discuss this in detail in the coming section. 

\subsection{Synergy between 295 km  and 1100 km baselines }
\begin{figure}[htb]
\begin{tabular}{cc}
\includegraphics[width=0.45\textwidth]{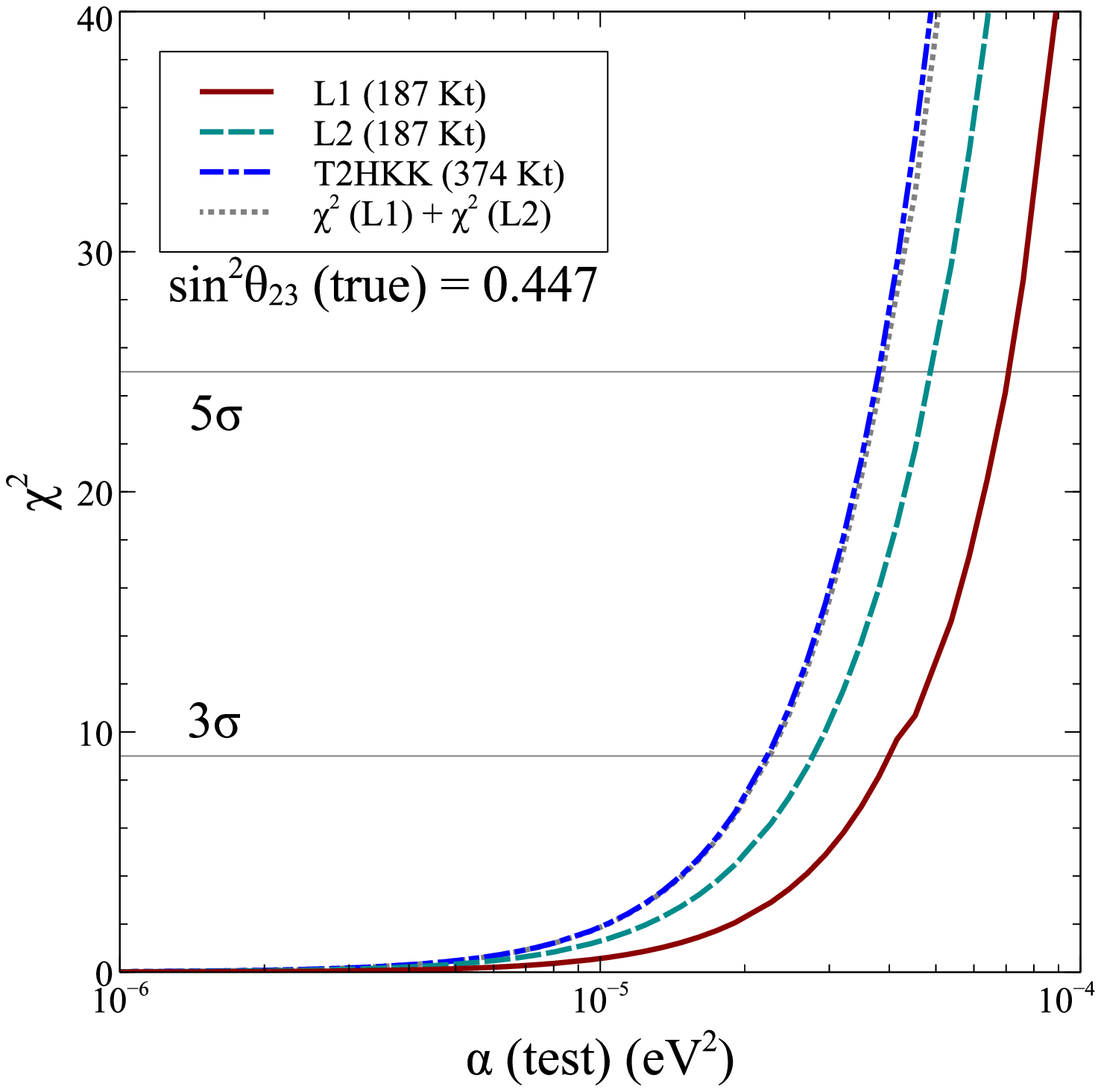}
\includegraphics[width=0.45\textwidth]{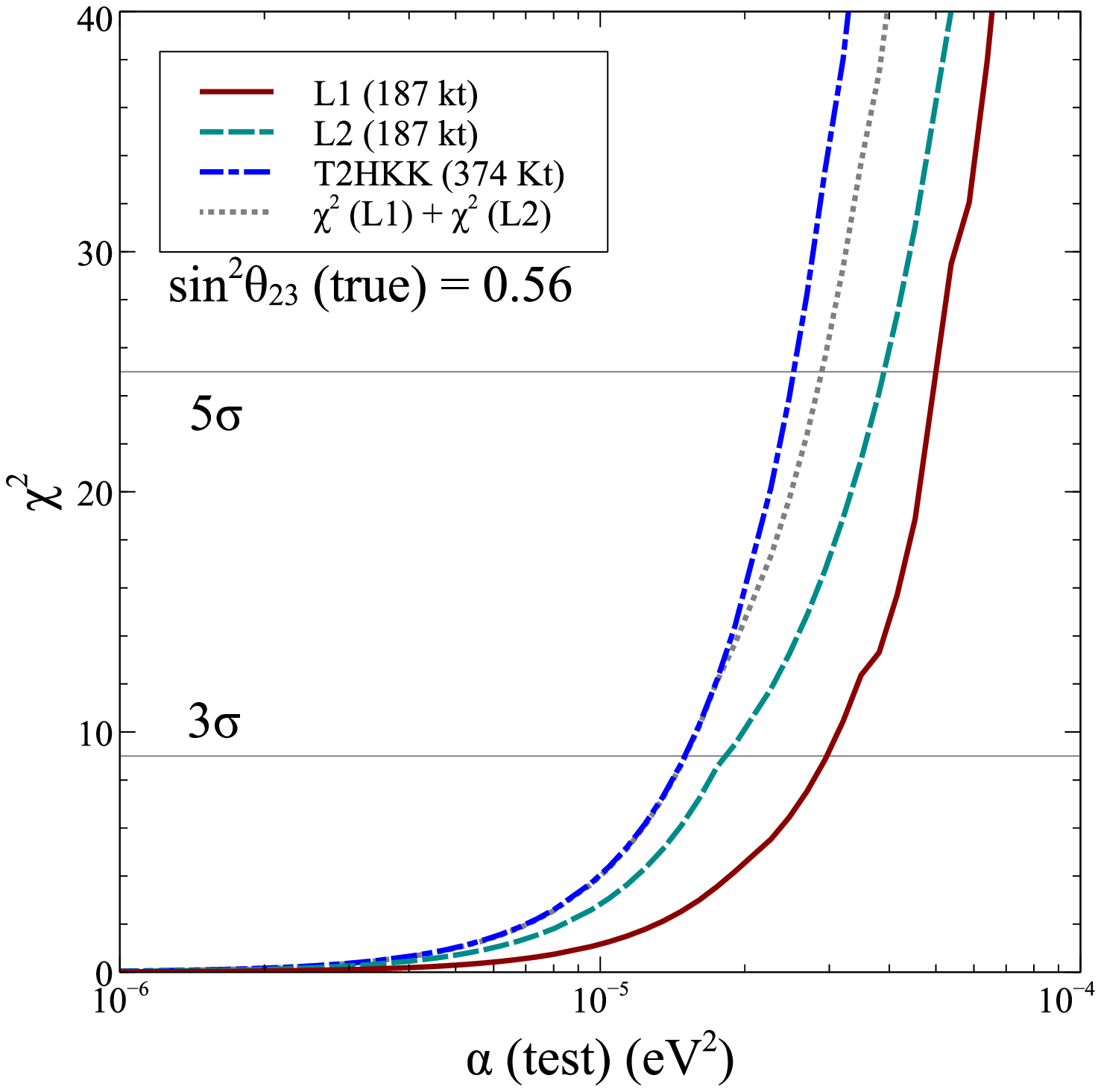}

\end{tabular}
\caption{\label{fig:syn} The $\chi^{2}$ as a function of $\alpha$, assuming no decay in the simulated data. The left (right) panel is for $\theta_{23}$ (true) in the lower (higher) octant. The dark red solid, cyan dashed, blue dashed-dotted curves are for a 187 kt detector at 295 Km, 187 kt detector at 1100 km and their combination respectively. The grey dotted curve shows the $\chi^{2}$ if we just add the $\chi^{2}$s of the dark red and cyan curves. }
\end{figure}

In this section we discuss a synergy between the two baselines $L1=295$ km and $L2=1100$ km
of T2HKK  in constraining neutrino decay. 
The first one corresponds to the Japanese detector (JD) and the second one corresponds to the Korean detector (KD). It can happen that 
the $\chi^{2}$ obtained by combining two or more experiments can be different from the naive sum of  their individual $\chi^{2}$s. 
We call such enhancement of $\chi^{2}$ as synergy between the experiments. 
In fig.~\ref{fig:syn}, we present the sensitivity to the $\alpha$ for one of the HK detectors situated at L1 (dark red solid)
 and L2 (cyan dashed). We also show their combined sensitivity ( blue dashed-dotted) which is nothing but the sensitivity of T2HKK. The grey dotted curve shows the naive summation of the $\chi^{2}$s 
i.e $\chi^2_{L1} + \chi^2_{L2}$. 
The left panel is for $\sin^{2}\theta_{23}=0.447$ and the right panel is for $\sin^{2}\theta_{23}=0.56$. 
We see that although there is no significant synergy for the left panel, there is some significant synergy for the right panel for $\alpha$ greater than $2.268\times10^{-5}$ eV$^{2}$, i.e., the $\chi^{2}_{L1+L2}$ of T2HKK is larger than the $\chi^{2}_{L1}+\chi^{2}_{L2}$.

\begin{figure}[htb]
\begin{tabular}{cc}
\includegraphics[width=0.45\textwidth]{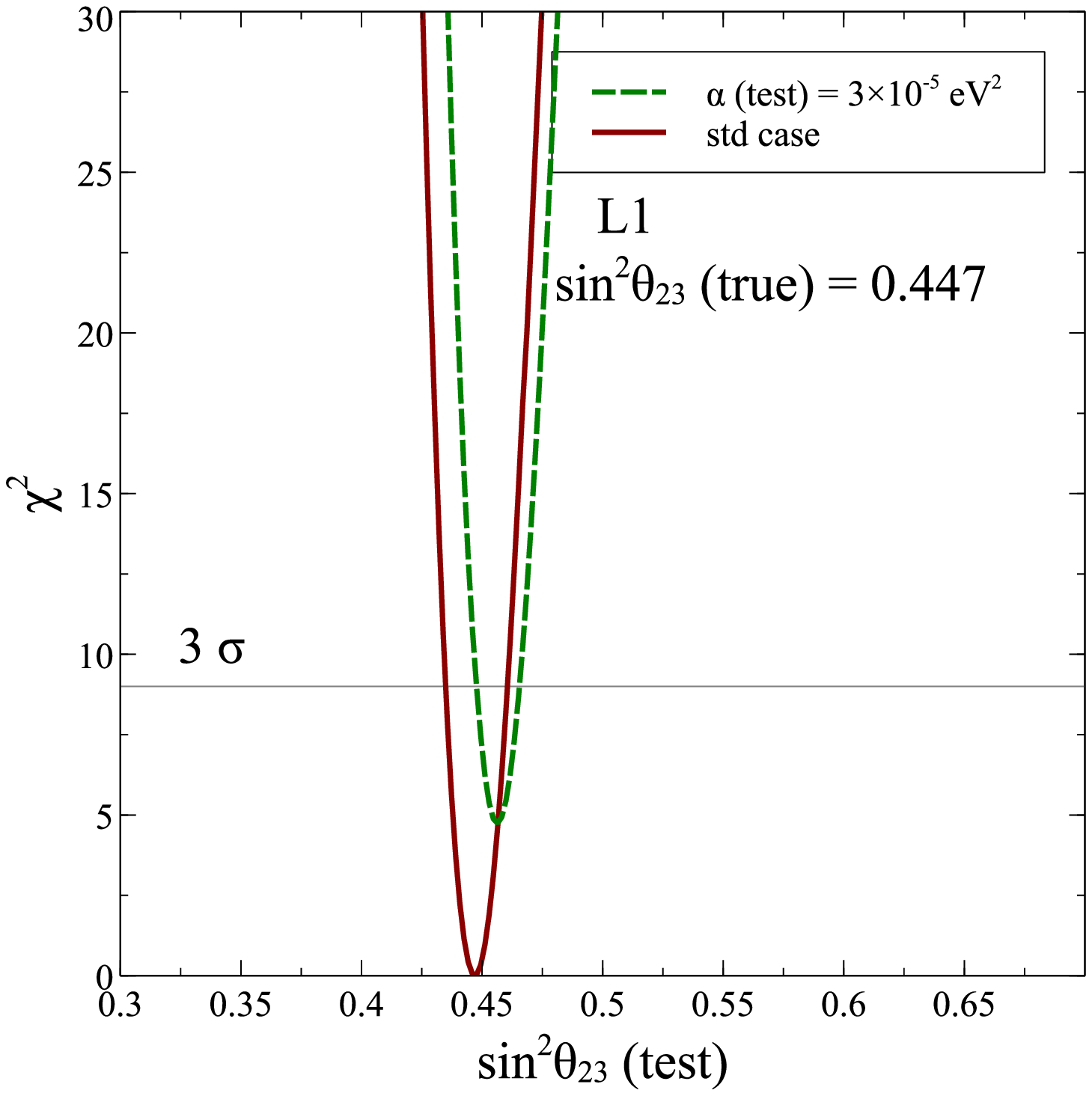}
\includegraphics[width=0.45\textwidth]{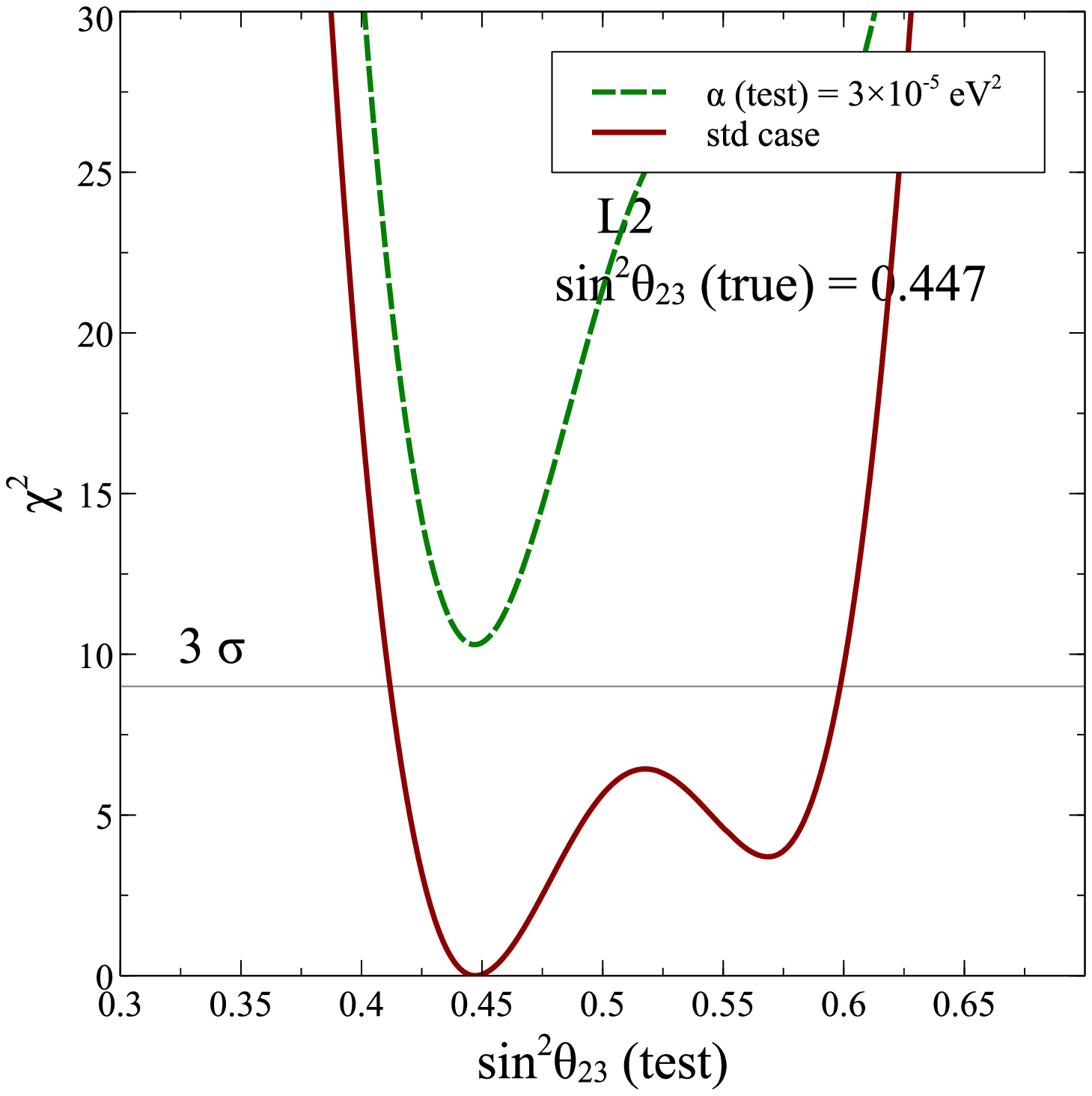}

\end{tabular}
\caption{\label{fig:th23_lo} The $\chi^{2}$ as a function of $\theta_{23}$ assuming the true $\sin^{2}\theta_{23}=0.447$. The left (right) panels are for baselines 295 (1100) km respectively. The dark red solid curves are for the standard cases and the green dashed curves are for the case when $\alpha = 0$ in the simulated data but assumed to be $3\times10^{-5}$ eV$^{2}$ in the fit. }

\end{figure}

\begin{figure}[htb]
\begin{tabular}{cc}
\includegraphics[width=0.45\textwidth]{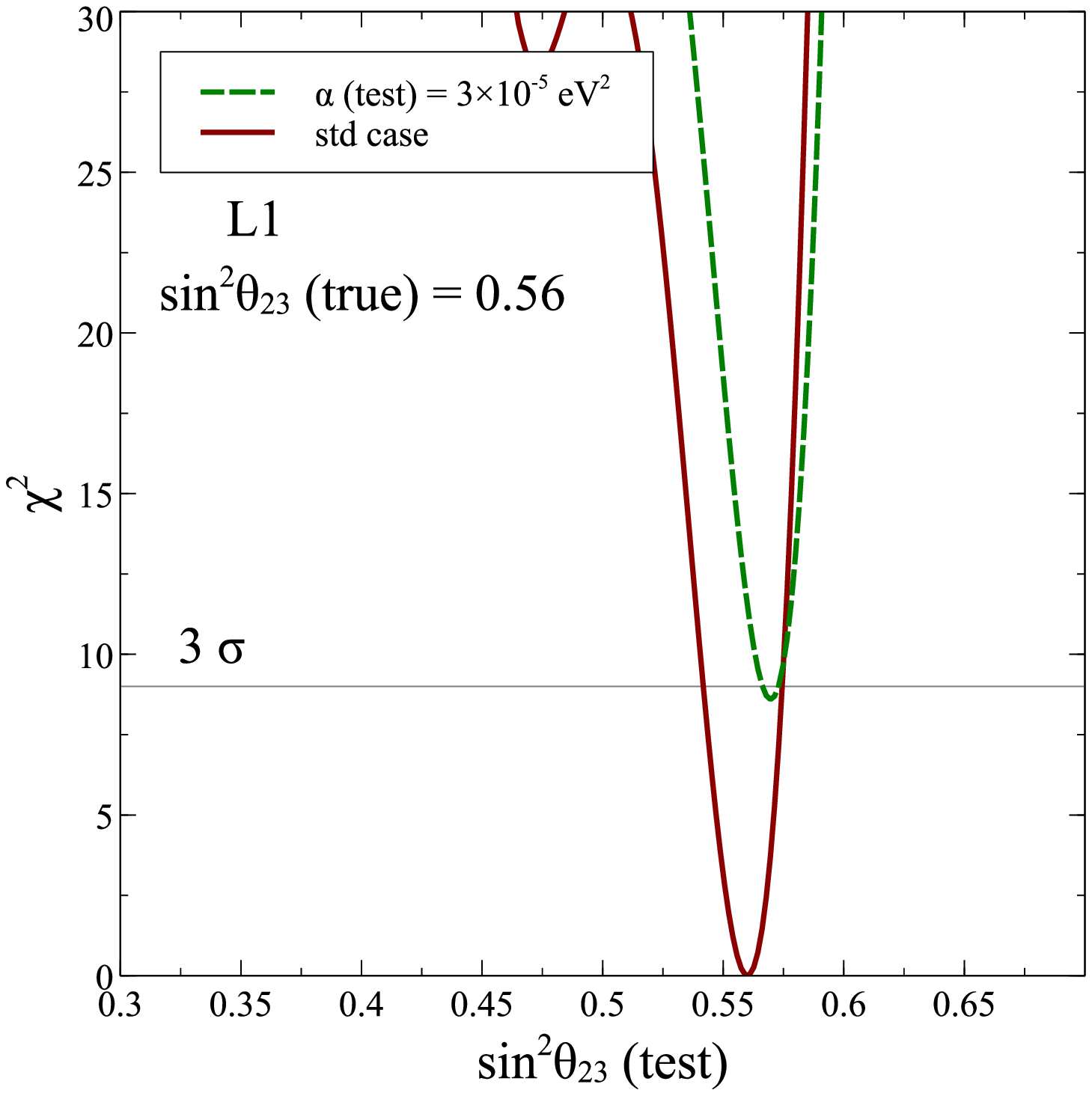}
\includegraphics[width=0.45\textwidth]{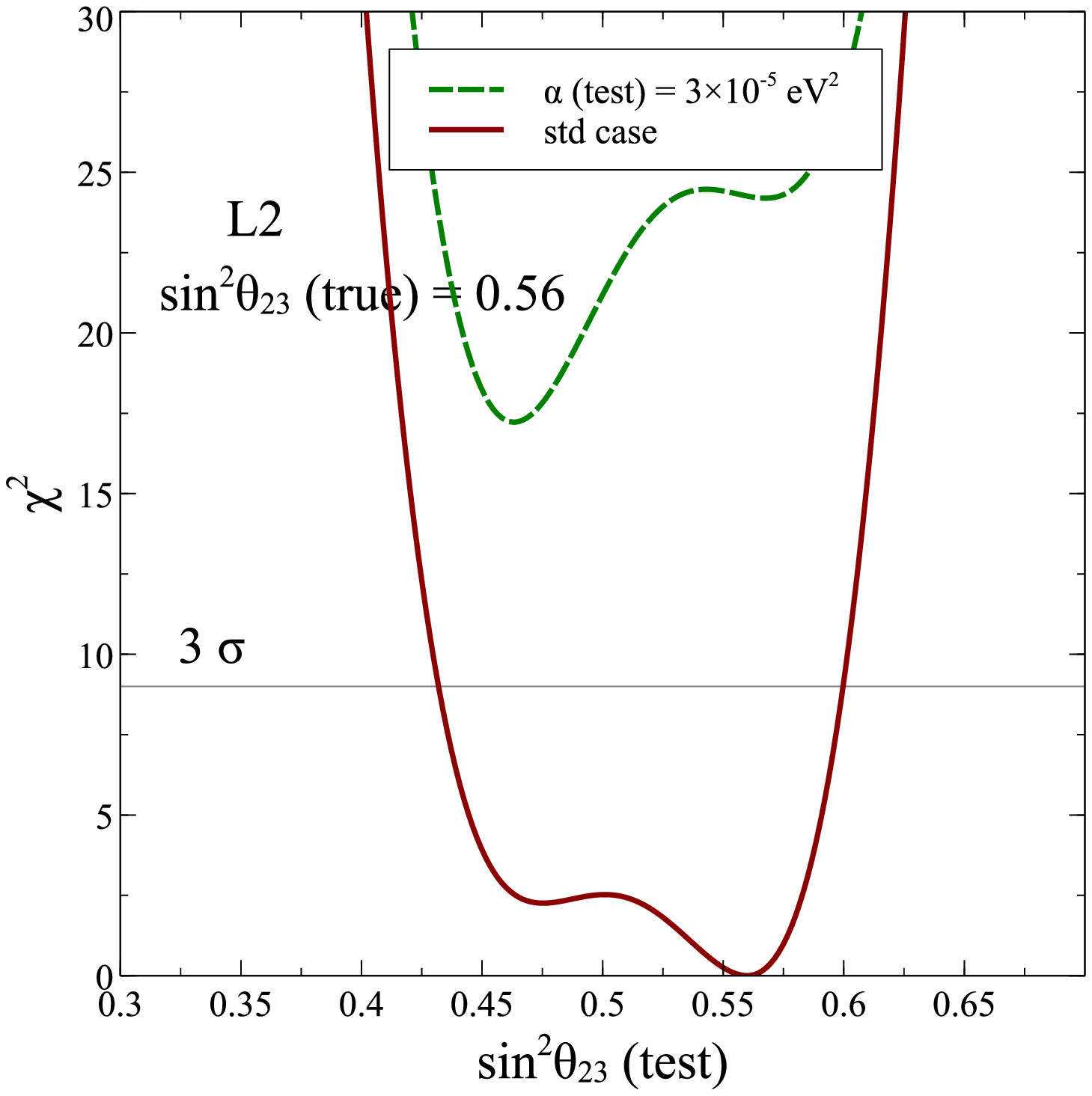}

\end{tabular}
\caption{\label{fig:th23_ho} The $\chi^{2}$ as a function of $\theta_{23}$ assuming true $\sin^{2}\theta_{23}=0.56$. The left (right) panels are for baselines 295 (1100) km respectively. The dark red solid curves are for the standard cases and the green dashed curves are for the case when $\alpha = 0$ in the simulated data but assumed to be $3\times10^{-5}$ eV$^{2}$ in the fit. }

\end{figure}

This synergy can be understood from the fig.~\ref{fig:th23_lo} and fig.~\ref{fig:th23_ho}. These figures give the $\chi^{2}$ for a given value of $\theta_{23}$ in the fit with a fixed value of $\alpha$ in the fit. The decay constant  $\alpha=0$ while simulating the data and $|\Delta m^2_{31}|$ 
and  $\delta_{CP}$ are marginalized for all the cases. 
Therefore global minima of the curves in these figures give the $\chi^{2}$ values for the 
corresponding $\alpha$ (test) in fig.~\ref{fig:syn}. Fig.~\ref{fig:th23_lo} is for the case 
when $\sin^{2}\theta_{23}=0.447$ and fig.\ref{fig:th23_ho} is for $\sin^{2}\theta_{23}=0.56$ 
in the simulated data. The dark red solid curves assume $\alpha=0$ in the fit and green dashed curves 
assume $\alpha=3\times10^{-5}$ eV$^{2}$ in the fit.


First let us consider the fig.~\ref{fig:th23_lo}. Here, $\sin^{2}\theta_{23} (true) =0.447$. We see that for L1, the global minimum,  for a test value of  $\alpha=3\times10^{-5}$ eV$^{2}$  in the fit, 
 shifts slightly towards a higher $\sin^{2}\theta_{23}$ but stays in the same octant as the $\alpha=0$. The shift towards a higher value can be understood by looking at the curve for $P_{\mu \mu}$. 
It is seen that at the oscillation maxima (i.e SPMIN),  the no-decay 
probability (black) curve is lower than that with decay (blue curve)
for lower octant. Hence when we try to 
fit data with no-decay with decay in theory,  the probabilities can come 
closer  
by increasing  $\theta_{23}$ slightly.  
Similar feature is also observed at L2 in the right panel though. the shift  is almost un-noticeable. 
Sensitivity to the parameter $\alpha$ for the combination L1 and L2 will be given by the global minimum of the sum of $\chi^{2}$s of two green dashed curves of fig.~\ref{fig:th23_lo}. As the global minima of these two curves approximately coincide (42$^{\circ}$ for L2 and 42.5$^{\circ}$ for L1), the global minimum of the sum of their $\chi^{2}$s will be approximately equal to the sum of $\chi^{2}$s at the global minima of L1 and L2. Therefore, the sensitivity to the combination of L1 and L2 will be approximately equal to the sum of the sensitivities for L1 and L2. Thus there is not significant synergy for this case.

Next we consider the fig.~\ref{fig:th23_ho}. Here, $\sin^{2}\theta_{23}=0.56$. We see that for L1, like previous case, the global minimum for $\alpha=3\times10^{-5}$ eV$^{2}$ (test) shifts a little towards higher value since   
at SPMIN, no-decay probability is larger, it can be 
matched by increased value of $\theta_{23}$ (cf. \ref{ fig:prob_th23_dep}). 


However, for L2, we see that the position of the global minimum drastically changes and now it is in the opposite octant, i.e., lower octant. 
Again, like the previous case, the sensitivity to the combination of L1 and L2 will be determined by  the sum of $\chi^{2}$s of two green dashed curves. But here, the positions of the global minima for L1 and L2 are completely different. Therefore, the position of the global minimum of the sum of $\chi^{2}$s of L1 and L2 curves is non-trivial and is not equal to the sum of the $\chi^{2}$s at the global minima of L1 and L2 curves. Hence, we see the synergy in this case. The shift of the global minimum can be attributed to the interplay between the appearance and disappearance channels. 

\begin{figure}[htb]
\begin{tabular}{cc}
\includegraphics[width=0.45\textwidth]{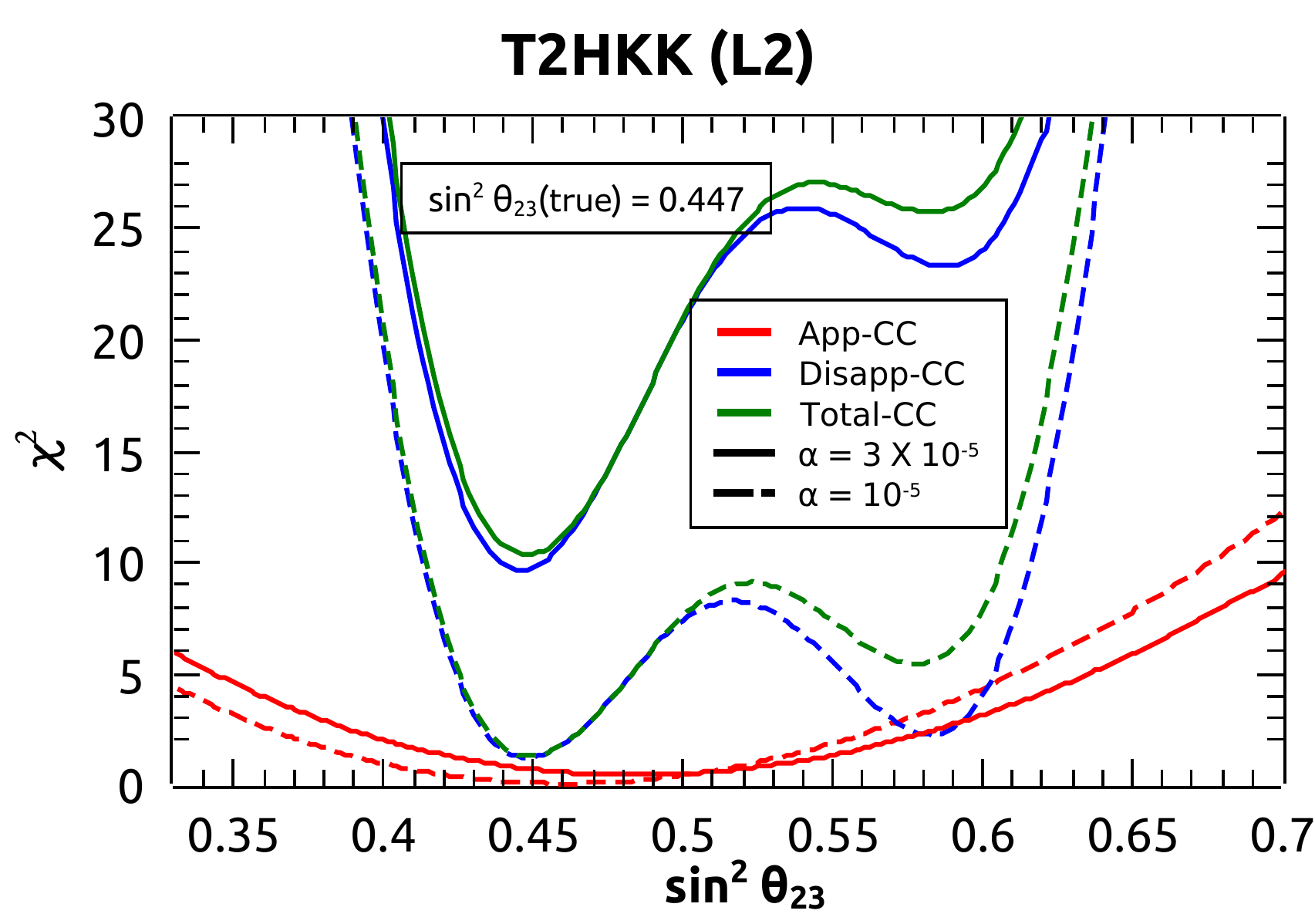}
\includegraphics[width=0.47\textwidth]{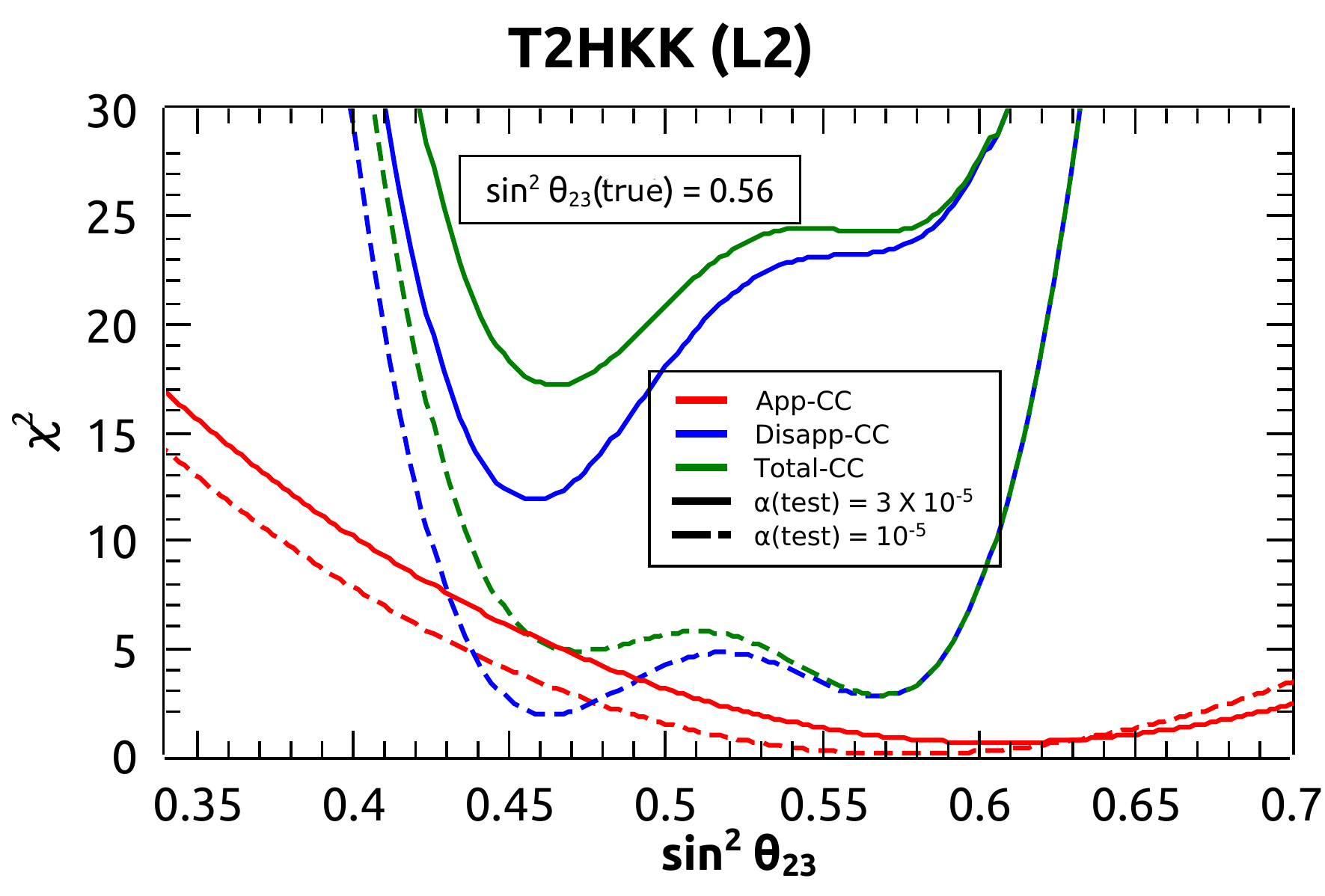}

\end{tabular}
\caption{\label{fig:t2hkkl2-testalp} The $\chi^{2}$ as a function of $\theta_{23}$. The left (right) panels are for true $\sin^{2}\theta_{23}=0.447$ ($\sin^{2}\theta_{23}=0.56$) respectively. The red and blue curve are for appearance and disappearance channels respectively. While the total contribution from both the channels is represented by green curves. The solid curves are for $\alpha \rm(test) = 3 \times 10^{-5} eV^{2}$ and the dashed curves for $\alpha \rm(test) = 10^{-5} eV^{2}$ in the fit. The data have been generated for $\alpha = 0$.} 

\end{figure}

In order to understand this interplay in fig.~\ref{fig:t2hkkl2-testalp} 
we have plotted the appearance and disappearance $\chi^2$ s for two different 
values of the decay constant $\alpha$ for L2. 
From the left panel of the figure it is seen that for true $\sin^2\theta_{23}$ in the lower octant,
the minima for both appearance and disappearance channels stay in the
correct octant. The octant sensitivity is seen to come from both disappearance and appearance channel. 
However, for appearance channels, the octant bands are not very widely 
separated for experiments close to second oscillation maxima, 
and with increase in $\alpha$ the effect of disappearance channel becomes more 
pronounced and there is a sharp rise in the octant sensitivity coming from the disappearance channel.

The right panel depicts the situation for the higher octant. 
In this case when the no-decay in data is fitted with decay, the 
minima for the appearance channel stays in the same octant, shifting slightly 
towards higher $\theta_{23}$. This behaviour can be understood from the appearance probability since for this HO and no-decay  gives a higher probability and hence if we fit with decay, 
the $\theta_{23}$ tends to shift to a larger value to get closer to the data. 
As the decay constant increases,
the shift in the value of $\theta_{23}$ to get closer to the no-decay curve will be higher. 

On the other hand the disappearance minima is seen to shift to the lower octant.
For a lower value of the decay constant, the difference between the disappearance $\chi^2$ minimum in the two octants is not very high. On the other hand, the 
appearance $\chi^2$ rises steeply in the opposite octant and when the appearance $\chi^2$ is added to the disappearance $\chi^2$, the overall minima comes in the correct octant. But for a higher value of the decay constant, the minima in the wrong octant is much lower than the minima in the correct octant and thus the overall minima comes in the wrong octant, being driven by disappearance channel. 
The reason for disappearance channel preferring lower octant can 
be understood from the probability figure \ref{fig:prob_oct},
by noticing that for peak flux energies of around 0.6 GeV, 
the red-curves (no-decay, HO) are closer to the blue curves (decay, lower octant). 
Hence, when data is generated in the higher octant with no decay, disappearance channel prefers the lower octant if $\alpha\neq0$ in the fit, the shift being higher 
as the decay constant increases. 
This effect is not present at the detector tuned for the first oscillation maximum at 295 km. From the top-left panel of Fig.~\ref{fig:prob_oct}, we see that the octant bands are wide enough such that the appearance channel gives high octant sensitivity and therefore the appearance channel always dominates and we do not get any false octant solution.

The above discussion can be now used to explain the kink in the sensitivity of ESS$\nu$SB (fig.~\ref{fig:sens} right panel). For ESS$\nu$SB there is only a single detector. For lower values of $\alpha$, the disappearance channel is still weak and the appearance channel dominates and we get the octant in the higher octant. However, as the $\alpha$ becomes more than some critical value, the disappearance channel begins to dominate and the octant flips to the wrong side. Thus the $\chi^{2}$ is abruptly increased, giving a lower sensitivity.







\subsection{Role of decay in the $\theta_{23}$ determination}

\begin{figure}[htb]
\begin{tabular}{ccc}
\includegraphics[width=0.3\textwidth]{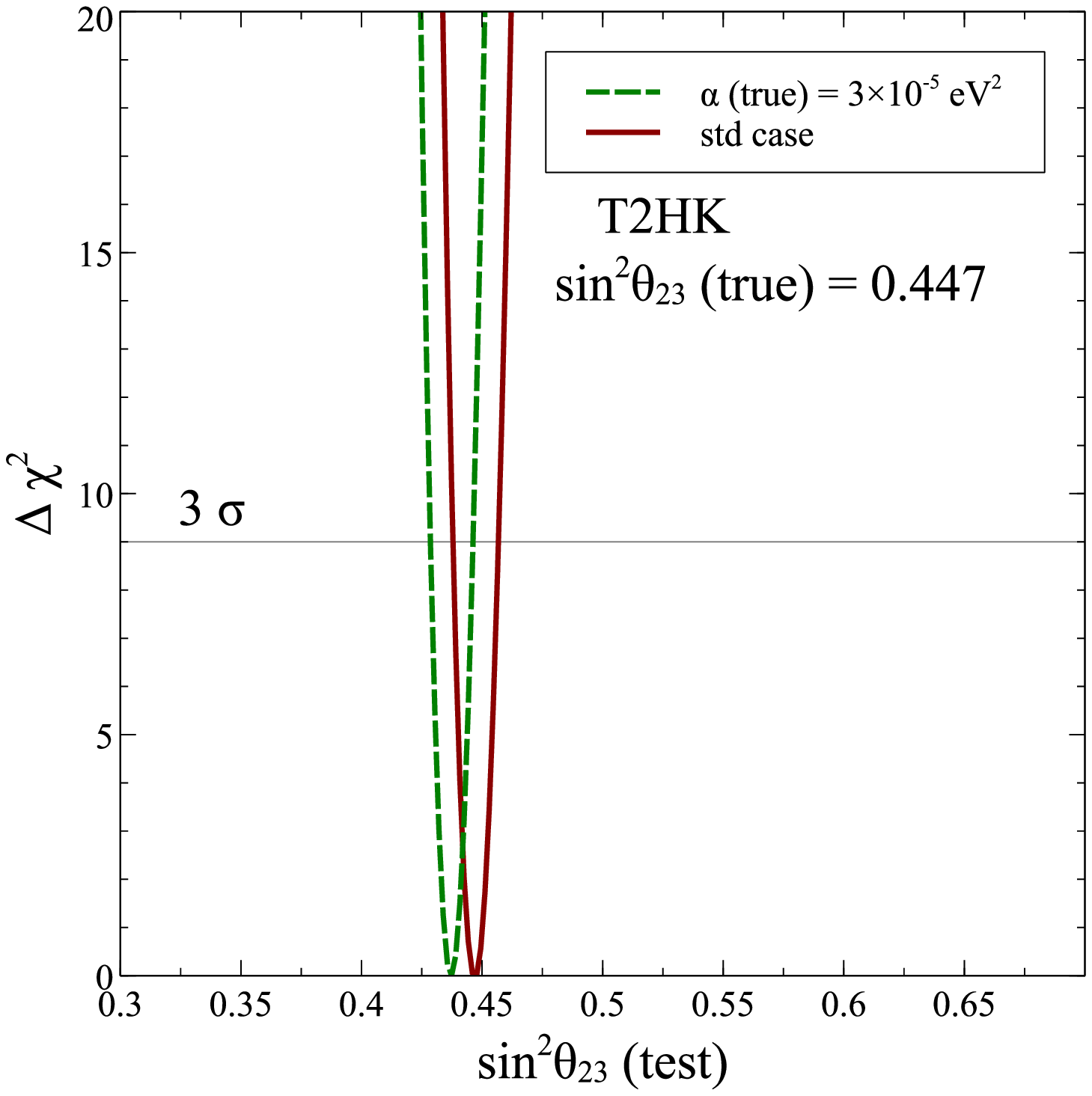}
\includegraphics[width=0.3\textwidth]{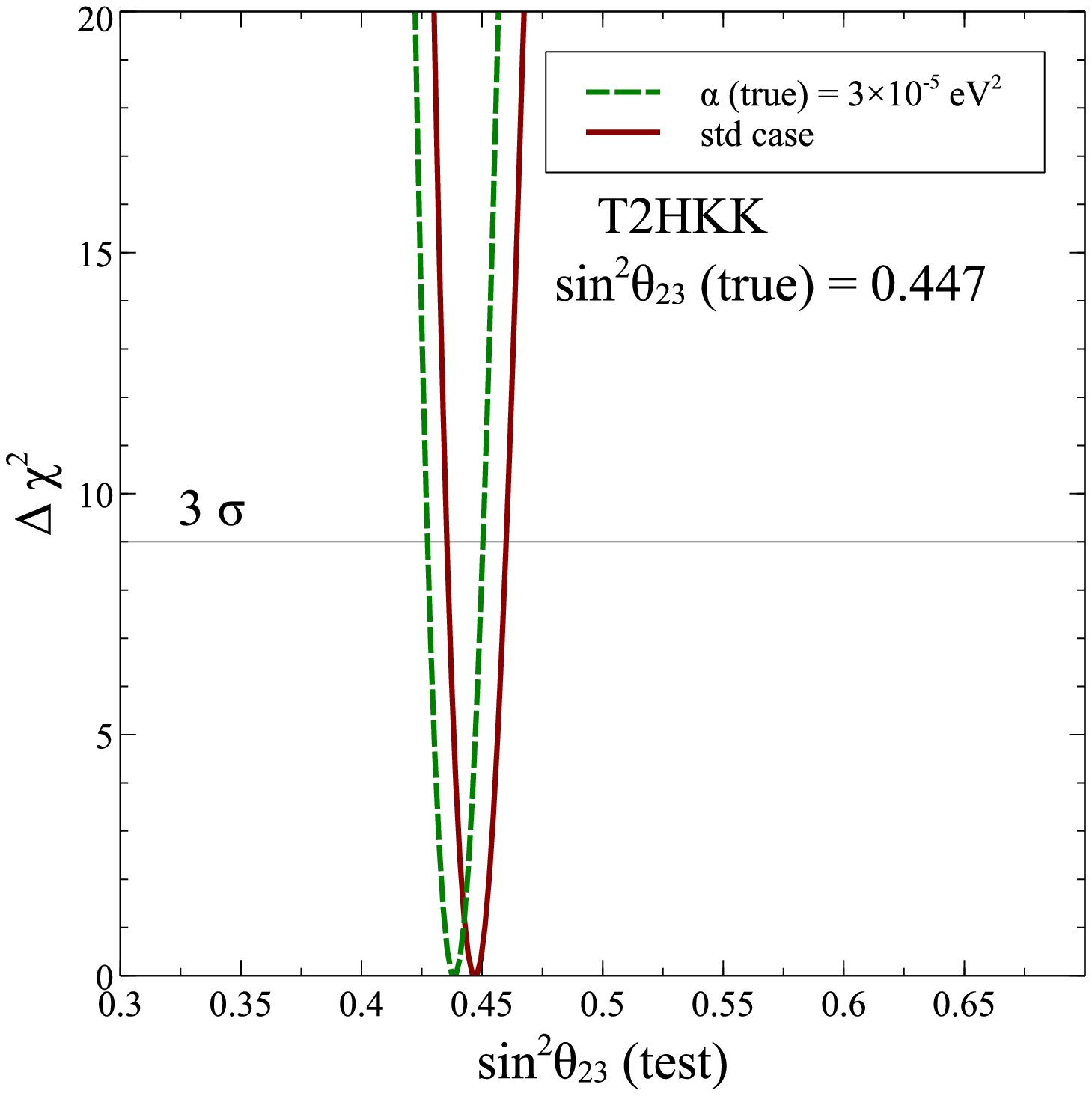}
\includegraphics[width=0.3\textwidth]{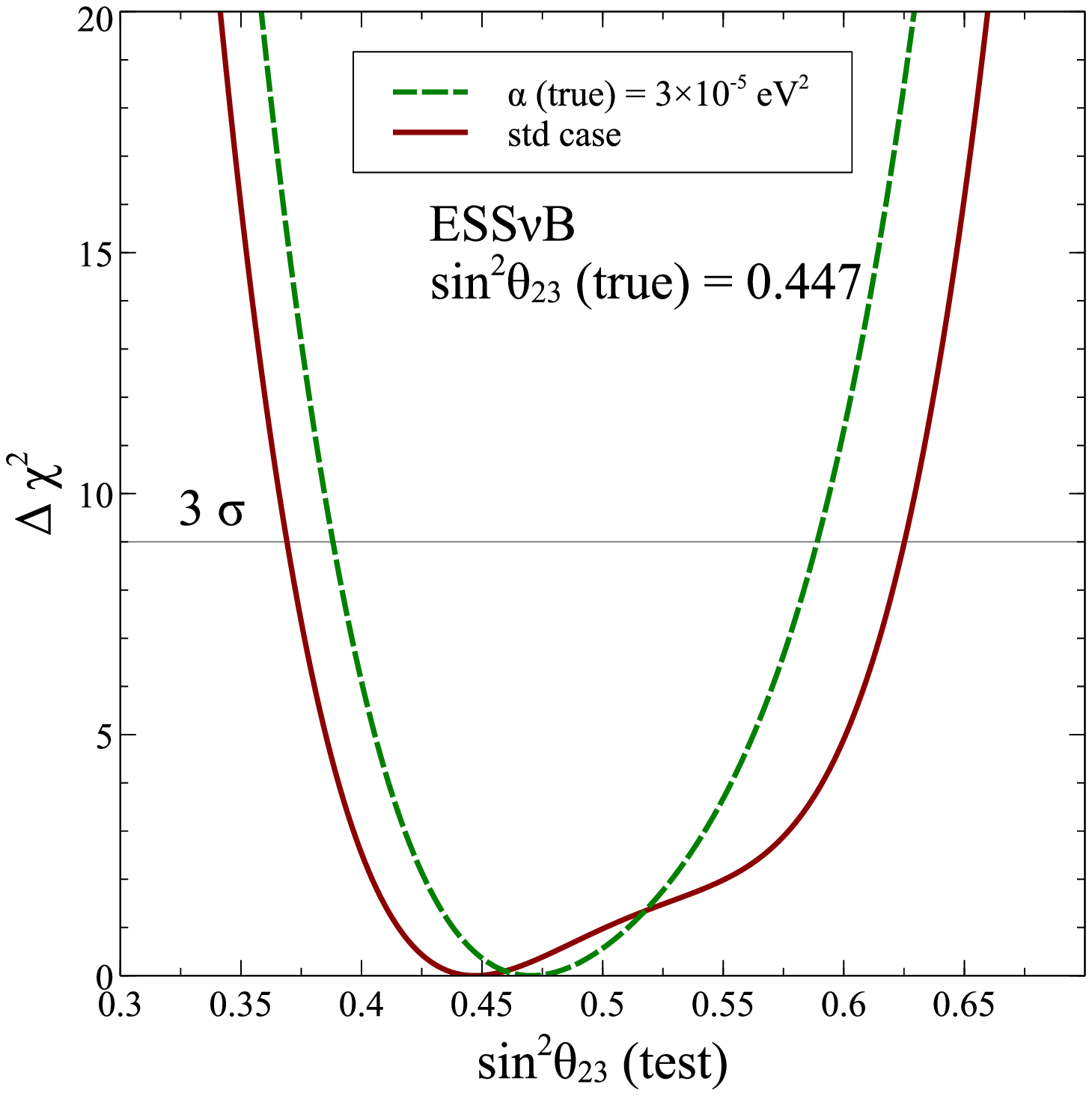}

\end{tabular}
\caption{\label{fig:th23_tr_lo} The $\Delta \chi^{2}$ as a function of $\theta_{23}$ (test) assuming $\sin^{2}\theta_{23}$(true) $ = 0.447$. The left, middle and the right panels are for T2HK, T2HKK and ESS$\nu$B respectively. The dark red solid curves are for standard cases and the green dashed curves are for the cases where $\alpha$ is assumed to be $3\times10^{-5}$ eV$^{2}$ in the simulated data but decay is not considered in the fit.  }
\end{figure}

In this section, we will see how presence of invisible neutrino decay can 
affect the measurement of $\theta_{23}$ in T2HK, T2HKK and ESS$\nu$B experiments. 

Fig.~\ref{fig:th23_tr_lo} gives $\Delta \chi^{2}$ as a function of $\theta_{23}$ (test). Here, we assumed$\sin^{2}\theta_{23}=0.447$ in the data. The left, middle and right panels are for T2HK, T2HKK and ESS$\nu$B respectively. In the fit, we marginalized over, $|\Delta m^2_{31}|$, $\delta_{CP}$ and kept $\alpha$ fixed at zero. For dark-red solid curves, we assumed stable neutrino in the data and for green dashed curves, we assumed $\alpha=3\times10^{-5}$ eV$^{2}$ in the data. 

 We see that for the three cases, the best-fit values for data generated for $\alpha=3\times10^{-5}$ eV$^{2}$ and $\alpha=0$ are different. We notice that for  true $\alpha=3\times10^{-5}$ eV$^{2}$ the best-fit of $\theta_{23}$ is shifted towards lower values for T2HK and T2HKK. However, for ESS$\nu$SB, the shift is in the opposite direction.
The shift towards lower values  for T2HK and T2HKK 
 is governed by the behaviour of $P_{\mu \mu}$  
at the oscillation maxima i.e at SPMIN, where the flux peaks. 
The data is generated for decay which gives a higher probability than no-decay in the 
lower octant.  
Therefore when data is fitted with $\alpha=0$, a reduced value of $\theta_{23}$
gives a better fit since the probability increases with decreasing $\theta_{23}$.

The reason for different behaviour for ESS$\nu$SB is more complicated. 
Here, resolution plays a big role. 
The resolution of the detector smears the imprint of probability over all energy bins. 
As a result, the feature near the oscillation minima is lost due to the larger bin width of
0.1 GeV. A bin from 0.3 GeV to 0.4 GeV contains events from the survival probability minima as well as the features of the probability around 0.3 GeV where the probabilities corresponding to decay and no decay curves 
show opposite behaviour than at $\sim$ 0.35 GeV. 
As an upshot if one plots the events then the no-decay events are higher than the events with 
decay in most bins.  
 When the simulated event spectrum is fitted without decay,
 $\theta_{23}$ is shifted towards higher value to bring the probability down. 
The  effect of smearing affects the experiments close to second oscillation maxima more since 
the variation with probability is sharper compared to the first oscillation maxima.

\begin{figure}[htb]
\begin{tabular}{ccc}
\includegraphics[width=0.3\textwidth]{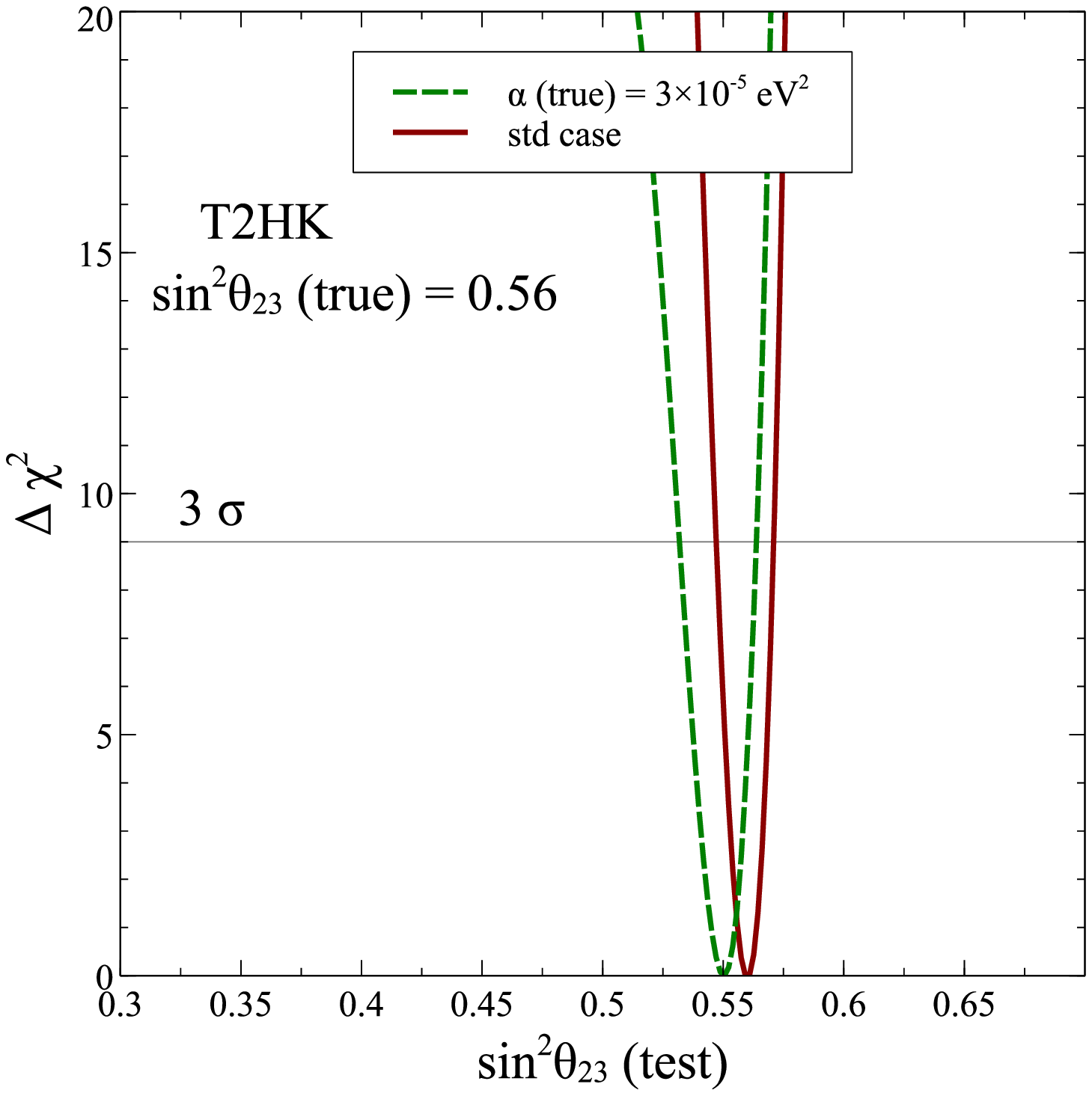}
\includegraphics[width=0.3\textwidth]{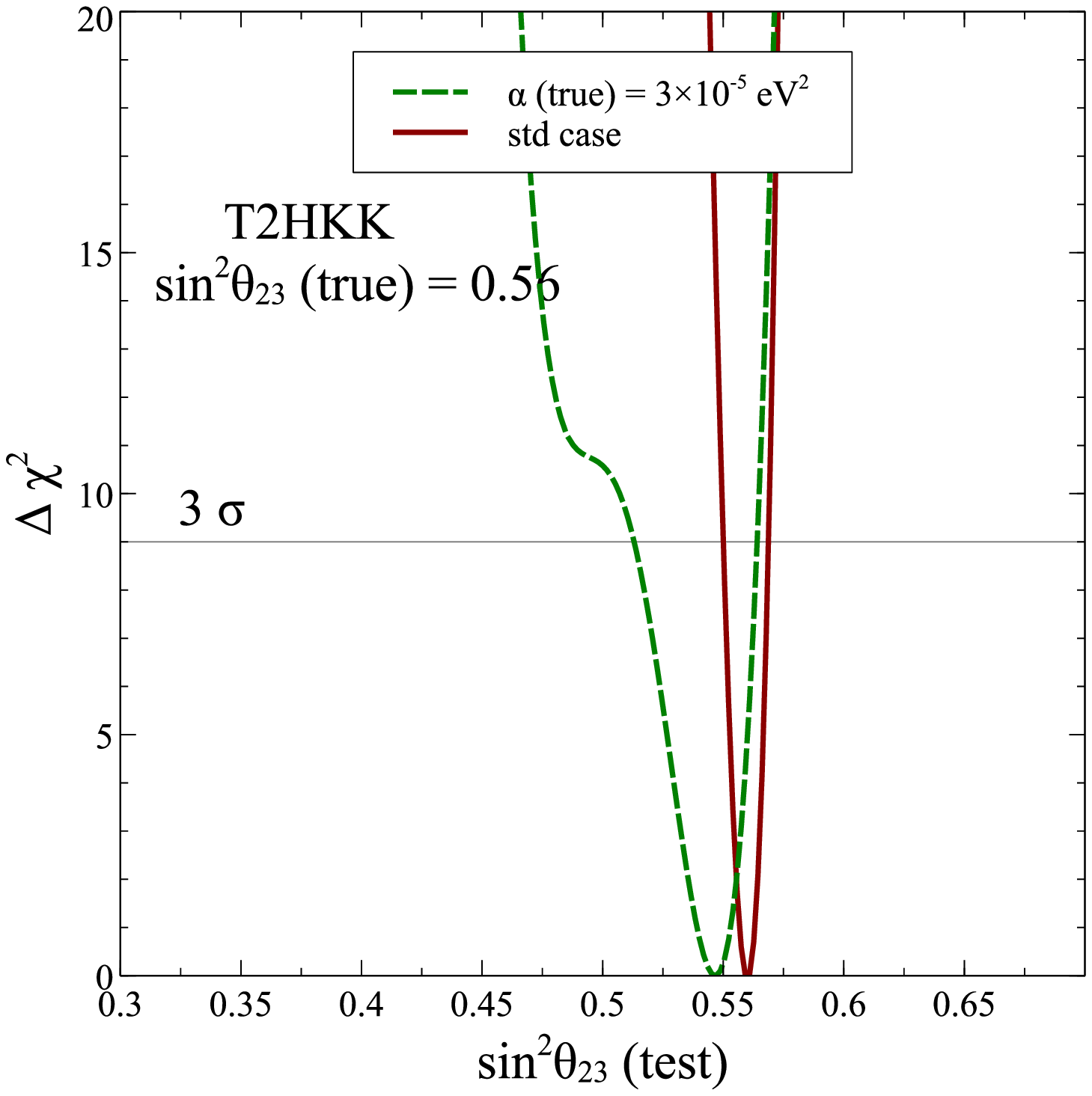}
\includegraphics[width=0.3\textwidth]{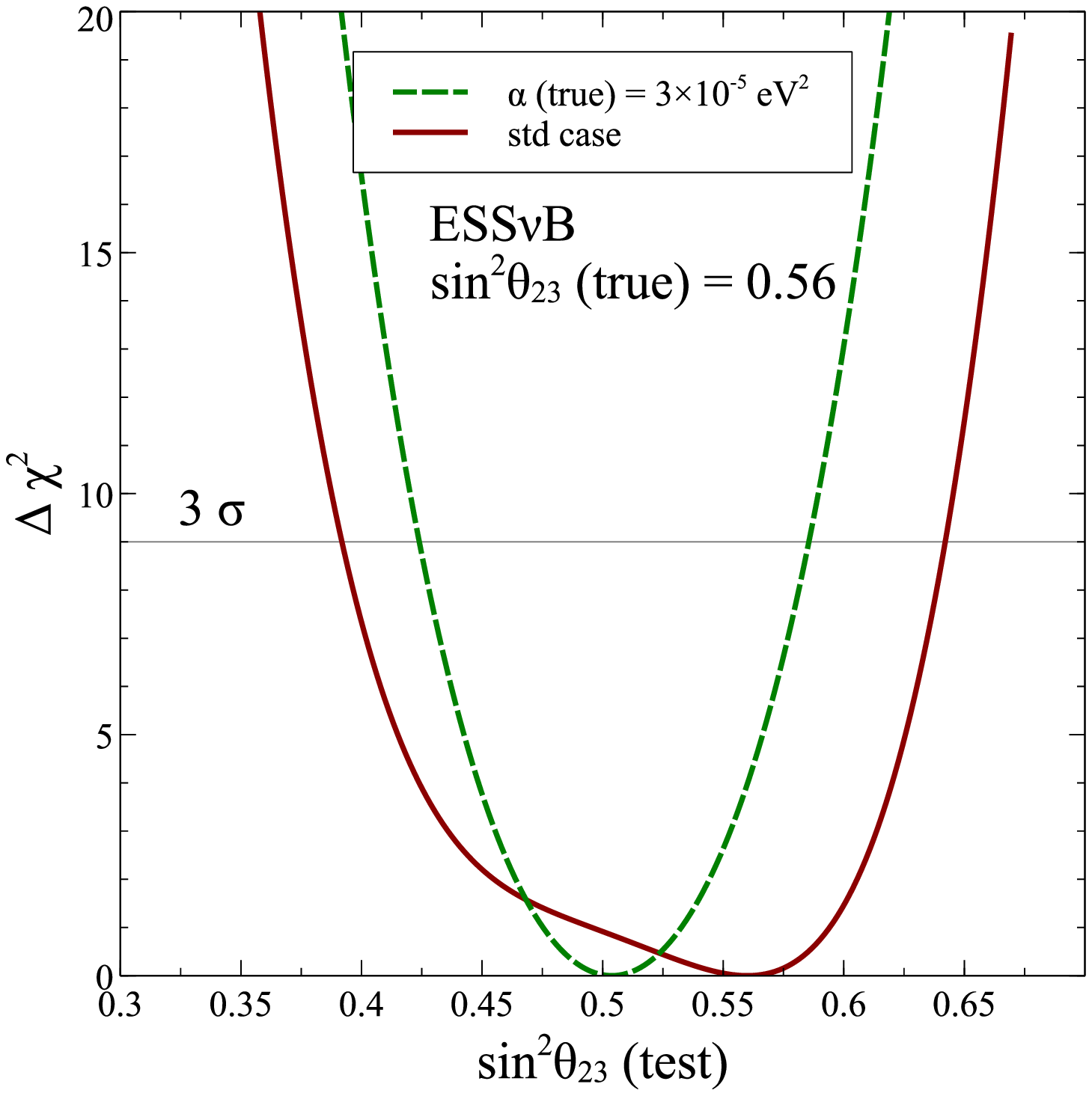}

\end{tabular}
\caption{\label{fig:th23_tr_ho} The $\Delta \chi^{2}$ as a function of $\theta_{23}$ (test) assuming $\sin^{2}\theta_{23}$(true) $ = 0.56$. The left, middle and the right panels are for T2HK, T2HKK and ESS$\nu$B respectively. The dark red solid curves are for standard cases and the green dashed curves are for the cases where $\alpha$ is assumed to be $3\times10^{-5}$ eV$^{2}$ in the simulated data but decay is not considered in the fit.  }

\end{figure}

Fig.~\ref{fig:th23_tr_ho} is similar to fig.~\ref{fig:th23_tr_lo}, but here, we assume $\sin^{2}\theta_{23}=0.56$ in the true data. We see like the previous figure, the best-fit values of $\theta_{23}$ is changed but in this case, for all three experiments, the shifts of $\theta_{23}$ are towards a lower value. 
This can be explained since,  
for $\theta_{23}$ in the higher octant the probability for non-zero $\alpha$
is smaller than the probability with $\alpha=0$ for both SPMIN and SPMAX 
and hence no-decay gives larger number of events. 
For $\theta_{23}$ in 
higher-octant, lowering $\theta_{23}$ reduces the probability and hence 
gives a better fit when we fit decay in data with no-decay.


\begin{figure}
\begin{tabular}{lr}
\includegraphics[scale=0.5]{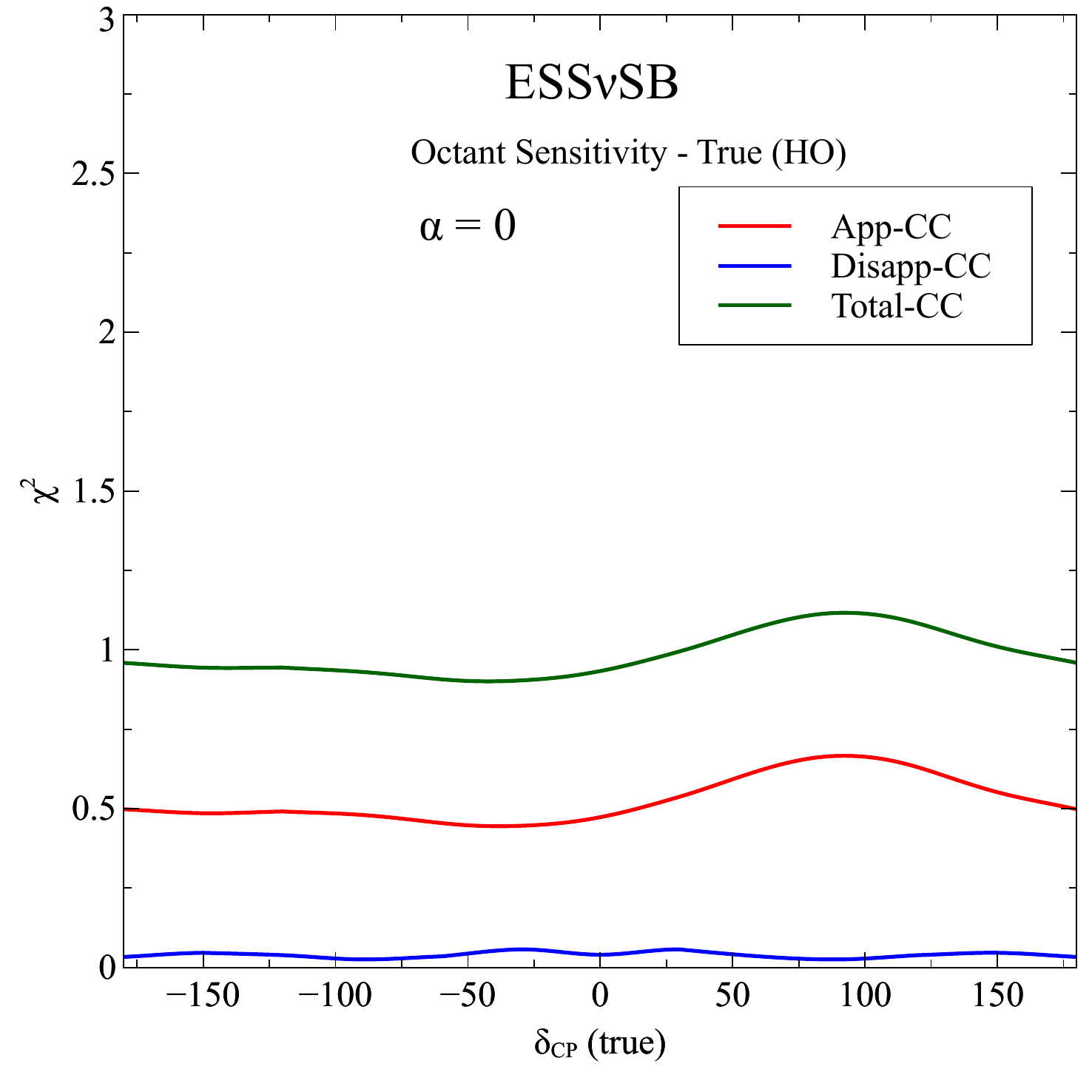}
\includegraphics[scale=0.5]{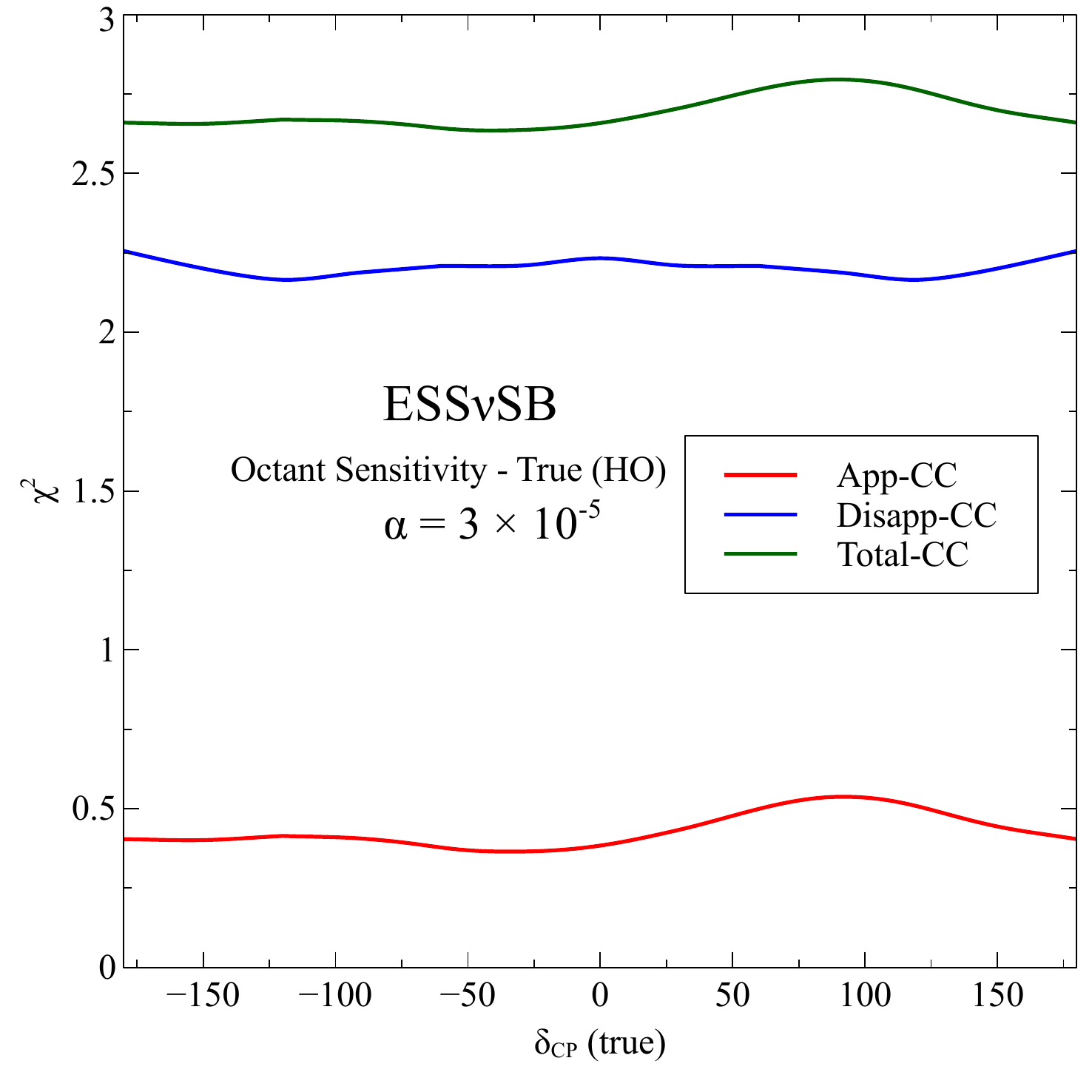} \\
\includegraphics[scale=0.5]{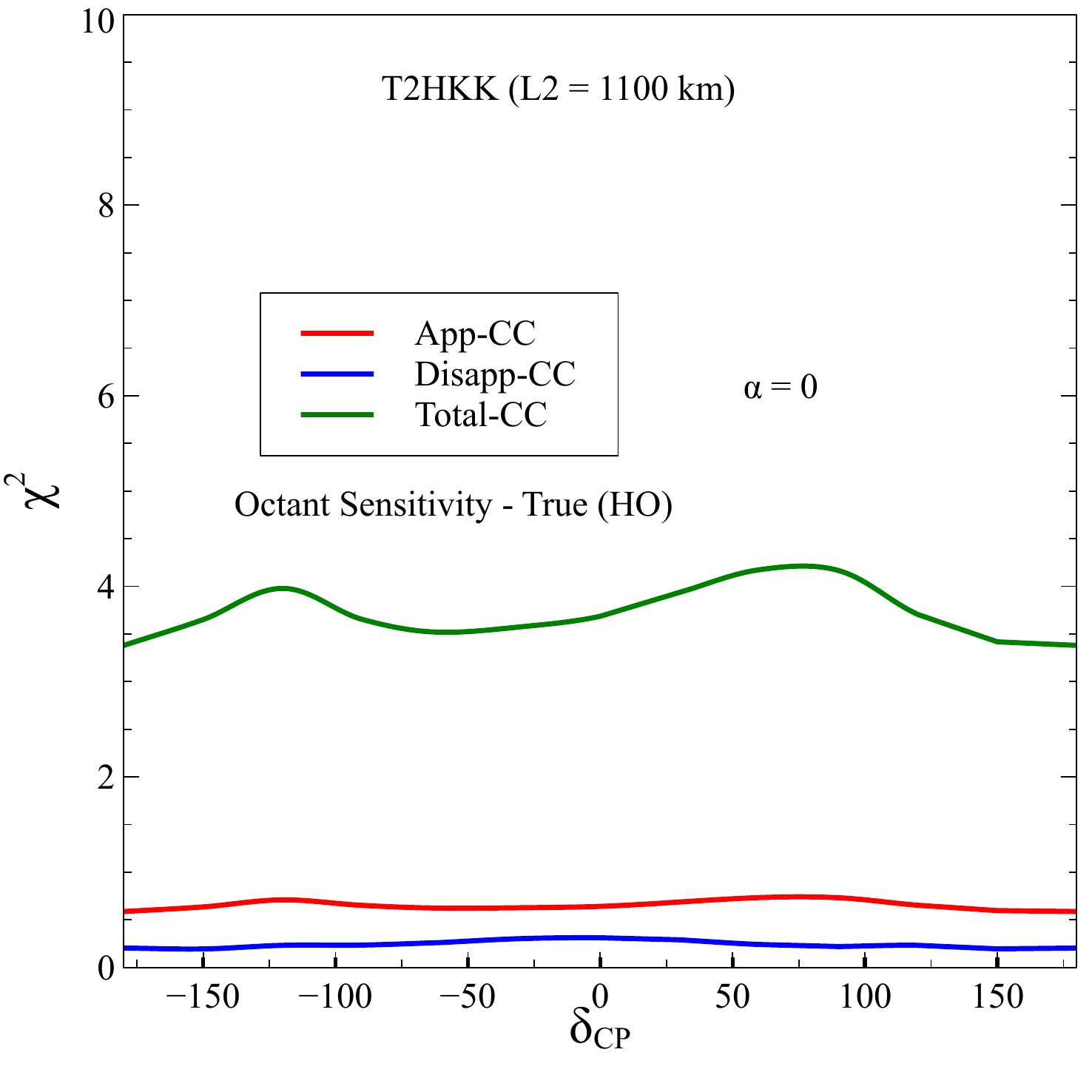}
\includegraphics[scale=0.5]{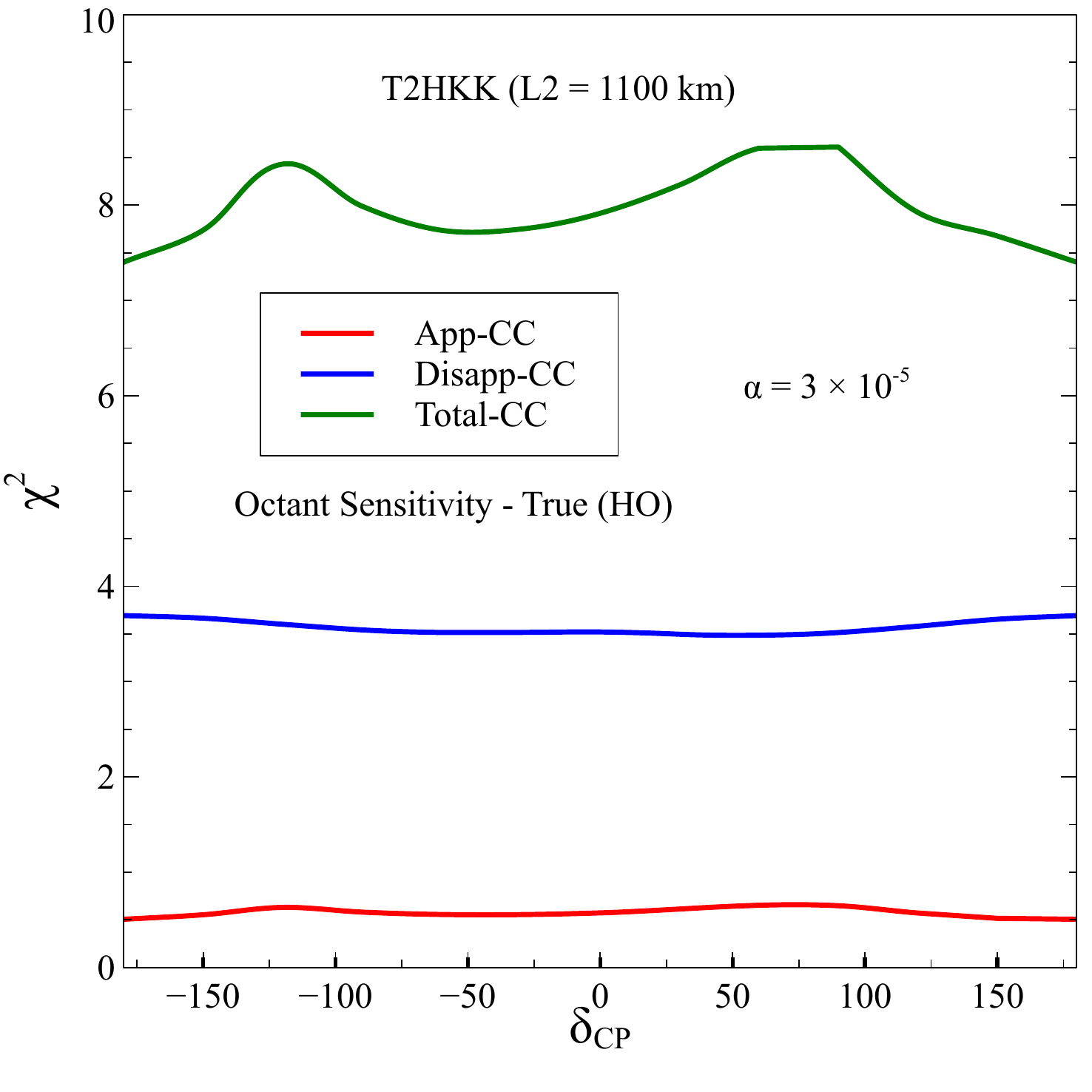}
\end{tabular}
\caption{\label{fig:ess_oct-sens} The octant sensitivity $\chi^{2}$ as a function of $\delta_{CP}$(true) for ESS$\nu$SB(top row) and T2HKK-L2 (bottom row). The left panel represents the octant sensitivity for the case without decay i.e. $\alpha=0$ and the right panel for the case with decay $\alpha=3\times10^{-5}$. The $\alpha$ values are same for both data and fit in each plot. The red and blue plots are for appearance-CC, disappearance-CC respectively, while the green plots shows the contribution from both appearance-CC and disappearance-CC. The true $\sin^{2}\theta_{23} = 0.56$ and the test $\theta_{23}$ is varied in the lower octant in the range $40^\circ < \theta_{23} < 45^\circ$.}

\end{figure}

In Fig.~\ref{fig:ess_oct-sens}, we show the octant sensitivities as a function of the 
CP phase $\delta_{CP}$ for experiments which are tuned to
the second oscillation maxima in presence and absence of decay. 
The top panels are for ESS$\nu$SB and the bottom panels are for a hypothetical T2HKK experiment whose both detectors are at 1100 km. 
The left panels are for no decay and the right panels are for the case when $\alpha\neq0$.
The red lines are for appearance channels, the blue lines are for disappearance channels and the green lines are for both appearance and disappearance channels together.
In absence of decay disappearance channel does not have any octant sensitivity. However when 
appearance $\chi^2$ is added to disappearance $\chi^2$, then the appearance probability being 
dependent on $\sin^2\theta_{23}$ rises monotonically in the opposite octant and gives a large 
octant-sensitive contribution. In presence of decay, the disappearance channel itself has 
octant sensitivity, which enhances the overall octant sensitivity.

\subsection{Combined Analysis}
\begin{figure}
\includegraphics[width=0.45\textwidth]{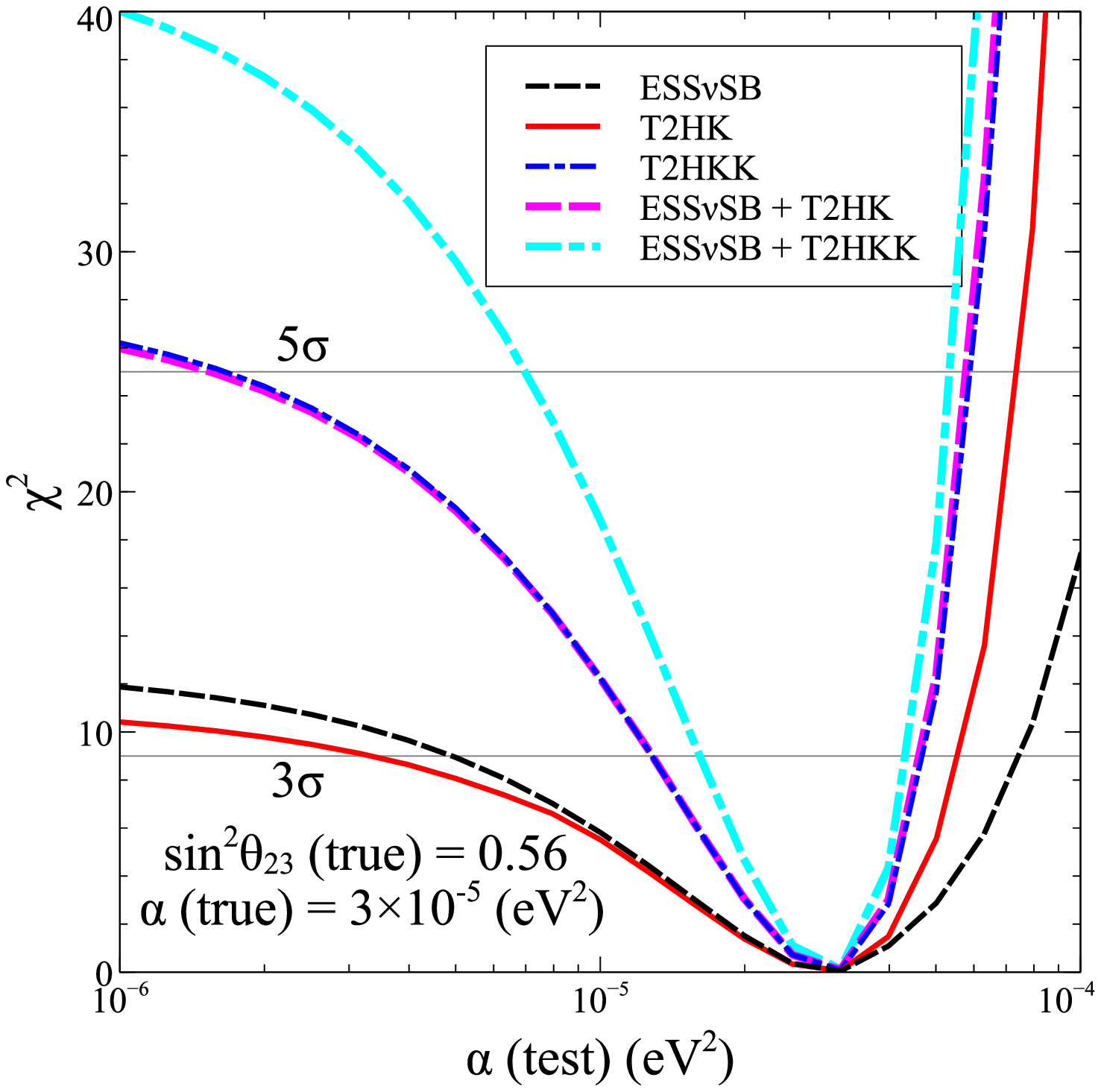}
\includegraphics[width=0.45\textwidth]{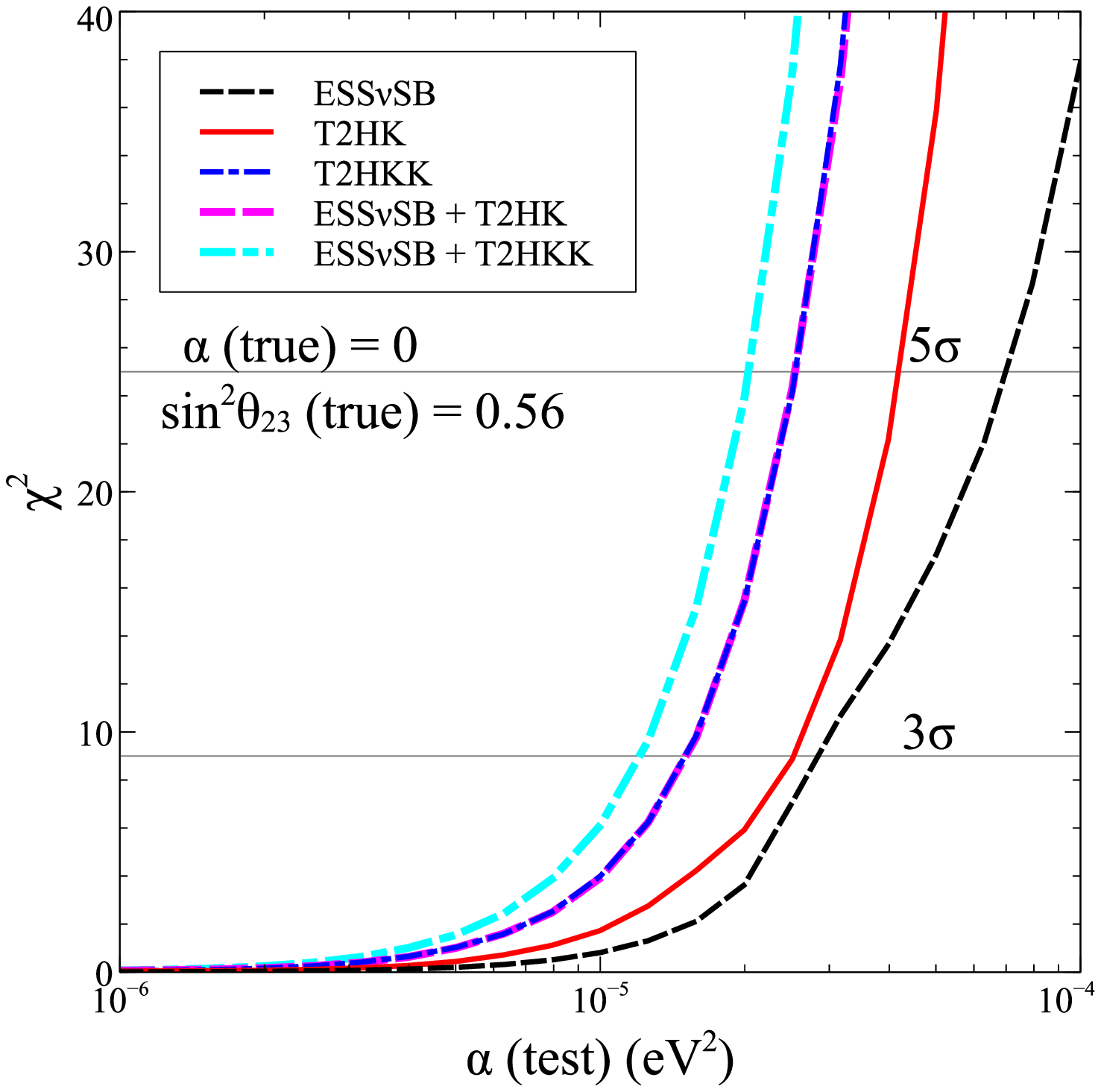}
\caption{\label{fig:combsens} The $\chi^{2}$ as a function of test $\alpha$ for various combination of experiments. The left panel is for the case where $\alpha=3\times10^{-5}$ (eV$^{2}$) is assumed for the data and the right panel is for the case where $\alpha=0$ in the decay. The black dashed curves are for ESS$\nu$SB, red solid curves are for T2HK, blue dashed-dotted curves for T2HKK, the magenta dashed for combination of ESS$\nu$SB and T2HK and the cyan curves are for combination of ESS$\nu$SB and T2HKK.
}
\end{figure}

Here in Fig.~\ref{fig:combsens}, we have shown sensitivities of different experiments to constrain the decay parameter $\alpha$(test) with (left) and without (right) decay in `data' assuming the higher octant as the true octant. The plots in the left panel show how precisely these experiments can measure $\alpha$ and hence for a given true value how they can exclude the no decay scenarios. On the other hand, plots in the right panel show the constraints these experiments can put on invisible neutrino decay. The results are shown for all the three experiments ESS$\nu$SB, T2HK, T2HKK and the combination of ESS$\nu$SB with both T2HK and T2HKK. In the Fig.~\ref{fig:combsens}, the black dashed line shows the sensitivity of ESS$\nu$SB, the blue dashed-dotted line is for T2HKK, while the red solid line represent the same for T2HK. The pink and cyan dashed line stands for the combination of ESS$\nu$SB with T2HK and T2HKK respectively.

 It is observed from the left panel where we have assumed decay in the simulated data, 
that T2HKK has the highest precision out of the three and at 3$\sigma$ it can precisely measure $\alpha$ in the range $1.278\times10^{-5} < \alpha < 4.614\times10^{-5}$ eV$^{2}$ for the given true value of $\alpha = 3\times 10^{-5} eV^{2}$. For the same true value of $\alpha$ T2HK gives better precision than ESS$\nu$SB and at 3$\sigma$ the range is $3.408\times10^{-6} < \alpha < 5.524\times10^{-5}$ eV$^{2}$ and $4.945\times10^{-6} < \alpha < 7.303\times10^{-5}$ eV$^{2}$ respectively. Combining T2HKK with ESS$\nu$SB improves the bounds further and at 3$\sigma$ the range is $1.597\times10^{-5} < \alpha < 4.231\times10^{-5}$ eV$^{2}$. Although the combination of T2HK and ESS$\nu$SB improves the bound, still it comes to be similar to that of T2HKK. In the right panel of Fig.~\ref{fig:combsens}, we observe that by combining two experiments it is possible to improve the constraint on $\alpha$. Specially, the combination of T2HKK and ESS$\nu$SB gives the highest limit on $\alpha$ and at 3$\sigma$ the constraint is $\alpha\leq1.193\times10^{-5}$ eV$^{2}$. Similarly, T2HK+ ESS$\nu$SB also improves the constrain and the derived limit in this case is very much equal to that of T2HKK. 
\begin{figure}
\includegraphics[width=0.45\textwidth]{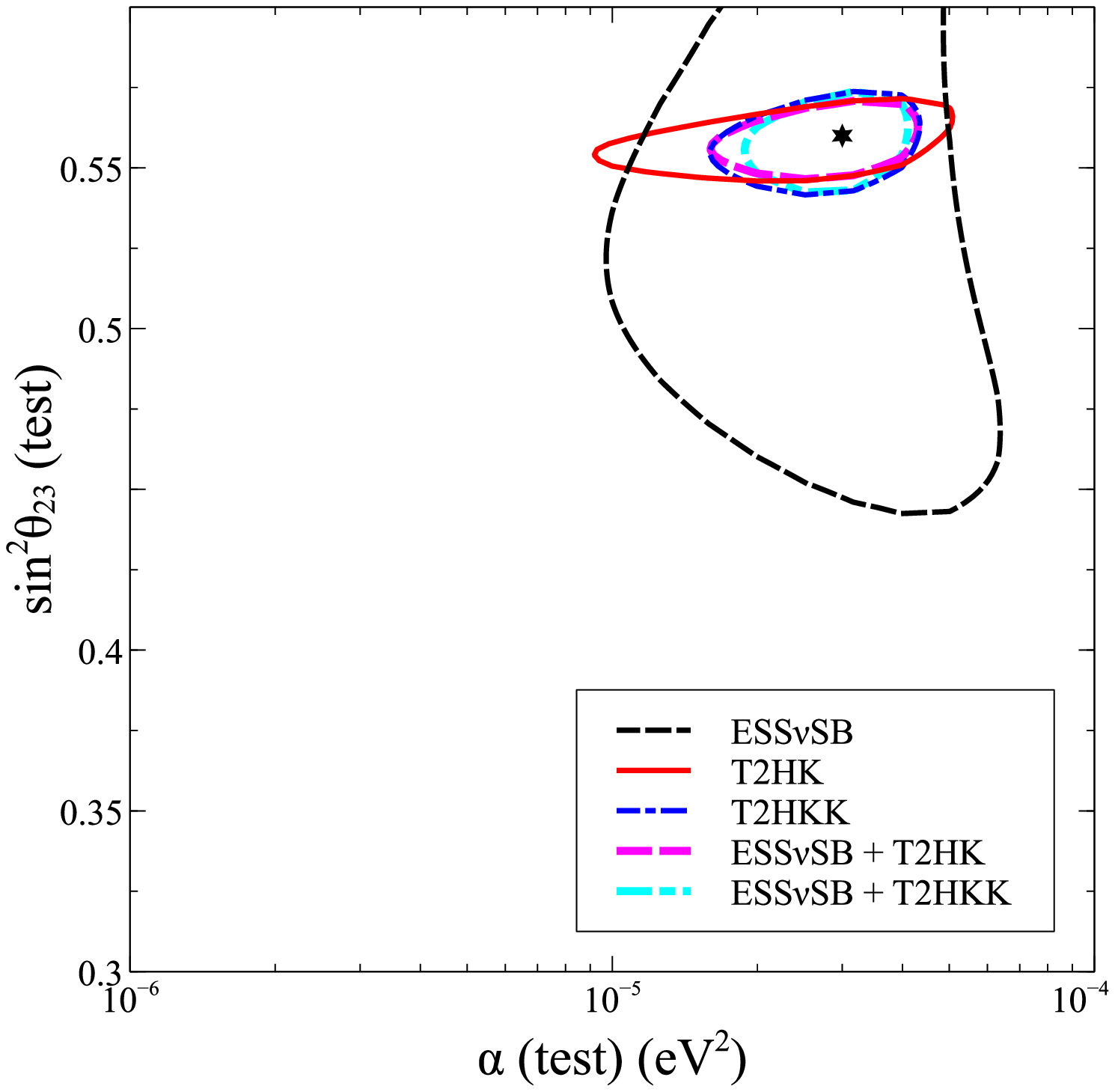}
\includegraphics[width=0.45\textwidth]{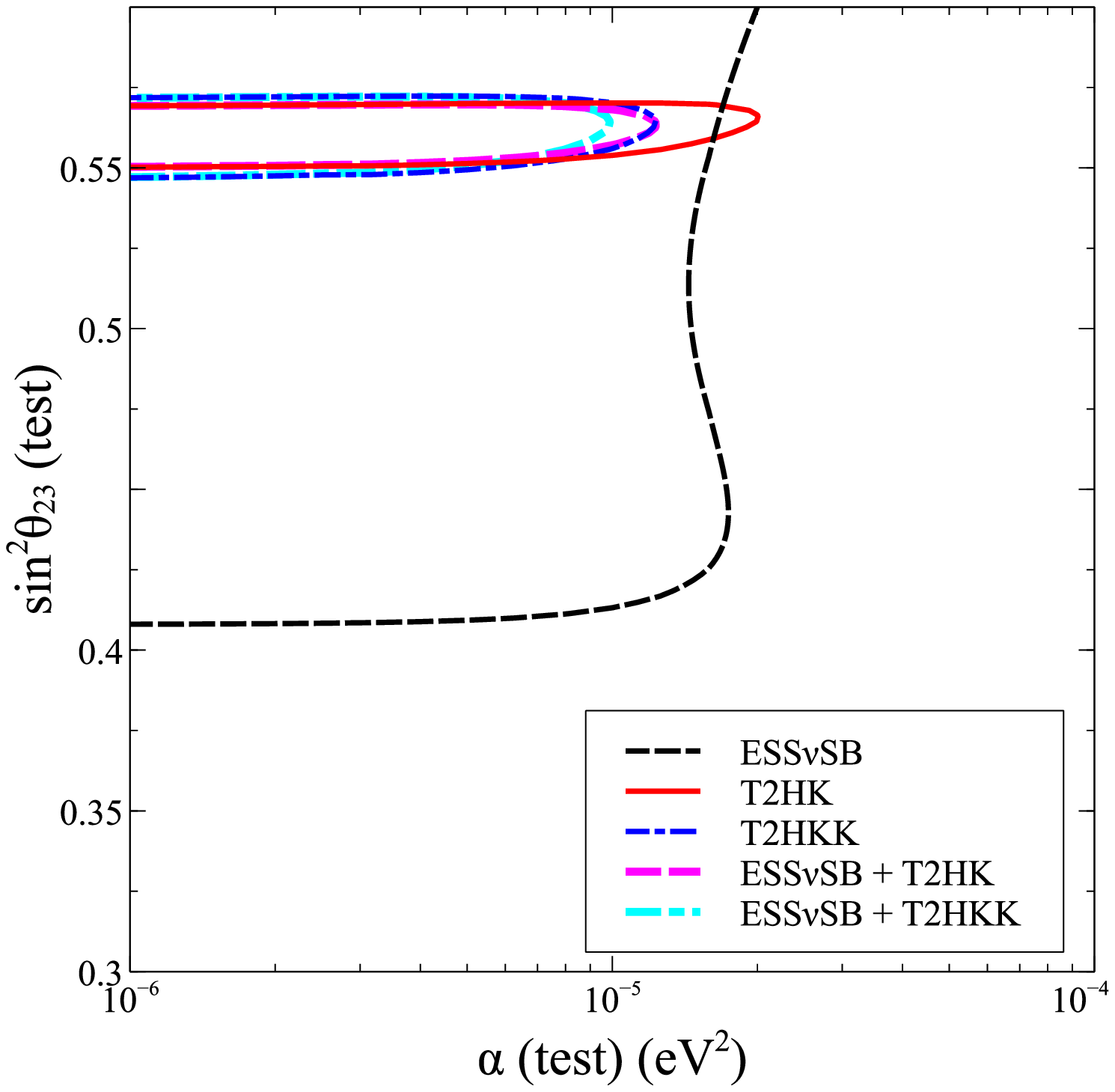}
\caption{\label{fig:combcont} The 95 \% confidence contours in the $\theta_{23}$-$\alpha$ plane for various combination of experiments. The left panel is for the case where $\alpha=3\times10^{-5}$ (eV$^{2}$) is assumed for the data and the right panel is for the case where $\alpha=0$ in the decay. The black dashed curves are for ESS$\nu$SB, red solid curves are for T2HK, blue dashed-dotted curves for T2HKK, the magenta dashed for combination of ESS$\nu$SB and T2HK and the cyan curves are for combination of ESS$\nu$SB and T2HKK.
}
\end{figure}

In Fig.~\ref{fig:combcont} we present two complementary plots to Fig.~\ref{fig:combsens}. If neutrino decays in nature in addition to oscillation, then how it affects the measurements of $\theta_{23}$ is shown in the left panel of Fig.~\ref{fig:combcont}. We have shown the results in the $\theta_{23}$(test)-$\alpha$(test) parameter space at 95$\%$ C.L. We observe that the parameter space shrinks noticeably when we combine two experiments. Again in the right panel, we show the allowed region in $\theta_{23}$(test)-$\alpha$(test) parameter space assuming no decay scenario in the simulated data. Here also, the combination of experiments gives better results that the individual experiments.

\section{Summary \& Conclusions}\label{sec:concl}

We  have examined the  physics potential of future 
long-baseline experiments T2HK/T2HKK and ESS$\nu$SB in the
 context of invisible neutrino decay. 
We performed our study for the case of 
normal hierarchy where $\nu_{3}$ is the highest mass state and this 
is assumed to be unstable, decaying into lighter sterile states.
We compared and contrasted the sensitivities to decay of these experiments 
with special emphasis on the  location of the experiments 
at first and second oscillation maximum and  the various 
factors that can affect the sensitivities. 
 

The effect of decay appears as $\exp{(-\alpha L/E)}$ where, 
$\alpha$ is the decay constant, $L$ is the distance traveled and $E$ is the energy. 
Hence, it is expected that second oscillation maxima experiments having 
higher baselines and/or lower energy than the first 
effect of decay depends on the $L/E$ factor and second oscillation maximum
occurs for lower energy it can be expected that effect of decay can be
more pronounced for second oscillation maxima.
However,
the experimental sensitivities for the decay rate also depend on other
vital factors like neutrino flux, resolution of the detector and
the sensitivity of the experiment to resolve $\theta_{23}$.
We expounded in detail which factors play important roles in 
determining the sensitivity to decay at the first and second oscillation 
maximum. 
Among the three experiments T2HK is designed to have its flux peak at the first oscillation maxima, T2HKK has one detector at the first oscillation maximum 
and one near the second oscillation maximum, while ESS$\nu$SB is an experiment 
at the second oscillation maximum.  
T2HK has a lower L/E as compared to the other experiments and has less sensitivity to decay. On the other hand being at the first oscillation maxima it has a 
better sensitivity to $\theta_{23}$ and its octant. 
In comparison, ESS$\nu$SB  has a higher L/E, but because of  
lower energy resolution as well as poor sensitivity to $\theta_{23}$, 
it has a reduced sensitivity to the decay constant $\alpha$.
The two detector set up of T2HKK helps since the detector at the first oscillation maximum helps in determining $\theta_{23}$ while the second helps
in better constraining $\alpha$.
We have studied the sensitivity of these experiments to the decay rate  $\alpha$
for true $\theta_{23}$ in both lower and higher octant.

We have also presented the octant sensitivity as a function of $\delta_{CP}$ 
for the three experiments. 
We have obtained an interesting result that in presence of decay, the 
overall octant sensitivity is enhanced. This can be attributed to the octant 
sensitive contribution coming from the disappearance channel in presence of 
decay. 
Since T2HK is  at the first oscillation maxima, the appearance 
channel already gives a good sensitivity and the effect of disappearance channel 
is not very significant. But since ESS$\nu$SB is at the second oscillation maximum
the octant sensitivity coming from appearance channel is not 
very high and   the disappearance channel 
plays a consequential role in enhancing the octant sensitivity in presence of 
decay.  Similar feature is also observed for the detector  
at 1100 km baseline 
with flux peak near the second oscillation maxima. 


The octant sensitivity coming from the disappearance channel  
increases with the decay constant $\alpha$ and also  
leads to an interesting interplay with the appearance channel 
in determining the overall minimum for experiments at 
second oscillation maxima. At the first oscillation maxima on the other hand, the appearance channel dominates. As a result of this different behaviour, 
for T2HKK, which has one detector at first and one at second oscillation 
maxima, a synergy is observed  for true $\theta_{23}$ in the higher octant
leading to a $\chi^2$ greater
than the naive sum of the $\chi^2$s at two different baselines.  
Since at second oscillation maximum the appearance channel does not
have very good octant sensitivity
the $\chi^2$ minimum for the second detector
comes in the  wrong octant driven by the disappearance channel.
The $\chi^2$ minima for the detector at the first oscillation maximum
on the other hand stays in the correct octant (the wrong octant solution
being disfavoured by the appearance channel).
As an upshot, the global minima while combining the two baselines come in
a position different from the individual minima resulting in the synergy.
On the other hand, for ESS$\nu$SB,  the first detector is not present 
and the leaning of the disappearance channel towards the wrong octant, 
weaken the sensitivity after a certain value of the decay constant $\alpha$.


Given the correlation between $\alpha$ and $\theta_{23}$, the 
determination of $\theta_{23}$ can also get affected, if there 
is decay in nature but it is ignored in the fit resulting in a wrong 
determination of $\theta_{23}$.  We have examined this issue in detail and found that the shift of  $\theta_{23}$ from its true value, can be  
in either direction, depending on the octant of true $\theta_{23}$ 
and  whether the contribution is coming from bins centered at 
the maxima or minima of the survival probability, 
which in turn depends on the smearing and bin-width. 
This effect is more for the second oscillation maxima experiments since 
the probability band is narrower. 

Apart from the sensitivity study for individual experiments we have also performed sensitivity studies for the two combinations of T2HK+ESS$\nu$SB and T2HKK+ESS$\nu$SB.
We found that the sensitivity increases significantly if we combine the experiments. 
The sensitivity to the decay-rate attained by the experiments T2HK, T2HKK, ESS$\nu$SB and their combinations of T2HK+ESS$\nu$SB, T2HKK+ESS$\nu$SB have been summarized in the Tab.\ref{tab:limit-alphatau} .

\begin{table}[h!]
\begin{center}
\begin{tabular}{|c|c|c|}
\hline
Experiment(s) & $\alpha~ \rm{(eV^2)}$  & $\tau_{3}/m_3~ \rm{(s/eV)}$ \\
\hline
\hline
T2HK & $\geq ~2.42 \times 10^{-5}$  & $\leq ~2.72 \times 10^{-11}$    \\
\hline
T2HKK & $\geq ~1.51 \times 10^{-5}$     & $\leq~ 4.36 \times 10^{-11}$  \\
\hline
ESS$\nu$SB & $\geq ~2.71 \times 10^{-5}$         & $\leq~ 2.43 \times 10^{-11}$    \\
\hline
T2HK+ESS$\nu$SB & $\geq ~1.51 \times 10^{-5}$         & $\leq~ 4.36 \times 10^{-11}$   \\
\hline
T2HKK+ESS$\nu$SB & $\geq ~1.19 \times 10^{-5}$     & $\leq~ 5.53 \times 10^{-11}$    \\
\hline
\hline
\end{tabular}
\end{center} 
\caption{{\footnotesize The sensitivity to the decay rate $\alpha$ and the neutrino decay lifetime at 3$\sigma$ CL for the standalone experiments T2HK, T2HKK, ESS$\nu$SB and also for the combinations of T2HK+ESS$\nu$SB, T2HKK+ESS$\nu$SB.}}
\label{tab:limit-alphatau}
\end{table} 

We have also done a precision study assuming $\alpha=3\times10^{-5}$ eV$^{2}$ in the data and found  that the precision is maximum for the combination of T2HKK+ESS$\nu$SB while standalone T2HKK has the best precision. 
This also reiterates the fact that combination of first and second oscillation maximum can give the best sensitivity to decay. 
Although combination of the experiments improve the sensitivity to $\alpha$, 
no significant improvement is observed for $\theta_{23}$.

In conclusion, we studied the sensitivity to decay at experiments 
at the first and second oscillation maxima and discussed the salient 
features. Overall the combination of first and second oscillation maxima 
is best for decay since at the first oscillation maximum the precision 
of $\theta_{23}$ is more and at the second oscillation maximum the 
sensitivity to decay could be more because of enhanced baseline (for same energy)
and/or lower energy (for same baseline). 
Our study underscores the 
importance of the disappearance channel in giving enhanced 
octant sensitivity for second oscillation maxima experiments in presence of 
decay and the importance 
of better energy resolution in obtaining a higher sensitivity to decay.

{\it Note Added}: While this work was being written a paper \cite{Choubey:2020dhw}
came on arXiv, which also considered neutrino decay in the context 
of ESS$\nu$SB experiment. Our study contains a comparative analysis 
of T2HK, T2HKK and ESS$\nu$SB experiment. 
 
\section*{Acknowledgment}
The authors would like to thank Enrique Fernandez-Martinez for providing them with the GLoBES files used for the simulation of the ESS$\nu$SB experiment. DD and DP would like to acknowledge the organisers of WHEPP-XVI for their hospitality where some initial discussions about the work took place.

\bibliographystyle{JHEP}
\bibliography{ref}

\providecommand{\href}[2]{#2}\begingroup\raggedright\begin{thebibliography}{10}

\bibitem{Esteban:2020cvm}
I.~Esteban, M.~Gonzalez-Garcia, M.~Maltoni, T.~Schwetz and A.~Zhou, \emph{{The
  fate of hints: updated global analysis of three-flavor neutrino
  oscillations}},  \href{https://arxiv.org/abs/2007.14792}{{\ttfamily
  2007.14792}}.

\bibitem{deSalas:2020pgw}
P.~de~Salas, D.~Forero, S.~Gariazzo, P.~Mart\'\i{}nez-Mirav\'e, O.~Mena,
  C.~Ternes et~al., \emph{{2020 Global reassessment of the neutrino oscillation
  picture}},  \href{https://arxiv.org/abs/2006.11237}{{\ttfamily 2006.11237}}.

\bibitem{Capozzi:2017ipn}
F.~Capozzi, E.~Di~Valentino, E.~Lisi, A.~Marrone, A.~Melchiorri and A.~Palazzo,
  \emph{{Global constraints on absolute neutrino masses and their ordering}},
  \href{https://doi.org/10.1103/PhysRevD.95.096014}{\emph{Phys. Rev. D}
  {\bfseries 95} (2017) 096014}
  [\href{https://arxiv.org/abs/2003.08511}{{\ttfamily 2003.08511}}].

\bibitem{Acciarri:2015uup}
{\scshape DUNE} collaboration, \emph{{Long-Baseline Neutrino Facility (LBNF)
  and Deep Underground Neutrino Experiment (DUNE)}: {Conceptual Design Report,
  Volume 2: The Physics Program for DUNE at LBNF}},
  \href{https://arxiv.org/abs/1512.06148}{{\ttfamily 1512.06148}}.

\bibitem{Acciarri:2016crz}
{\scshape DUNE} collaboration, \emph{{Long-Baseline Neutrino Facility (LBNF)
  and Deep Underground Neutrino Experiment (DUNE)}: {Conceptual Design Report,
  Volume 1: The LBNF and DUNE Projects}},
  \href{https://arxiv.org/abs/1601.05471}{{\ttfamily 1601.05471}}.

\bibitem{Acciarri:2016ooe}
{\scshape DUNE} collaboration, \emph{{Long-Baseline Neutrino Facility (LBNF)
  and Deep Underground Neutrino Experiment (DUNE)}: {Conceptual Design Report,
  Volume 4 The DUNE Detectors at LBNF}},
  \href{https://arxiv.org/abs/1601.02984}{{\ttfamily 1601.02984}}.

\bibitem{Abi:2018dnh}
{\scshape DUNE} collaboration, \emph{{The DUNE Far Detector Interim Design
  Report Volume 1: Physics, Technology and Strategies}},
  \href{https://arxiv.org/abs/1807.10334}{{\ttfamily 1807.10334}}.

\bibitem{Strait:2016mof}
{\scshape DUNE} collaboration, \emph{{Long-Baseline Neutrino Facility (LBNF)
  and Deep Underground Neutrino Experiment (DUNE)}: {Conceptual Design Report,
  Volume 3: Long-Baseline Neutrino Facility for DUNE June 24, 2015}},
  \href{https://arxiv.org/abs/1601.05823}{{\ttfamily 1601.05823}}.

\bibitem{Abe:2015zbg}
{\scshape Hyper-Kamiokande Proto-} collaboration, \emph{{Physics potential of a
  long-baseline neutrino oscillation experiment using a J-PARC neutrino beam
  and Hyper-Kamiokande}},
  \href{https://doi.org/10.1093/ptep/ptv061}{\emph{PTEP} {\bfseries 2015}
  (2015) 053C02} [\href{https://arxiv.org/abs/1502.05199}{{\ttfamily
  1502.05199}}].

\bibitem{Abe:2016ero}
{\scshape Hyper-Kamiokande} collaboration, \emph{{Physics potentials with the
  second Hyper-Kamiokande detector in Korea}},
  \href{https://doi.org/10.1093/ptep/pty044}{\emph{PTEP} {\bfseries 2018}
  (2018) 063C01} [\href{https://arxiv.org/abs/1611.06118}{{\ttfamily
  1611.06118}}].

\bibitem{Baussan:2013zcy}
{\scshape ESSnuSB} collaboration, \emph{{A very intense neutrino super beam
  experiment for leptonic CP violation discovery based on the European
  spallation source linac}},
  \href{https://doi.org/10.1016/j.nuclphysb.2014.05.016}{\emph{Nucl. Phys. B}
  {\bfseries 885} (2014) 127}
  [\href{https://arxiv.org/abs/1309.7022}{{\ttfamily 1309.7022}}].

\bibitem{An:2015jdp}
{\scshape JUNO} collaboration, \emph{{Neutrino Physics with JUNO}},
  \href{https://doi.org/10.1088/0954-3899/43/3/030401}{\emph{J. Phys. G}
  {\bfseries 43} (2016) 030401}
  [\href{https://arxiv.org/abs/1507.05613}{{\ttfamily 1507.05613}}].

\bibitem{Kumar:2017sdq}
{\scshape ICAL} collaboration, \emph{{Physics Potential of the ICAL detector at
  the India-based Neutrino Observatory (INO)}},
  \href{https://doi.org/10.1007/s12043-017-1373-4}{\emph{Pramana} {\bfseries
  88} (2017) 79} [\href{https://arxiv.org/abs/1505.07380}{{\ttfamily
  1505.07380}}].

\bibitem{Aartsen:2014oha}
{\scshape IceCube PINGU} collaboration, \emph{{Letter of Intent: The Precision
  IceCube Next Generation Upgrade (PINGU)}},
  \href{https://arxiv.org/abs/1401.2046}{{\ttfamily 1401.2046}}.

\bibitem{Margiotta:2014gza}
{\scshape KM3NeT} collaboration, \emph{{The KM3NeT deep-sea neutrino
  telescope}}, \href{https://doi.org/10.1016/j.nima.2014.05.090}{\emph{Nucl.
  Instrum. Meth. A} {\bfseries 766} (2014) 83}
  [\href{https://arxiv.org/abs/1408.1392}{{\ttfamily 1408.1392}}].

\bibitem{Bahcall:1972my}
J.N.~Bahcall, N.~Cabibbo and A.~Yahil, \emph{{Are neutrinos stable
  particles?}}, \href{https://doi.org/10.1103/PhysRevLett.28.316}{\emph{Phys.
  Rev. Lett.} {\bfseries 28} (1972) 316}.

\bibitem{Acker:1993sz}
A.~Acker and S.~Pakvasa, \emph{{Solar neutrino decay}},
  \href{https://doi.org/10.1016/0370-2693(94)90663-7}{\emph{Phys. Lett. B}
  {\bfseries 320} (1994) 320}
  [\href{https://arxiv.org/abs/hep-ph/9310207}{{\ttfamily hep-ph/9310207}}].

\bibitem{Berezhiani:1991vk}
Z.~Berezhiani, G.~Fiorentini, M.~Moretti and A.~Rossi, \emph{{Fast neutrino
  decay and solar neutrino detectors}},
  \href{https://doi.org/10.1007/BF01559483}{\emph{Z. Phys. C} {\bfseries 54}
  (1992) 581}.

\bibitem{Berezhiani:1992xg}
Z.G.~Berezhiani, M.~Moretti and A.~Rossi, \emph{{Matter induced neutrino decay
  and solar anti-neutrinos}},
  \href{https://doi.org/10.1007/BF01557699}{\emph{Z. Phys. C} {\bfseries 58}
  (1993) 423}.

\bibitem{Choubey:2000an}
S.~Choubey, S.~Goswami and D.~Majumdar, \emph{{Status of the neutrino decay
  solution to the solar neutrino problem}},
  \href{https://doi.org/10.1016/S0370-2693(00)00608-0}{\emph{Phys. Lett. B}
  {\bfseries 484} (2000) 73}
  [\href{https://arxiv.org/abs/hep-ph/0004193}{{\ttfamily hep-ph/0004193}}].

\bibitem{Bandyopadhyay:2001ct}
A.~Bandyopadhyay, S.~Choubey and S.~Goswami, \emph{{MSW mediated neutrino decay
  and the solar neutrino problem}},
  \href{https://doi.org/10.1103/PhysRevD.63.113019}{\emph{Phys. Rev. D}
  {\bfseries 63} (2001) 113019}
  [\href{https://arxiv.org/abs/hep-ph/0101273}{{\ttfamily hep-ph/0101273}}].

\bibitem{Joshipura:2002fb}
A.S.~Joshipura, E.~Masso and S.~Mohanty, \emph{{Constraints on decay plus
  oscillation solutions of the solar neutrino problem}},
  \href{https://doi.org/10.1103/PhysRevD.66.113008}{\emph{Phys. Rev. D}
  {\bfseries 66} (2002) 113008}
  [\href{https://arxiv.org/abs/hep-ph/0203181}{{\ttfamily hep-ph/0203181}}].

\bibitem{Bandyopadhyay:2002qg}
A.~Bandyopadhyay, S.~Choubey and S.~Goswami, \emph{{Neutrino decay confronts
  the SNO data}},
  \href{https://doi.org/10.1016/S0370-2693(03)00044-3}{\emph{Phys. Lett. B}
  {\bfseries 555} (2003) 33}
  [\href{https://arxiv.org/abs/hep-ph/0204173}{{\ttfamily hep-ph/0204173}}].

\bibitem{Picoreti:2015ika}
R.~Picoreti, M.~Guzzo, P.~de~Holanda and O.~Peres, \emph{{Neutrino Decay and
  Solar Neutrino Seasonal Effect}},
  \href{https://doi.org/10.1016/j.physletb.2016.08.007}{\emph{Phys. Lett. B}
  {\bfseries 761} (2016) 70}
  [\href{https://arxiv.org/abs/1506.08158}{{\ttfamily 1506.08158}}].

\bibitem{Berryman:2014qha}
J.M.~Berryman, A.~de~Gouvea and D.~Hernandez, \emph{{Solar Neutrinos and the
  Decaying Neutrino Hypothesis}},
  \href{https://doi.org/10.1103/PhysRevD.92.073003}{\emph{Phys. Rev. D}
  {\bfseries 92} (2015) 073003}
  [\href{https://arxiv.org/abs/1411.0308}{{\ttfamily 1411.0308}}].

\bibitem{Frieman:1987as}
J.A.~Frieman, H.E.~Haber and K.~Freese, \emph{{Neutrino Mixing, Decays and
  Supernova Sn1987a}},
  \href{https://doi.org/10.1016/0370-2693(88)91120-3}{\emph{Phys. Lett. B}
  {\bfseries 200} (1988) 115}.

\bibitem{Huang:2018nxj}
G.-Y.~Huang and S.~Zhou, \emph{{Constraining Neutrino Lifetimes and Magnetic
  Moments via Solar Neutrinos in the Large Xenon Detectors}},
  \href{https://doi.org/10.1088/1475-7516/2019/02/024}{\emph{JCAP} {\bfseries
  02} (2019) 024} [\href{https://arxiv.org/abs/1810.03877}{{\ttfamily
  1810.03877}}].

\bibitem{LoSecco:1998cd}
J.~LoSecco, \emph{{What the atmospheric neutrino anomaly is not}},
  \href{https://arxiv.org/abs/hep-ph/9809499}{{\ttfamily hep-ph/9809499}}.

\bibitem{Barger:1998xk}
V.D.~Barger, J.~Learned, S.~Pakvasa and T.J.~Weiler, \emph{{Neutrino decay as
  an explanation of atmospheric neutrino observations}},
  \href{https://doi.org/10.1103/PhysRevLett.82.2640}{\emph{Phys. Rev. Lett.}
  {\bfseries 82} (1999) 2640}
  [\href{https://arxiv.org/abs/astro-ph/9810121}{{\ttfamily
  astro-ph/9810121}}].

\bibitem{Lipari:1999vh}
P.~Lipari and M.~Lusignoli, \emph{{On exotic solutions of the atmospheric
  neutrino problem}},
  \href{https://doi.org/10.1103/PhysRevD.60.013003}{\emph{Phys. Rev. D}
  {\bfseries 60} (1999) 013003}
  [\href{https://arxiv.org/abs/hep-ph/9901350}{{\ttfamily hep-ph/9901350}}].

\bibitem{Fogli:1999qt}
G.L.~Fogli, E.~Lisi, A.~Marrone and G.~Scioscia, \emph{{Super-Kamiokande data
  and atmospheric neutrino decay}},
  \href{https://doi.org/10.1103/PhysRevD.59.117303}{\emph{Phys. Rev. D}
  {\bfseries 59} (1999) 117303}
  [\href{https://arxiv.org/abs/hep-ph/9902267}{{\ttfamily hep-ph/9902267}}].

\bibitem{Choubey:1999ir}
S.~Choubey and S.~Goswami, \emph{{Is neutrino decay really ruled out as a
  solution to the atmospheric neutrino problem from Super-Kamiokande data?}},
  \href{https://doi.org/10.1016/S0927-6505(00)00106-7}{\emph{Astropart. Phys.}
  {\bfseries 14} (2000) 67}
  [\href{https://arxiv.org/abs/hep-ph/9904257}{{\ttfamily hep-ph/9904257}}].

\bibitem{Barger:1999bg}
V.D.~Barger, J.~Learned, P.~Lipari, M.~Lusignoli, S.~Pakvasa and T.J.~Weiler,
  \emph{{Neutrino decay and atmospheric neutrinos}},
  \href{https://doi.org/10.1016/S0370-2693(99)00887-4}{\emph{Phys. Lett. B}
  {\bfseries 462} (1999) 109}
  [\href{https://arxiv.org/abs/hep-ph/9907421}{{\ttfamily hep-ph/9907421}}].

\bibitem{Ashie:2004mr}
{\scshape Super-Kamiokande} collaboration, \emph{{Evidence for an oscillatory
  signature in atmospheric neutrino oscillation}},
  \href{https://doi.org/10.1103/PhysRevLett.93.101801}{\emph{Phys. Rev. Lett.}
  {\bfseries 93} (2004) 101801}
  [\href{https://arxiv.org/abs/hep-ex/0404034}{{\ttfamily hep-ex/0404034}}].

\bibitem{GonzalezGarcia:2008ru}
M.~Gonzalez-Garcia and M.~Maltoni, \emph{{Status of Oscillation plus Decay of
  Atmospheric and Long-Baseline Neutrinos}},
  \href{https://doi.org/10.1016/j.physletb.2008.04.041}{\emph{Phys. Lett. B}
  {\bfseries 663} (2008) 405}
  [\href{https://arxiv.org/abs/0802.3699}{{\ttfamily 0802.3699}}].

\bibitem{Gomes:2014yua}
R.~Gomes, A.~Gomes and O.~Peres, \emph{{Constraints on neutrino decay lifetime
  using long-baseline charged and neutral current data}},
  \href{https://doi.org/10.1016/j.physletb.2014.12.014}{\emph{Phys. Lett. B}
  {\bfseries 740} (2015) 345}
  [\href{https://arxiv.org/abs/1407.5640}{{\ttfamily 1407.5640}}].

\bibitem{Choubey:2018cfz}
S.~Choubey, D.~Dutta and D.~Pramanik, \emph{{Invisible neutrino decay in the
  light of NOvA and T2K data}},
  \href{https://doi.org/10.1007/JHEP08(2018)141}{\emph{JHEP} {\bfseries 08}
  (2018) 141} [\href{https://arxiv.org/abs/1805.01848}{{\ttfamily
  1805.01848}}].

\bibitem{Denton:2018aml}
P.B.~Denton and I.~Tamborra, \emph{{Invisible Neutrino Decay Could Resolve
  IceCube's Track and Cascade Tension}},
  \href{https://doi.org/10.1103/PhysRevLett.121.121802}{\emph{Phys. Rev. Lett.}
  {\bfseries 121} (2018) 121802}
  [\href{https://arxiv.org/abs/1805.05950}{{\ttfamily 1805.05950}}].

\bibitem{Abrahao:2015rba}
T.~Abrahão, H.~Minakata, H.~Nunokawa and A.A.~Quiroga, \emph{{Constraint on
  Neutrino Decay with Medium-Baseline Reactor Neutrino Oscillation
  Experiments}}, \href{https://doi.org/10.1007/JHEP11(2015)001}{\emph{JHEP}
  {\bfseries 11} (2015) 001}
  [\href{https://arxiv.org/abs/1506.02314}{{\ttfamily 1506.02314}}].

\bibitem{Beacom:2002vi}
J.F.~Beacom, N.F.~Bell, D.~Hooper, S.~Pakvasa and T.J.~Weiler, \emph{{Decay of
  High-Energy Astrophysical Neutrinos}},
  \href{https://doi.org/10.1103/PhysRevLett.90.181301}{\emph{Phys. Rev. Lett.}
  {\bfseries 90} (2003) 181301}
  [\href{https://arxiv.org/abs/hep-ph/0211305}{{\ttfamily hep-ph/0211305}}].

\bibitem{Maltoni:2008jr}
M.~Maltoni and W.~Winter, \emph{{Testing neutrino oscillations plus decay with
  neutrino telescopes}},
  \href{https://doi.org/10.1088/1126-6708/2008/07/064}{\emph{JHEP} {\bfseries
  07} (2008) 064} [\href{https://arxiv.org/abs/0803.2050}{{\ttfamily
  0803.2050}}].

\bibitem{Pakvasa:2012db}
S.~Pakvasa, A.~Joshipura and S.~Mohanty, \emph{{Explanation for the low flux of
  high energy astrophysical muon-neutrinos}},
  \href{https://doi.org/10.1103/PhysRevLett.110.171802}{\emph{Phys. Rev. Lett.}
  {\bfseries 110} (2013) 171802}
  [\href{https://arxiv.org/abs/1209.5630}{{\ttfamily 1209.5630}}].

\bibitem{Pagliaroli:2015rca}
G.~Pagliaroli, A.~Palladino, F.~Villante and F.~Vissani, \emph{{Testing
  nonradiative neutrino decay scenarios with IceCube data}},
  \href{https://doi.org/10.1103/PhysRevD.92.113008}{\emph{Phys. Rev. D}
  {\bfseries 92} (2015) 113008}
  [\href{https://arxiv.org/abs/1506.02624}{{\ttfamily 1506.02624}}].

\bibitem{Bustamante:2016ciw}
M.~Bustamante, J.F.~Beacom and K.~Murase, \emph{{Testing decay of astrophysical
  neutrinos with incomplete information}},
  \href{https://doi.org/10.1103/PhysRevD.95.063013}{\emph{Phys. Rev. D}
  {\bfseries 95} (2017) 063013}
  [\href{https://arxiv.org/abs/1610.02096}{{\ttfamily 1610.02096}}].

\bibitem{Choubey:2017dyu}
S.~Choubey, S.~Goswami and D.~Pramanik, \emph{{A study of invisible neutrino
  decay at DUNE and its effects on $\theta_{23}$ measurement}},
  \href{https://doi.org/10.1007/JHEP02(2018)055}{\emph{JHEP} {\bfseries 02}
  (2018) 055} [\href{https://arxiv.org/abs/1705.05820}{{\ttfamily
  1705.05820}}].

\bibitem{Ghoshal:2020hyo}
A.~Ghoshal, A.~Giarnetti and D.~Meloni, \emph{{Neutrino Invisible Decay at
  DUNE: a multi-channel analysis}},
  \href{https://arxiv.org/abs/2003.09012}{{\ttfamily 2003.09012}}.

\bibitem{Tang:2018rer}
J.~Tang, T.-C.~Wang and Y.~Zhang, \emph{{Invisible neutrino decays at the
  MOMENT experiment}},
  \href{https://doi.org/10.1007/JHEP04(2019)004}{\emph{JHEP} {\bfseries 04}
  (2019) 004} [\href{https://arxiv.org/abs/1811.05623}{{\ttfamily
  1811.05623}}].

\bibitem{Choubey:2017eyg}
S.~Choubey, S.~Goswami, C.~Gupta, S.~Lakshmi and T.~Thakore, \emph{{Sensitivity
  to neutrino decay with atmospheric neutrinos at the INO-ICAL detector}},
  \href{https://doi.org/10.1103/PhysRevD.97.033005}{\emph{Phys. Rev. D}
  {\bfseries 97} (2018) 033005}
  [\href{https://arxiv.org/abs/1709.10376}{{\ttfamily 1709.10376}}].

\bibitem{Mohan:2020tbi}
L.~Mohan, \emph{{Probing the sensitivity to leptonic $\delta_{CP}$ in presence
  of invisible decay of $\nu_3$ using atmospheric neutrinos}},
  \href{https://arxiv.org/abs/2006.04233}{{\ttfamily 2006.04233}}.

\bibitem{deSalas:2018kri}
P.~de~Salas, S.~Pastor, C.~Ternes, T.~Thakore and M.~Tórtola,
  \emph{{Constraining the invisible neutrino decay with KM3NeT-ORCA}},
  \href{https://doi.org/10.1016/j.physletb.2018.12.066}{\emph{Phys. Lett. B}
  {\bfseries 789} (2019) 472}
  [\href{https://arxiv.org/abs/1810.10916}{{\ttfamily 1810.10916}}].

\bibitem{Acker:1991ej}
A.~Acker, S.~Pakvasa and J.T.~Pantaleone, \emph{{Decaying Dirac neutrinos}},
  \href{https://doi.org/10.1103/PhysRevD.45.R1}{\emph{Phys. Rev. D} {\bfseries
  45} (1992) 1}.

\bibitem{Gelmini:1980re}
G.~Gelmini and M.~Roncadelli, \emph{{Left-Handed Neutrino Mass Scale and
  Spontaneously Broken Lepton Number}},
  \href{https://doi.org/10.1016/0370-2693(81)90559-1}{\emph{Phys. Lett. B}
  {\bfseries 99} (1981) 411}.

\bibitem{Chikashige:1980ui}
Y.~Chikashige, R.N.~Mohapatra and R.~Peccei, \emph{{Are There Real Goldstone
  Bosons Associated with Broken Lepton Number?}},
  \href{https://doi.org/10.1016/0370-2693(81)90011-3}{\emph{Phys. Lett. B}
  {\bfseries 98} (1981) 265}.

\bibitem{Pakvasa:1999ta}
S.~Pakvasa, \emph{{Do neutrinos decay?}},
  \href{https://doi.org/10.1063/1.1336244}{\emph{AIP Conf. Proc.} {\bfseries
  542} (2000) 99} [\href{https://arxiv.org/abs/hep-ph/0004077}{{\ttfamily
  hep-ph/0004077}}].

\bibitem{Kim:1990km}
C.~Kim and W.~Lam, \emph{{Some remarks on neutrino decay via a Nambu-Goldstone
  boson}}, \href{https://doi.org/10.1142/S0217732390000354}{\emph{Mod. Phys.
  Lett. A} {\bfseries 5} (1990) 297}.

\bibitem{Acker:1992eh}
A.~Acker, A.~Joshipura and S.~Pakvasa, \emph{{A Neutrino decay model, solar
  anti-neutrinos and atmospheric neutrinos}},
  \href{https://doi.org/10.1016/0370-2693(92)91520-J}{\emph{Phys. Lett. B}
  {\bfseries 285} (1992) 371}.

\bibitem{Lindner:2001fx}
M.~Lindner, T.~Ohlsson and W.~Winter, \emph{{A Combined treatment of neutrino
  decay and neutrino oscillations}},
  \href{https://doi.org/10.1016/S0550-3213(01)00237-1}{\emph{Nucl. Phys. B}
  {\bfseries 607} (2001) 326}
  [\href{https://arxiv.org/abs/hep-ph/0103170}{{\ttfamily hep-ph/0103170}}].

\bibitem{Coloma:2017zpg}
P.~Coloma and O.L.G.~Peres, \emph{{Visible neutrino decay at DUNE}},
  \href{https://arxiv.org/abs/1705.03599}{{\ttfamily 1705.03599}}.

\bibitem{Gago:2017zzy}
A.M.~Gago, R.A.~Gomes, A.L.G.~Gomes, J.~Jones-Perez and O.L.G.~Peres,
  \emph{{Visible neutrino decay in the light of appearance and disappearance
  long baseline experiments}},
  \href{https://doi.org/10.1007/JHEP11(2017)022}{\emph{JHEP} {\bfseries 11}
  (2017) 022} [\href{https://arxiv.org/abs/1705.03074}{{\ttfamily
  1705.03074}}].

\bibitem{Ascencio-Sosa:2018lbk}
M.~Ascencio-Sosa, A.~Calatayud-Cadenillas, A.~Gago and J.~Jones-Pérez,
  \emph{{Matter effects in neutrino visible decay at future long-baseline
  experiments}},
  \href{https://doi.org/10.1140/epjc/s10052-018-6276-0}{\emph{Eur. Phys. J. C}
  {\bfseries 78} (2018) 809}
  [\href{https://arxiv.org/abs/1805.03279}{{\ttfamily 1805.03279}}].

\bibitem{Porto-Silva:2020gma}
Y.P.~Porto-Silva, S.~Prakash, O.~Peres, H.~Nunokawa and H.~Minakata,
  \emph{{Constraining visible neutrino decay at KamLAND and JUNO}},
  \href{https://arxiv.org/abs/2002.12134}{{\ttfamily 2002.12134}}.

\bibitem{Abdullahi:2020rge}
A.~Abdullahi and P.B.~Denton, \emph{{Visible Decay of Astrophysical Neutrinos
  at IceCube}}, \href{https://doi.org/10.1103/PhysRevD.102.023018}{\emph{Phys.
  Rev. D} {\bfseries 102} (2020) 023018}
  [\href{https://arxiv.org/abs/2005.07200}{{\ttfamily 2005.07200}}].

\bibitem{Escudero:2019gfk}
M.~Escudero and M.~Fairbairn, \emph{{Cosmological Constraints on Invisible
  Neutrino Decays Revisited}},
  \href{https://doi.org/10.1103/PhysRevD.100.103531}{\emph{Phys. Rev. D}
  {\bfseries 100} (2019) 103531}
  [\href{https://arxiv.org/abs/1907.05425}{{\ttfamily 1907.05425}}].

\bibitem{Fukuda:2002uc}
{\scshape Super-Kamiokande} collaboration, \emph{{The Super-Kamiokande
  detector}}, \href{https://doi.org/10.1016/S0168-9002(03)00425-X}{\emph{Nucl.
  Instrum. Meth. A} {\bfseries 501} (2003) 418}.

\bibitem{Kajita:2006bt}
T.~Kajita, H.~Minakata, S.~Nakayama and H.~Nunokawa, \emph{{Resolving
  eight-fold neutrino parameter degeneracy by two identical detectors with
  different baselines}},
  \href{https://doi.org/10.1103/PhysRevD.75.013006}{\emph{Phys. Rev. D}
  {\bfseries 75} (2007) 013006}
  [\href{https://arxiv.org/abs/hep-ph/0609286}{{\ttfamily hep-ph/0609286}}].

\bibitem{globes1}
P.~Huber, M.~Lindner and W.~Winter, \emph{{Simulation of long-baseline neutrino
  oscillation experiments with GLoBES}},
  \href{https://doi.org/10.1016/j.cpc.2005.01.003}{\emph{Comput. Phys. Commun.}
  {\bfseries 167} (2005) 195}
  [\href{https://arxiv.org/abs/hep-ph/0407333}{{\ttfamily hep-ph/0407333}}].

\bibitem{globes2}
P.~Huber, J.~Kopp, M.~Lindner, M.~Rolinec and W.~Winter, \emph{{New features in
  the simulation of neutrino oscillation experiments with GLoBES 3.0}},
  \href{https://doi.org/10.1016/j.cpc.2007.05.004}{\emph{Comput. Phys. Commun.}
  {\bfseries 177} (2007) 432}
  [\href{https://arxiv.org/abs/hep-ph/0701187}{{\ttfamily hep-ph/0701187}}].

\bibitem{Agostino:2012fd}
{\scshape MEMPHYS} collaboration, \emph{{Study of the performance of a large
  scale water-Cherenkov detector (MEMPHYS)}},
  \href{https://doi.org/10.1088/1475-7516/2013/01/024}{\emph{JCAP} {\bfseries
  01} (2013) 024} [\href{https://arxiv.org/abs/1206.6665}{{\ttfamily
  1206.6665}}].

\bibitem{nufit}
N.~5.0. \url{www.nu-fit.org}, 2020.

\bibitem{Choubey:2020dhw}
S.~Choubey, M.~Ghosh, D.~Kempe and T.~Ohlsson, \emph{{Exploring invisible
  neutrino decay at ESSnuSB}},
  \href{https://arxiv.org/abs/2010.16334}{{\ttfamily 2010.16334}}.

\end{thebibliography}\endgroup

\end{document}